\documentclass[%
preprint, 11pt, tightenlines, 
preprintnumbers,
nofootinbib,
 amsmath,amssymb,
 aps,
notitlepage
]{revtex4-2}

\usepackage{graphicx}
\usepackage{here}
\usepackage{multirow}
\usepackage{dcolumn,longtable}
\usepackage{bm,xfrac}
\usepackage[dvipsnames]{xcolor}
\usepackage{array}
\usepackage{bbold}
\usepackage{ulem,cancel}
\usepackage{hyperref}
\hypersetup{
    colorlinks=true,
    linkcolor=blue,
    filecolor=magenta,      
    urlcolor=cyan,
    pdftitle={Overleaf Example},
    pdfpagemode=FullScreen,
    }
\usepackage{comment}
\newcolumntype{C}[1]{>{\vspace{1.5pt}\centering\let\newline\\\arraybackslash\hspace{0pt}}m{#1}}

\newcommand \beq{\begin{eqnarray}}
\newcommand \eeq{\end{eqnarray}}
\newcommand{\nn}{\nonumber\\ }
\newcommand{\rme}{{\mathrm e}}
\newcommand{\rmd}{{\mathrm d}}
\newcommand{\del}{\partial}

\def\x{{\boldsymbol x}}
\def\r{{\boldsymbol r}}

\def\bra#1{\langle#1\vert}
\def\ket#1{\vert#1\rangle}
\def\L{\mathcal{L}}

\begin{document}


\title{Quarkonium dynamics in the quantum Brownian regime with non-abelian quantum master equations}

\author{Stéphane Delorme}
\email{stephane.delorme@ifj.edu.pl}
\affiliation{Institute of Nuclear Physics, Polish Academy of Sciences, ul. Radzikowskiego 152, 31-342, Krak\'ow, Poland\\IMT Atlantique, Nantes Université, CNRS/IN2P3, SUBATECH, 44307, Nantes, France}%
\author{Roland Katz}%
\author{Thierry Gousset}%
 \author{Pol Bernard Gossiaux}%
\email{gossiaux@subatech.in2p3.fr}
\affiliation{IMT Atlantique, Nantes Université, CNRS/IN2P3, SUBATECH, 44307, Nantes, France}%
\author{Jean-Paul Blaizot}
\email{jean-paul.blaizot@ipht.fr}
\affiliation{Universit\'e Paris-Saclay, CNRS, CEA, Institut de physique th\'eorique, 91191, Gif-sur-Yvette, France}

\date{\today}

\begin{abstract}
We present numerical solutions in a one-dimensional setting of quantum master equations that have been recently derived. We focus on the dynamics of a single heavy quark-antiquark pair in a Quark-Gluon Plasma in thermal equilibrium, in the so-called  quantum Brownian regime where the temperature of the plasma is large in comparison with the spacing between the energy levels of the $Q\bar Q$ system. The one-dimensional potential used in the calculations has been adjusted so as to produce numbers that are relevant for the phenomenology of the charmonium. The  equations are solved using different initial states and medium configurations. Various temperature regimes are studied and the effects of screening and collisions thoroughly analyzed.  Technical features of the equations are analyzed. The contributions of the different operators that control the evolution are discussed as a function of the temperature.  Some phenomenological consequences are addressed.

\end{abstract}

\preprint{IFJPAN-IV-2024-3}
\maketitle
\newpage

\section{\label{sec:intro}Introduction}
The quarkonia, bound states of heavy quark-antiquark pairs, have been used for a long time to diagnose the properties of matter produced in ultrarelativistic heavy ions collisions \cite{Andronic:2024}.
The dynamics of bound states are indeed very much affected by the presence of a medium in which they propagate: They can in particular disappear due to screening or collisional effects if the temperature of the medium is high enough; If the number of $Q\bar Q$ pairs produced in a given collision is large, one may anticipate that these pairs recombine into quarkonia during the evolution of the system. Thus, the measurements of the properties of quarkonia produced in ultrarelativistic heavy ion collisions can provide information on the state of the medium.  And indeed many experimental data provide clear evidence that quarkonia are strongly affected by the matter produced in high energy nuclear collisions \cite{Andronic_2016}. 

Extracting significant information from these data is however not a simple task. We are facing a rather complicated many-body problem, which requires going beyond the elementary approaches which have been used in the past. In recent years, numerous efforts have been made to adapt to this complicated problem tools that have been developed in the study of open quantum systems~\cite{Young:2010jq,Borghini:2011ms,Akamatsu:2011se,Akamatsu:2012vt,Katz:2013rpa,Akamatsu:2014qsa,Katz:2015qja,Blaizot:2015hya,Brambilla:2016wgg,Brambilla:2017zei,DeBoni:2017ocl,Blaizot:2017ypk,Blaizot:2018oev,Yao:2018sgn}. Such approaches generically lead to Quantum Master Equations (QME) that capture the most important effects  of the medium on the dynamics of the $Q\bar Q$ heavy quark pairs (for recent reviews see \cite{Rothkopf_2020,Yao:2021lus,Akamatsu:2020ypb}. Generically two regimes can be identified, depending on whether the bound states of the $Q\bar Q$ pairs are an important part of the dynamics, or not. Which regime occurs depends on the temperature $T$. The so-called ``Quantum Optical'' regime corresponds typically  to ``low'' temperatures, $T\lesssim \Delta E$, where $\Delta E$ denotes the typical energy difference between two eigenstates of the (in-medium) Hamiltonian. In this regime, the effect of the medium can be seen as a perturbation applied to the bound states of the system, modifying their binding energies and contributing  (small) widths to the states. This is a regime where other approaches, such as transport approaches  based on rate equations can also prove useful \cite{Grandchamp:2003uw,Zhao:2010nk,Du:2017qkv,Zhou:2014kka,Liu:2010ej,Yao:2020xzw,Yao:2018sgn}. In the other regime, which takes place at ``high'' temperatures, $T\gtrsim \Delta E$, the binding forces represent a small perturbation of the dynamics of nearly independent heavy quarks subjected to the influence of stochastic forces generated by the collisions with the plasma constituents. This regime is commonly referred to as the Quantum Brownian Motion (QBM), and this is the regime we shall be concerned with in this paper. 

These two regimes are usually  described by two different types of master equations, and the way to analyze the corresponding physics refers usually to two different basis. In the optical regime, the basis of localized bound states is the most appropriate one. In the QBM regime however, the analysis is better carried out in coordinate space, which plays the role of a ``preferred basis''. This is because the operators that represent the effects of the collisions of the heavy quarks with the plasma constituents are local operators, and are (nearly) diagonal in the coordinate space basis. In this basis, the collisions bring the density matrix of the $Q\bar Q$ system to a diagonal form in a relatively short time scale, which is not necessarily the case in other basis. This entails simplifications in the description of the dynamics, suggesting in particular the usefulness of semi-classical approximations\cite{Young:2008he,Blaizot:2017ypk}. The dynamics of the color degrees of freedom remain fully quantum however, and they are treated as equations that couple the different color components (singlet and octet) of the $Q\bar Q$ density matrix. 

Common to all the derivations of the QME in the present context is the fact that the mass of the heavy quarks is large, which implies that their motion is slow on the time scale of collective plasma dynamics. This is used to obtain Markovian equations, which in most approaches take the form of Lindblad equations. In the case of quantum chromodynamics (QCD), further simplifications have been implemented, relying on the use of effective field theories. In addition to the large quark mass, such approaches focus on the dipolar nature of the microscopic interactions, and  exploit various scale hierarchies that allow for systematic and controlled approximations.  This is the case of the non-relativistic QCD (NRQCD) approach~\cite{Akamatsu:2012vt,Akamatsu:2014qsa,Blaizot:2015hya,Blaizot:2017ypk},  and  the so-called potential non-relativistic QCD (pNRQCD) approach~\cite{Brambilla:2016wgg,Brambilla:2017zei,Yao:2018sgn,Brambilla:2020qwo,Brambilla:2021wkt,Brambilla:2022ynh}. Various strategies have been followed to solve the resulting equations for one single $Q\bar{Q}$ pair: by transforming the quantum master equations into Schrödinger equations with a complex potential~\cite{Kajimoto:2017rel,Sharma:2019xum,Akamatsu:2021vsh} or into stochastic Schrödinger equations~\cite{Miura:2019ssi,Miura:2022arv}, by expanding them to the first spherical harmonics~\cite{Brambilla:2016wgg,Brambilla:2017zei,Brambilla:2020qwo,Brambilla:2021wkt,Brambilla:2022ynh}, by performing a direct resolution in one dimension and limited to the quantum electrodynamics (QED) case~\cite{DeBoni:2017ocl,Alund:2020ctu} or by reducing them to semi-classical Langevin equations~\cite{Blaizot:2015hya, Blaizot:2017ypk,Blaizot:2018oev}. The latter is of particular interest, as it allows to treat multiple quark-antiquark pairs simultaneously, which has not been done yet with a full quantum equation, except in the abelian case \cite{Blaizot:2015hya}.

In this paper, we present the  full numerical solution of the QME derived in~\cite{Blaizot:2017ypk}, in the QBM regime and for the particular case of a single $Q\bar Q$ pair, using a potential tuned specifically for such one-dimensional studies.  It is the first time that the complete  solution of these particular  equations is obtained. We note however that these equations have essentially the same physics content as those studied by different means in Ref.~\cite{Miura:2022arv}. A one-dimensional setting limits some aspects of the dynamics, but as we shall see, it provides interesting information on the evolution of a $Q \bar Q$ in a medium in thermal equilibrium. It allows us to identify what we consider as robust features that are likely to survive in a more realistic three-dimensional setting, while being much less computationally demanding.  

The paper is organized as follows. The next section, Sec.~\ref{sec:1},  is an introductory section where we present  a brief summary of the QME that have been derived in~\cite{Blaizot:2017ypk}. These equations describe the evolution of the reduced density matrix of heavy quark-antiquark pairs coupled to a quark gluon plasma in thermal equilibrium. These equations are written in the coordinate space representation, and take the form of a Lindblad equation, involving a set of four Liouvillian superoperators $\L_i$.  We  discuss the various regimes that these equations can describe as a function of the temperature of the plasma, and we provide estimates of the characteristic temperatures that delineate these various regimes, based on the parameters of the specific one-dimensional potential that we use. The main body of the paper is divided into two sections. In the first one, Sec.~\ref{sec:2}, we present the results obtained by solving the QME for a single charm-anticharm pair with different initial conditions for the pair, and various temperatures of the plasma. Assuming a medium with fixed temperature, we first study the evolution and late stage steady state of a pair initially in a color singlet eigenstate of the screened Hamiltonian. The evolution of the density matrices, the marginal distributions and the instantaneous projections on the vacuum eigenstates are analyzed in detail. We then investigate more systematically the effect of the medium temperature by either starting from a 1S-like vacuum eigenstate or from a more compact Gaussian singlet state. The impact of the initial state on the evolution toward equilibrium is further analyzed by testing an initial P-like color octet state. Most of the analysis of this section involves a medium of fixed temperature. At the end of the section we consider the modification of the dynamics that take place in an expanding plasma, assuming a simple one-dimensional Bjorken expansion~\cite{Bjorken:1982qr}. The second major section  of the paper, Sec.~\ref{sec:3},  addresses several more ``technical" issues related to generic properties of the QME. In Sec.~\ref{sec:3.A}, we inspect the role of the various superoperators of the Lindblad equation at various stages of the evolution and for the various temperature regimes introduced in Sec.~\ref{sec:1}. A particular discussion concerns the higher order superoperator $\mathcal{L}_4$ in ensuring the positivity of the density matrix evolution. Sec.~\ref{sec:3.B} is dedicated to the study  of the asymptotic distributions toward which the ${\rm c\bar{c}}$ pair relaxes at late time. We discuss the similarities and differences between the  QCD case and the corresponding abelian  case, and compare the steady states of the QME to thermal equilibrium distributions. The last section of the paper summarizes our conclusions. Seven appendices contain technical material and additional developments. The explicit expressions of the Lindblad superoperators for the case of a single $Q\bar{Q}$ pair are given in appendices~\ref{app:positivity}, \ref{app:L4}, and  \ref{app:transition_operators}, where an explicit check of the positivity in the evolution is carried out, as well as a discussion on the regularization adopted for the imaginary part of the potential. The main features of the complex one-dimensional potential from \cite{Katz:2022fpb} are recalled in App.~\ref{app:numerics}. The corresponding dissociation rates are calculated  in App.~\ref{app:dissocrates}. The following appendix~\ref{app:L2alone} is a standalone study of the evolution under the single superoperator $\L_2$. The last appendix~\ref{app:equilibrium} presents the details of the deviation of the asymptotic momentum distribution obtained solving the QME from a thermal one. 

\section{Equations for the $Q\bar{Q}$ density matrix}
\label{sec:1}
In this section, we start by reviewing briefly the quantum master equations (QME) that were established in~\cite{Blaizot:2017ypk}, and that we shall be solving in this paper in a simplified one-dimensional setting. We recall the assumptions under which these equations were derived and comment on their range of validity. In particular we emphasize that they are appropriate in the so-called Quantum Brownian Motion (QBM) regime.\footnote{In particular we ignore the effect of the energy gap in the various transitions between states \cite{Blaizot:2021xqa}.} Although this will not be pursued here we also note that they constitute a natural step toward semi-classical approximations,\footnote{Such approximations will be discussed more thoroughly in a forthcoming paper~\cite{Daddi:2024}.} which have been explored in details in the abelian case in \cite{Blaizot:2015hya}. In fact, the effect of the collisions, which are part of the QME, is  essentially classical, as already pointed out in~\cite{Laine:2006ns}.

The calculations to be presented in this paper concern the time evolution of a single $Q\bar Q$ pair immersed in a quark-gluon plasma which is in thermal equilibrium at a given temperature. In the second part of this section, we identify the characteristic temperatures that delineate the various regimes that can be identified in this particular context. The calculations are based on a simplified complex one-dimensional potential that has been adjusted so that numbers that come out of the calculations are in a physically relevant range for the corresponding three-dimensional systems, in particular for the physics of the charmonium~\cite{Katz:2022fpb}. 

\subsection{\label{sec:1.1b} Master equations}

The equations that we want to solve describe the time evolution of the density matrix of a $Q\bar Q$ pair in a quark-gluon plasma in equilibrium at temperature $T$. In the position representation and in one spatial dimension, the reduced density matrix, obtained by integrating out the plasma degrees of freedom, takes the form 
\beq\label{eq:sDs}
\bra{r_1 ,r_2}{\mathcal D}\ket{r_1', r_2'}= \bra{R+{s}/{2},R-{s}/{2} }{\mathcal D}\ket{R'+s'/2,R'-s'/2}
\eeq
where we have introduced center of mass ($R,R'$)  and relative ($s,s'$)  coordinates of the two members of the $Q\bar Q$ pair. In fact we shall only study here the relative motion, and reduce further the density matrix by integrating out the center of mass. Thus, in the space of relative coordinates, and with a slight abuse of notation, we set
\beq\label{eq:sDsp}
\bra{s}{\mathcal D}\ket{s'}=\int_R\bra{R+{s}/{2},R-{s}/{2} }{\mathcal D}\ket{R+s'/2,R-s'/2}. 
\eeq
Thus defined, the reduced density matrix $\mathcal D$ remains a matrix in color space which we may conveniently decompose into singlet and octet components \cite{Blaizot:2017ypk}
\beq\label{eq:DsDodef}
\mathcal{D} = \mathcal{D}_{\mathrm s}\left|\mathrm s\right>\left<\mathrm s\right| + \mathcal{D}_{\mathrm o}\sum_{i=1}^{N_c^2-1}\left|\mathrm o^{i}\right>\left<\mathrm o^{i}\right|,
\eeq
where $\left|\mathrm s\right>$ denotes a color singlet state and $\left|\mathrm o^{i}\right>$ a projection of a color octet state in the Hilbert space of the quark-antiquark pair.

The equations satisfied by $\mathcal{D}_{\mathrm s}$ and $\mathcal{D}_{\mathrm o}$ can be written in the form~\cite{Blaizot:2017ypk}
\begin{equation}\label{eq:2.1}
    \frac{\rmd}{\rmd t}\begin{pmatrix}\mathcal{D}_{\mathrm s}\\\mathcal{D}_{\mathrm o}\end{pmatrix} = \mathcal{L}\begin{pmatrix}\mathcal{D}_{\mathrm s}\\\mathcal{D}_{\mathrm o}\end{pmatrix}.
\end{equation}
The Liouville operator $\mathcal{L} = \sum_{i=0}^4 \mathcal{L}_{i} $ (also referred to as a superoperator)  is  a $2\times 2$ matrix in the space spanned by $\mathcal{D}_{\mathrm s}$ and $\mathcal{D}_{\mathrm o}$.\footnote{The matrix elements of $\mathcal{L}$ in this two-dimensional space will be denoted $\mathcal{L}^{\rm ss}$, $\mathcal{L}^{\rm so}$, $\mathcal{L}^{\rm os}$ and $\mathcal{L}^{\rm oo}$.} The expressions of  the  superoperators $\mathcal{L}_i$,  before color projection and center of mass elimination, are given by
\begin{align}
    &\mathcal{L}_{0}\mathcal{D} = - i\left[H_{Q},\mathcal{D}\right] \nonumber\\
    &\mathcal{L}_{1}\mathcal{D} = -\frac{i}{2}\int_{xx'}V(x-x')\left[n_{x}^{a}n_{x'}^{a},\mathcal{D}\right] \nonumber\\
    &\mathcal{L}_{2}\mathcal{D} = \frac{1}{2}\int_{xx'}W(x-x')\left(\left\{n_{x}^{a}n_{x'}^{a},\mathcal{D}\right\} - 2n_{x}^{a}\mathcal{D}n_{x'}^{a}\right) \nonumber\\
    & \mathcal{L}_{3}\mathcal{D} = -\frac{i}{4T}\int_{xx'}W(x-x')\left(\dot{n}_{x}^{a}\mathcal{D}n_{x'}^{a} - n_{x}^{a}\mathcal{D}\dot{n}_{x'}^{a} + \frac{1}{2} \left\{\mathcal{D},\left[\dot{n}_{x}^{a},n_{x'}^{a}\right]\right\}\right)\nn
    & \mathcal{L}_{4}\mathcal{D} = \frac{1}{32 T^2}\int_{xx'}W(x-x')\left(\{\dot{n}_{x}^{a}\dot n_{x'}^{a},\mathcal{D}\} -  2\dot n_{x}^{a}\mathcal{D}\dot n_{x'}^{a}\right).
   \label{eq:2.2} 
\end{align}
In these equations, $H_{Q}$ is the free quark Hamiltonian (i.e., the kinetic energy), $V$ and $W$ are respectively the real and imaginary parts of a complex potential, and $n_{x}^{a}$ is the color charge density operator at location $x$: 
\beq\label{colordensity}
n^a_x=\delta(x-\hat r)\, t^a\otimes \mathbb{I} -\mathbb{I}\otimes\delta(x-\hat r) \, \tilde t^a,
\eeq
where $\hat r$ is the position operator, and  the two components of the tensor product refer respectively to the Hilbert spaces of the heavy quark for the first component and the heavy antiquark for the second component. In Eq.~(\ref{colordensity}), $t^a$ is a color matrix in the fundamental representation of SU(3) and $\tilde t^a$ denotes its transpose. The matrix elements of the time derivative of $n_x=\delta(x-\hat r)$  are given by \cite{Blaizot:2017ypk}
 \beq\label{divjx}
\bra{r}\dot n_x\ket{r'}=-\frac{1}{2iM}\left\{\left[\partial_r \delta(r-r')\right] \cdot\partial_x [\delta(x-r')+\delta(x-r)]\right\}, 
\eeq
where $M$ is the mass of the heavy quark. Thus the presence of a time derivative of the density in Eqs.~(\ref{eq:2.2}) brings in a factor $1/M$ and two spatial derivatives in the final expressions of the corresponding coordinate space representation of the superoperators (see App.~\ref{app:transition_operators} for details).

In writing Eqs.~(\ref{eq:2.2}) we have used a specific time discretization discussed in appendices A and B of~\cite{Blaizot:2017ypk}. This discretization allows us to deduce the sum of operators $\mathcal{L}_{2}+\mathcal{L}_{3}+\mathcal{L}_{4}$ by making, in the expression of $\mathcal{L}_{2}$ given in Eq.~(\ref{eq:2.2}), the following substitutions
\begin{align}\label{eq:substitutionbis}
    n_{\x}^{a}\longrightarrow \left(n_{\x}^{a} - \frac{i}{4T}\dot{n}_{\x}^{a}\right),\qquad  n_{\x'}^{a}\longrightarrow \left(n_{\x'}^{a} - \frac{i}{4T}\dot{n}_{\x'}^{a}\right).
\end{align}
The operators $\mathcal{L}_{3}$ and $\mathcal{L}_{4}$ that result from this substitution collect the terms that are respectively linear and quadratic in the time derivative of the density. This procedure has the advantage to put the evolution equation for the density matrix into a Lindblad form. This form is explicit in the expression of $\mathcal{L}_{2}$ (and that of $\mathcal{L}_{4}$), but not in the expression of $\mathcal{L}_{3}$. In~\cite{Blaizot:2017ypk} it was argued that the operator $\mathcal{L}_{4}$, which contains two derivatives of the density, is subleading (by a factor $1/MT$) and it was not considered in the various applications discussed there. This operator is however necessary to guarantee the positivity of the evolution (i.e. making sure that the density matrix remains a positive definite matrix at all times or, equivalently, that the probability to find the system in any arbitrary state is non-negative) and it will be included in the present work. The question of positivity and the role of this operator is discussed in Sec.~\ref{sec:3.A} and in App.~\ref{app:positivity}.  The explicit expressions of the operators $\mathcal{L}_{i}$ used in this paper are given in App.~\ref{app:transition_operators} in terms of derivatives of the complex potential.

In Eqs.~(\ref{eq:2.2}) the operator $\mathcal{L}_{0}$ describes the free motion of the quarks (i.e. their kinetic energy), while the operator $\mathcal{L}_{1}$ describes the attractive or repulsive effects of the real potential $V$. Together, the operators $\mathcal{L}_{0}$ and $\mathcal{L}_{1}$ describe a unitary evolution of the $Q\bar Q$ system. In contrast, the other operators, which involve the imaginary potential $W$, imply a non-unitary evolution.  The operator $\mathcal{L}_{2}$ captures  the effect of the random kicks due to collisions of the heavy quarks with the plasma constituents which lead to diffusion in momentum space and collisional decoherence. The operator $\mathcal{L}_{3}$, linear in the velocity, accounts for friction and dissipation. The operator $\mathcal{L}_{4}$, brings in corrections which are quadratic in the velocity and which guarantee the positivity of the evolution.\footnote{See remark after table \ref{table:structure2} in App.~\ref{app:L4}.}

In deriving these equations the fact that the mass of the heavy quarks is large, i.e. $M\gg T$, was exploited. This allowed to treat the heavy quarks non relativistically, and to simplify their mutual interaction, as well as their interactions with the medium constituents (essentially keeping the Coulomb interaction and ignoring the magnetic interactions). Aside from screening effects, which are included in the real potential $V$, the other interactions of the heavy quarks with the medium constituents can be viewed as collisions. These  involve essentially one-gluon exchanges that take place over a typical time $\tau_D\sim 1/m_D$, where the Debye screening mass $m_D$ characterizes the plasma collective dynamics, such as the screening effects included in $V$. An important step in the derivation rests on the fact that we are interested on the evolution of the $Q\bar Q$ pair on times scales that are large compared to $\tau_D$. Under such circumstances, the low-frequency part of the plasma response is filtered out, and one can treat the collisions as if they were essentially instantaneous (hence their representation by an imaginary ``potential''). 

\subsection{Various regimes and characteristic temperatures}
 
We shall now elaborate more on the specific features of the complex potential that we use in this paper~\cite{Katz:2022fpb}. As already mentioned, its  form and parameters are adjusted so as to produce numbers (for binding energies, damping rates, etc.) that are of comparable magnitudes with those one would obtain in a more realistic situation based on a three dimensional potential. Note that the constraints of the fitting procedure make the potential well defined only above some non-vanishing temperature ($T\gtrsim 120$ MeV), and still appears to present some non-physical artifact below $T=175~{\rm MeV}$. As we shall restrict our studies to $T\gtrsim 200$~MeV, which  corresponds to the range of validity of the master equations recalled in the previous subsection -- as we shall verify shortly --, these non-physical features will be fairly innocuous.

The potential to be used departs from that obtained in weak coupling, including HTL resummations, by taking into account in particular string breaking at large distance  or other non perturbative information obtained from lattice calculations. However, its main component remains inspired by the form of the weak  coupling potential  in three dimensions, whose general expression, for a quark-antiquark pair, reads\footnote{As written the complex potential of Eq.~(\ref{eq:VHTL3D}) is that of an abelian theory. In QCD, the actual potential depends on the color state of the $Q\bar Q$ pair on which it is acting. To get, for instance, the potential acting  on a color singlet $Q\bar Q$ pair, one needs to multiply $V+iW$ in Eq.~(\ref{eq:VHTL3D}) by $C_F=(N_c^2-1)/2N_c$. The precise color factors entering the superoperators of Eq.~(\ref{eq:2.2}) are given in App.~\ref{app:transition_operators}. See also App.~\ref{app:numerics}.} \cite{Laine:2006ns} (see also \cite{Beraudo:2007ky})
\begin{equation}\label{eq:VHTL3D}
V_{_{\!\mathrm{HTL}}}(r,T) +i W_{_{\!\mathrm{HTL}}}(r,T)= -\alpha \Biggl[m_D + \frac{e^{-m_{D}r}}{r}\Biggr]+i\alpha  T\phi(m_D r),     
\end{equation}
where 
\begin{equation}\label{eq:VHTL3DPhi}
\phi(x)=2\int_{0}^{\infty}\mathrm{d}z\;\frac{z}{\left(z^2+1\right)^2}\left(1-\frac{\sin(zx)}{zx}\right). 
\end{equation}
The function $\phi(x)$ increases monotonously from $0$ to $1$ as $x$ runs from $0$ to $\infty$, and it behaves as $\phi(x)\sim x^2$ at small $x$. This particular small distance behavior is characteristic of the underlying  dipolar transitions associated with gluon emission or absorption. 

The complex potential (\ref{eq:VHTL3D}) depends on the plasma properties only through the Debye screening mass $m_D$.  As mentioned in the previous subsection, in the regime where the master equations (\ref{eq:2.2}) are valid, only the static response of the plasma is needed, and this is indeed entirely controlled by the Debye mass. At weak coupling $m_D\sim gT$ where $g$ is the gauge coupling,  supposed to be small, so that $m_D\ll T$. However, since at the temperatures of interest $g$ can be of order unity, the magnitude of the Debye mass can be of the same order as the temperature. We shall therefore  set $m_D=CT$, with $C$ a coefficient of order unity (in practice, $C\simeq 2$).  

The real part of the potential $V_{_{\!\mathrm{HTL}}}(r,T)$ in Eq.~(\ref{eq:VHTL3D}) exhibits screening at a scale characterized by the Debye mass $m_D$. At very short distance, i.e. for $r$ such that $m_D r\ll 1$,  screening plays no role and the potential indeed identifies with a Coulomb potential, $V_{_{\!\mathrm{HTL}}}(r)\simeq -\alpha /r$. At large distance, the potential goes to a constant, $V_{_{\!\mathrm{HTL}}}(r)\to -\alpha m_D$. This constant can be associated to a screening correction to the heavy quark mass. It controls the edge of the continuum where a bound state becomes  degenerate with the energy of two independent quarks  with  mass $M-\alpha m_D/2\equiv M(T)$. 
\begin{figure}[htb!]
\centering
\includegraphics[width=0.6 \textwidth]{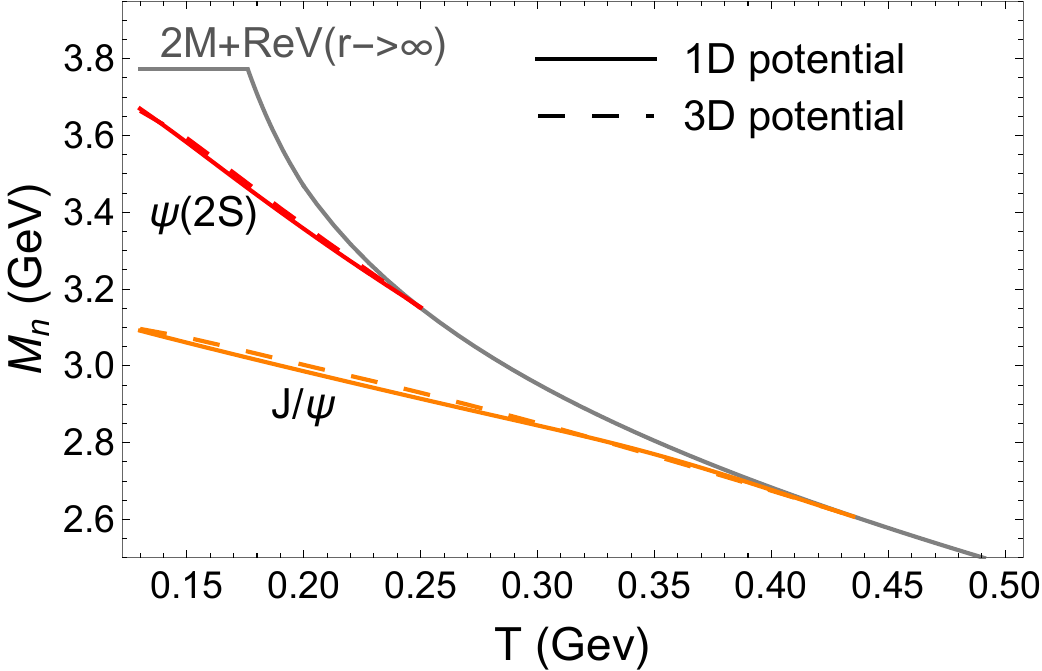}
\caption{Charmonium mass spectrum of the S states for the one-dimensional potential (solid lines) obtained from the expectation values of the Hamiltonian compared to the three-dimensional case (dashed lines). }
\label{fig:MassSpectra1D}
\end{figure}

\begin{figure*}[t]
\centering
\includegraphics[width=0.49\linewidth]{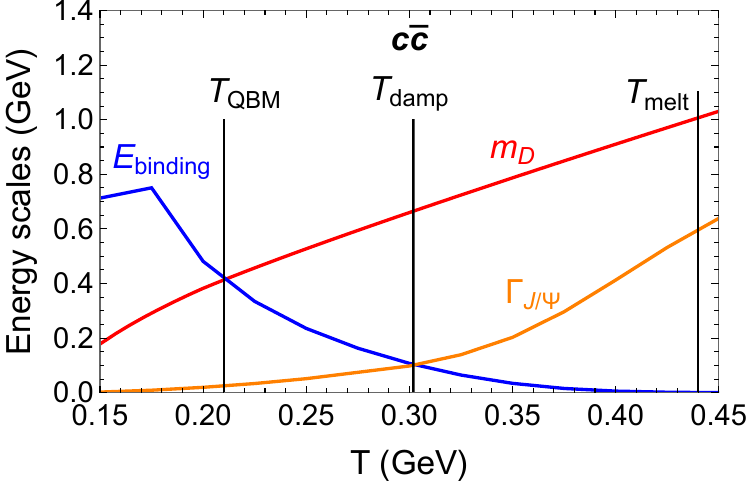}
\includegraphics[width=0.49\linewidth]{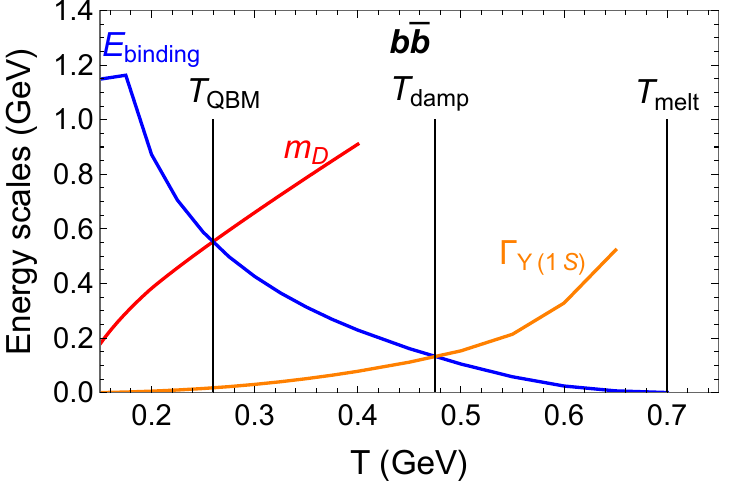}
\caption{Left: Various energy scales (in GeV) obtained from our 1D-potential construction~\cite{Katz:2022fpb}, in the case of a ${\rm c\bar{c}}$ pair in a 1S state. An estimation of the temperatures $T_{\mathrm{melt}}$, $T_{_\mathrm{QBM}}$ and $T_{\mathrm{damp}}$ is also provided (see text for details). Right: same for a ${\rm b\bar{b}}$ pair.}
    \label{fig:scaleanalysis}
\end{figure*}

These generic features of the real potential  of Ref.~\cite{Katz:2022fpb} are well illustrated in Fig.~\ref{fig:MassSpectra1D}. The Debye mass parameterization is taken directly from~\cite{Lafferty:2019jpr} and satisfies $m_D\approx C T $ with $C\approx 2$ for $T>200$~MeV.  Screening manifests itself by the decrease of the binding energy  of the various bound states with increasing temperature.\footnote{In fact, for the particular potential that we use, the parameter $V(T,r\to \infty)$ in Eq. (\ref{eq:massquarkonia}) is kept constant up to $T=175$ MeV. This results in a small artifact: The binding energies slowly increases with the temperature up to $T=175$ MeV (see Fig.~\ref{fig:scaleanalysis}), beyond which the binding energies decrease as expected.} The binding energy $E_n(T)$ is defined as the eigenvalue of the non-relativistic Hamiltonian
\begin{equation}
H=\frac{P^2}{M}+V(T,r)-V(T,r\to\infty),\qquad H\ket{\psi_n}=E_n\ket{\psi_n}
\label{eq:in-medium-hamiltonian}
\end{equation}
where the subtraction of $V(T,r\to\infty)$ removes the screening correction to the heavy quark mass. The total mass, $M_n$  of the $Q\bar Q$ system in an eigenstate $|\psi_n\rangle$ of the non-relativistic Hamiltonian $H$, is given by 
\beq
M_n&=& 2 M+\bra{\psi_n} \frac{P^2}{M}+V(T,r) \ket{\psi_n} \nn &=&E_n+2 M+V(T,r\to\infty)=E_n+2 M(T)
\label{eq:massquarkonia}
\eeq
When the binding energy vanishes, $M_n=2 M+V(r\to\infty)=2 M(T)$, generalizing the expression given above for the effective mass $M(T)$.  At the temperature for which the binding energy vanishes, the mass of the $Q\bar Q$ system is equal to twice the effective mass of two heavy quarks at the same temperature (and in the same overall color state). This temperature marks the disappearance of the bound state. It will be referred to as the ``melting temperature'',\footnote{In order to avoid dealing with too loosely bound states, $T_{\mathrm{melt}}$ will be operationally defined as the temperature at which $E_{\rm binding}\lesssim 1{\rm MeV}$.} and denoted $T_{\mathrm{melt}}$. Note that $T_{\mathrm{melt}}$ depends on the state considered: It decreases as the vacuum binding energy of the state decreases.
Thus, in Fig.~\ref{fig:MassSpectra1D}, the 2S melts at a lower temperature than the 1S ($T_{\mathrm{melt}}$ is approximately 440 MeV for the 1S and 250 MeV for the 2S).\footnote{As explained in App.~\ref{app:numerics} we refer to the bound states of the 1D potential  by using the three-dimensional conventions, although there is of course no angular momentum in one dimension.} Note that the screening correction to the heavy quark mass does not enter the calculation of the binding energy, and plays therefore no role in the determination of $T_{\mathrm{melt}}$.
 
The real part of the potential is involved in the unitary part of the evolution. In terms of the Liouville operators acting on the density matrix, this unitary evolution is driven by ${\mathcal L}_0+{\mathcal L}_1$. 
In the low temperature regime, where the bound states can be clearly identified, these  bound states form a convenient basis to analyze the density matrix.\footnote{Note however that the continuum states play an important role. Depending upon the initial conditions these continuum modes may be populated and be part of the unitary evolution described by ${\mathcal L}_0+{\mathcal L}_1$.} 
In this low temperature regime, the motion of the heavy quarks in the bound state is fast compared to the plasma collective dynamics. Let us characterize the time scale of the relative motion by $\tau_{_{\rm BS}}\sim 1/E_{_{\rm BS}}$, where $E_{_{\rm BS}}$ is the binding energy. As recalled earlier, the time scale of plasma dynamics is $\tau_D\sim 1/m_D$, with $m_D$ the Debye mass. 
The condition  $\tau_{_{\rm BS}}\sim \tau_D$, or equivalently $E_{_{\rm BS}} \sim m_D$,  characterizes the transition region between the ``low temperature'' and the ``high temperature'' regimes. We shall call $T_{_\mathrm{QBM}}$ the corresponding temperature (see Fig.~\ref{fig:scaleanalysis}). Below $T_{_\mathrm{QBM}}$, the bound states remain a dominant feature of the dynamics, which can be understood in terms of small modification of the bound state properties and the various transitions between these bound states. 
In this paper we shall be interested in the high temperature regime, i.e. in the regime where  
$T>T_{_\mathrm{QBM}}$. Then, $\tau_{_{\rm BS}}>\tau_D$, and the motion of the heavy quarks is slow on the scale of the plasma dynamics. The perturbations caused in the plasma by the motion of the heavy quarks are nearly instantaneously damped. This high temperature regime is often referred to in the literature as the QBM regime. This is the regime for which the equations~(\ref{eq:2.2}) are supposed to be valid. In the present case, with the one-dimensional potential that we use, $T_{_\mathrm{QBM}}\simeq 210$~MeV for a $J/\Psi$ state, as can be read from the left panel of Fig.~\ref{fig:scaleanalysis}.
The transition to the QBM regime is of course not a sharp one, and we could as well define $T_{_\mathrm{QBM}}$ as the temperature at which the binding energy and the temperature are approximately equal. From Fig.~\ref{fig:scaleanalysis}, one sees that the corresponding temperature is slightly higher, about 240 MeV. In this paper we shall stick to the value defined as above, i.e. 210 MeV,  keeping in mind that, at this temperature, effects of bound states can still be visible. We shall find that it is indeed the case with the simulations performed at $T$= 200 MeV.

We turn now to the effects of collisions which are responsible for the non-unitary part of the evolution. In weak coupling regime, the rate of such collisions is of the order of $\alpha T$, the order of magnitude of the imaginary potential of Eq.~(\ref{eq:VHTL3D}). Defining the collision time $\tau_R\sim 1/\alpha T$ and the plasma time $\tau_D\sim 1/m_D$, we get for the ratio
\beq 
\frac{\tau_R}{\tau_D}=\frac{C}{\alpha}\gg 1, \qquad  (m_D=C T).
\label{eq:taurlargeT}
\eeq 
Thus, independently of the value of the temperature, the collision time is large compared to $\tau_D$ (even in  strict weak coupling): Relaxation processes governed by collisions are slow on the time scale of the plasma collective dynamics. However, 
the damping rate of the bound state depends strongly on the size of the $Q\bar Q$ pair, in particular it vanishes for small size, due to the dipolar nature of the interaction. To take this into account, we calculate the decay width $\Gamma$ of the 1S eigenstate of the temperature dependent potential (see App.~\ref{app:dissocrates}) and  plot it in Fig.~\ref{fig:scaleanalysis}. One sees that the decay width increases with temperature. This increase is the combination of several effects: one is the increase of the overall magnitude of the imaginary potential, proportional to $\alpha T$, another is the temperature dependence of the function $\phi(m_D r)\simeq m_D^2 r^2$  of the imaginary potential (see Eq.~(\ref{eq:VHTL3DPhi})).   
 For a parametric estimate which takes approximately these two effects into account, we may use 
 \beq
 \Gamma\simeq \alpha T (a_0 m_D)^2\simeq 4C^2 T^3/(\alpha M^2), 
\label{eq:taursmallT}
\eeq 
where $a_0$ denotes the typical size of the bound state. 
Still another effect is the increase of this bound state size that accompanies the decrease of the binding energy.\footnote{Note that at low temperature ($T\lesssim T_{_\mathrm{QBM}}$) the calculation overestimates the decay width since it ignores the effect of the energy gap between levels. See~\cite{Blaizot:2021xqa} for a recent discussion of this effect.} Since the size of the bound state increases with the temperature (see e.g. Fig.~\ref{fig:spectrumD}),  the rate $\Gamma$ varies more rapidly with temperature than predicted by Eq.~(\ref{eq:taursmallT}), which appears to be the case as illustrated in Fig.~\ref{fig:decayrate2}. 

The point where the decay width becomes equal to the binding energy defines  another characteristic  temperature  $T=T_{\mathrm{damp}}$, with $T_{_\mathrm{QBM}}< T_{\rm damp}< T_{\rm melt}$. This temperature marks the beginning of an overdamped regime where one expects  the $Q\bar Q$ spectral function to become structureless. 

In summary, we have identified three important temperatures. The first one is $T_{_\mathrm{QBM}}$, defined as the temperature at which the binding energy equals the Debye mass. This temperature marks the onset of the Quantum Brownian Motion regime. The next temperature is $T_{\mathrm{damp}}(> T_{_\mathrm{QBM}})$ defined as the temperature at which the binding energy equals the decay width. Above that temperature, it is expected that the $Q\bar Q$ spectral function becomes structureless. Finally, above the temperature $T_{\mathrm{melt}}$ the binding energy vanishes.  Note that the two temperatures $T_{_\mathrm{QBM}}$ and $T_{\mathrm{melt}}$ depend only on the Debye mass and the real part of the potential. Only $T_{\mathrm{damp}}$ depends on collisions, i.e. on the imaginary part of the potential. Finally, although the results of the next section will concern solely the physics of charmonium, we present in the right panel of Fig.~\ref{fig:scaleanalysis} results corresponding to the bottomonium. It can be seen that the same pattern is observed and the same characteristic scales emerge, to within an overall  rescaling of the temperature. 

\section{Main physics results}
\label{sec:2}
In this section, we study the time evolution of the  density matrix $\langle s|\mathcal{D}|s'\rangle$ of a single ${\rm c\bar{c}}$ pair, in the form given in Eq.~(\ref{eq:sDsp}),  where the center of mass degrees of freedom are traced out and $s$ and $s'$ refer to relative coordinates.\footnote{Note that the center of mass momentum does not enter explicitly the expressions of the operators $\L_i$, except $\L_4$ (see App.~\ref{app:transition_operators}).} We examine various contexts. In Sec.~\ref{sec:2.A}, the initial state is chosen to be the 1S, color singlet, eigenstate of the in-medium Hamiltonian (\ref{eq:in-medium-hamiltonian}), and we provide an overview of various aspects of its time evolution in a medium at temperature $T=300$ MeV. At this temperature, the 1S state is the only bound state.  We also investigate the changes in the evolution that occur at the smaller temperature $T=200$ MeV at which three bound states exist, like in the vacuum. In order to study the overall evolution of the system at the higher temperature $T=600$ MeV, at which the 1D potential does not sustain any bound state, we choose, in Sec.~\ref{sec:2.B}, a vacuum 1S state as initial state. We then explore more completely the dependence of the medium temperature $T$, by varying $T$ from $200$ to $600$~MeV. In Sec.~\ref{sec:2.C}, we consider the evolution of a compact wave packet which mimics the wave packet created in the gluon fusion picture of ${\rm c\bar{c}}$ production~\cite{Mangano:1997sxc}. 
Finally, in Sec.~\ref{sec:2.E}, in order to get closer to the environment of heavy ion collisions,  we relax the fixed medium temperature assumption and consider the evolution of a ${\rm c\bar{c}}$ pair in a time-dependent environment with a temperature following (adiabatically) a Bjorken profile \cite{Bjorken:1982qr}. 

This section aims at providing an overall physical picture of the fate of a single ${\rm c\bar{c}}$ pair in a hot QGP medium whose temperatures are chosen so as to illustrate the various regimes introduced in the previous section. We recall that the one-dimensional potential used in this work was constructed with the aim to yield numerical estimates that are phenomenologically relevant. Hence, the various time and length scales that appear in this section are meant to be as realistic as can possibly be in a one-dimensional setting. The details of the potential and of the numerical implementation are presented in App.~\ref{app:numerics}. Let us recall here that the one-dimensional grid used in the calculations extends from -10 fm to +10 fm with respect to the origin of the relative distances. 

\subsection{\label{sec:2.A} In-medium 1S singlet initial state}

As a first step in our analysis, we consider as initial density matrix the projector on the 1S ground  state of the in-medium real potential in the color singlet channel, at the plasma temperature $T=300$~MeV.
 Starting from an eigenstate of the in-medium potential at the plasma temperature  leads to the cancellation of the contribution of $\mathcal{L}_0+\mathcal{L}_1$ at the initial time and  allows us  to focus on the role of the imaginary potential. Besides, only one singlet bound state exists for $T=300$~MeV, which simplifies somewhat the analysis.
\begin{figure*}[h]
\centering
\includegraphics[width=0.98\linewidth]{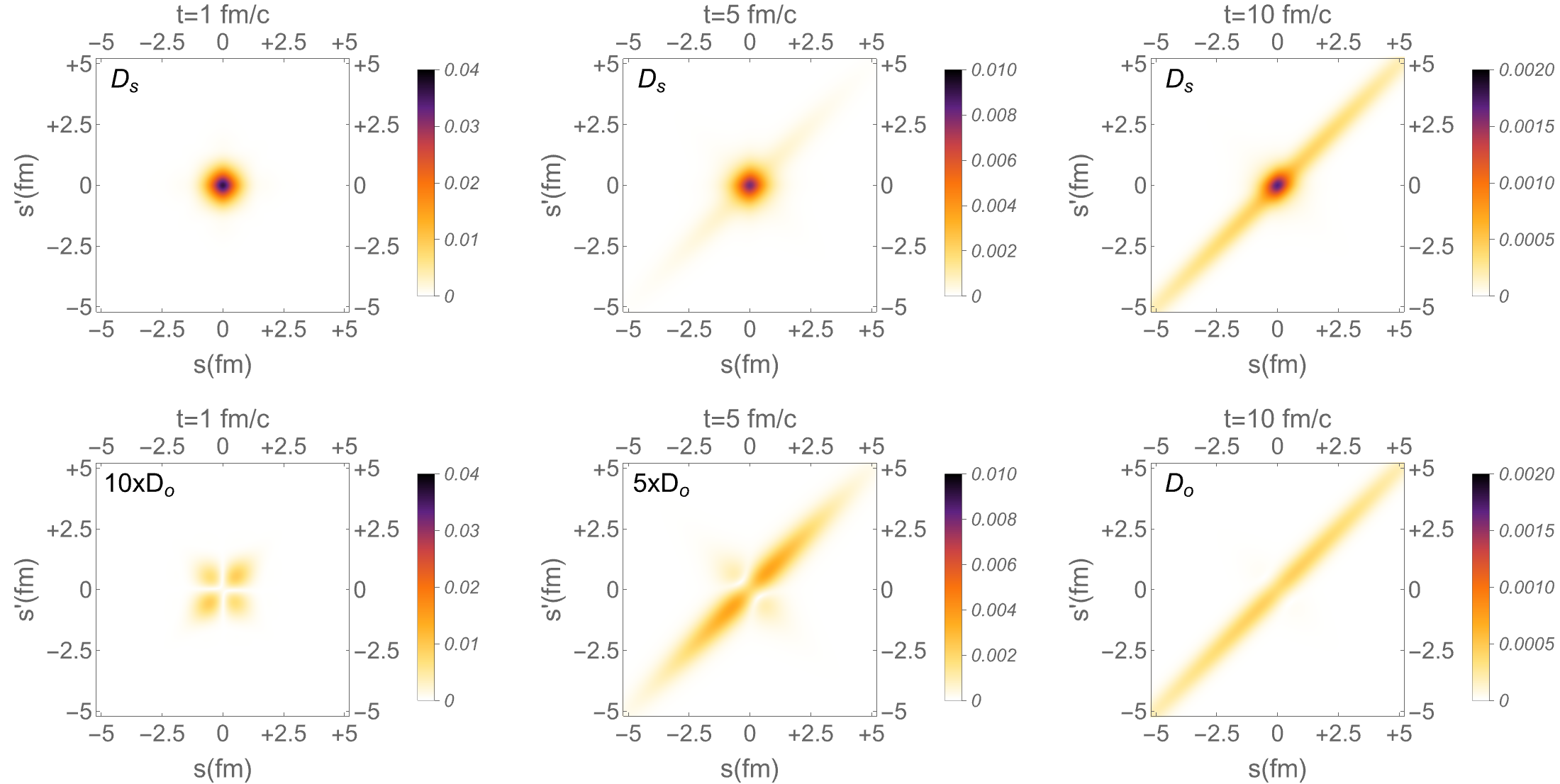}
\caption{Time evolution (from left to right: $t=1$, 5 and 10~fm/$c$) of the singlet density matrix $\bra{s}\mathcal{D}_{\mathrm{s}}\ket{s'}$ (top panels) and octet density matrix  $\bra{s}\mathcal{D}_{\mathrm{o}}\ket{s'}$ (bottom panels) in a medium at temperature $T = 300$~MeV. The initial state is an ``in-medium" 1S color singlet. For better readability, the color scale is different for each time and the octet densities have been rescaled by a numerical factor, 10 for $t=1\,\mathrm{fm}/c$ and 5 for $t=5\,\mathrm{fm}/c$.}
\label{fig:DsDo1Ssingletthermal}
\end{figure*}
Fig.~\ref{fig:DsDo1Ssingletthermal} shows the time evolution of the $s,s'$ matrix elements of the components $\mathcal{D}_{\mathrm{s}}$ and $\mathcal{D}_{\mathrm{o}}$ of the density operator (see Eq.~(\ref{eq:DsDodef})).  We see that the octet channel, initially absent,  is quickly populated (within 1 fm/$c$) as a P state (see bottom left panel) via dipolar  transitions from the 1S state  induced by the operator $\mathcal{L}^{\mathrm{os}}$. Since the potential is (weakly) repulsive in the octet configuration, the relative distance between c and $\bar{\rm c}$  increases with time, leading to a spreading of the density along the $s=s'$ axis. Since an octet state can be locally converted back into a singlet state via the $\mathcal{L}^{\mathrm{so}}$ operator,  the same spreading occurs for $\mathcal{D}_{\mathrm{s}}(s,s)$ as well, in spite of the attractive nature of the potential in the singlet channel. In fact, in the singlet channel, a  peak remains visible in the central region of the corresponding plots, i.e. for $s$ and $s'$ within the range of the singlet potential. This peak reflects the survival of bound states  until the late stage of the evolution. 

One also observes in Fig.~\ref{fig:DsDo1Ssingletthermal} that the reduced density matrix $\langle s|\mathcal{D}|s'\rangle$ tends to become diagonal at late time. This is the result of collisional decoherence induced by the operators $\mathcal{L}_2$, which brings the system closer and closer to a classical system with a diagonal density matrix $\langle s|\mathcal{D}|s'\rangle \propto \delta(s-s')$ (see App.~\ref{app:L2alone}).  The operator $\mathcal{L}_3$ introduces friction that eventually balances the effect of $\mathcal{L}_2$ so that the off-diagonal elements of $\langle s|\mathcal{D}|s'\rangle$ remain significant within a range of the order of the thermal wavelength,  $|s-s'|\lesssim \lambda_{\mathrm{th}}=\frac{1}{\sqrt{MT}}\approx 0.3$ fm.  The two  features of the reduced density matrix $\langle s|\mathcal{D}|s'\rangle$ that we have just discussed, namely the spreading along the $s+s'$ axis, and the shrinking along the $s-s'$ axis (decoherence) are found generically in all our simulations (in agreement with other results in the literature, see e.g. \cite{Akamatsu:2021vsh,Miura:2022arv}). 

\begin{figure*}[h]
\centering
\includegraphics[width=0.48\linewidth]{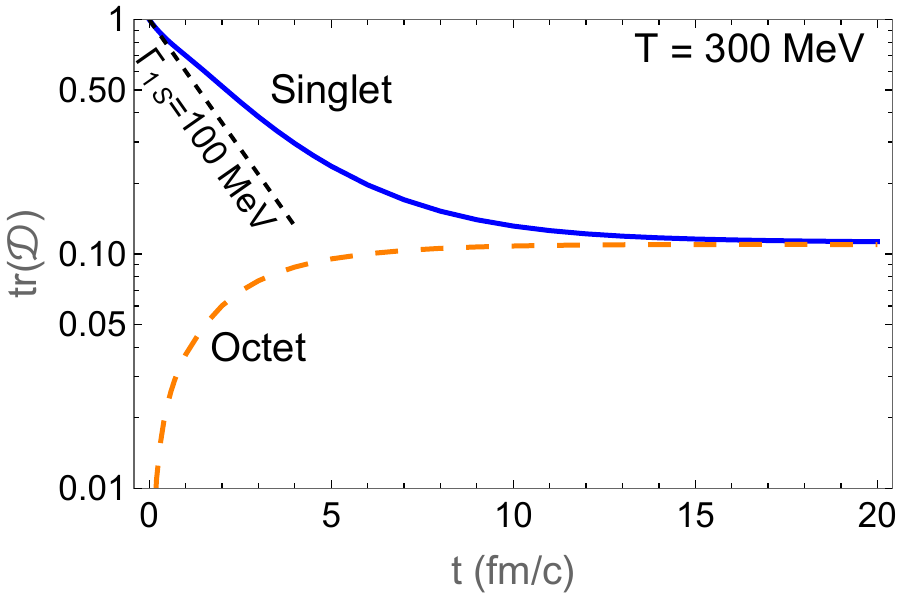}
\includegraphics[width=0.5\linewidth]{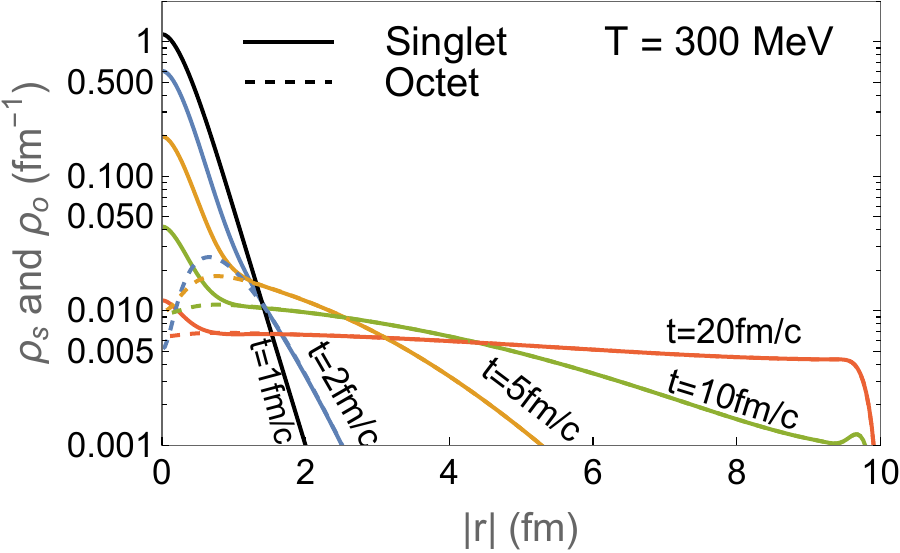}
\caption{Left: Time evolution of the singlet weight ($\mathrm{tr}\,\mathcal{D}_{\mathrm{s}} $, solid) and the octet one ($\mathrm{tr}\,\mathcal{D}_{\mathrm{o}} $, long-dashed). The dashed line represents $\exp(-\Gamma_{\!_{\rm 1S}} t)$. Same initial conditions as in Fig.~\ref{fig:DsDo1Ssingletthermal}. Right: Spatial densities $\rho_{\mathrm{s}}(r,t)$ (solid) and $\rho_{\mathrm{o}}(r,t)$ (dashed) for various times: 0~fm/$c$ (black), 2~fm/$c$ (blue), 5~fm/$c$ (orange), 10~fm/$c$ (green) and 20~fm/$c$ (red). The medium temperature is $T=300$ MeV.}
   \label{fig:singletoctetvstime}
\end{figure*}

Color equilibration is another  feature of the present QME. As a global measure, we consider the singlet and octet weights, defined respectively as $\mathrm{tr}\,\mathcal{D}_{\mathrm{s}} $ and $\mathrm{tr}\,\mathcal{D}_{\mathrm{o}}$. These weights are  plotted on the left panel of Fig.~\ref{fig:singletoctetvstime}, which shows that they become equal after a time  $t\approx 10$~fm/$c$. Note that since $1=\mathrm{tr}\,\mathcal{D}=\mathrm{tr}\,\mathcal{D}_{\mathrm{s}}+8\mathrm{tr}\,\mathcal{D}_{\mathrm{o}}$, the equality $\mathrm{tr}\,\mathcal{D}_{\mathrm{s}}=\mathrm{tr}\,\mathcal{D}_{\mathrm{o}} $ implies that there are,  at that time, eight times more octet pairs than singlet pairs in the system,\footnote{A word of caution about the wording is perhaps needed here in order to avoid confusion. There is only one pair in the system, but the results would be identical for an arbitrary number of independent pairs. It is often convenient to think about the system as a collection of pairs, as done implicitly here and repeatedly throughout, but of course what is referred to is a probability for one pair.} and according to Fig.~\ref{fig:DsDo1Ssingletthermal}, these pairs are to be found in continuum states. At very early time, the evolution of the density matrix is dominated by that of its major component, the in-medium 1S state. Indeed an exponential fit at very early time ($t\lesssim 0.5$ fm/$c$) yields a decay rate  $\Gamma_{\!_{\rm 1S}}\approx 0.1~\mathrm{GeV}$, which is in good agreement with the value estimated in App.~\ref{app:dissocrates} for the in-medium 1S state (see Fig.~\ref{fig:decayrate2}).  One notes however that the actual behavior of the singlet weight rapidly deviates from this initial exponential law, in agreement with expectations based on the analysis presented in  App.~\ref{app:L2alone} (see Eq.~(\ref{eq:trDsL2alonegaussian}))  of the evolution induced by the  operator $\L_2$ alone.

A more differential information can be obtained from the diagonal elements of the density matrix, i.e., the local densities 
\beq\label{eq:densityrhoso}
\rho_{\mathrm{s/o}}(r,t)=\mathcal{D}_{\mathrm{s/o}}(r,r,t).
\eeq
These are plotted in the right panel of Fig.~\ref{fig:singletoctetvstime}.  Color equilibration manifests itself as $\rho_{\mathrm{s}}(r,t)$ becoming equal to $\rho_{\mathrm{o}}(r,t)$, which occurs rapidly at large relative distance:   $\rho_{\mathrm{s}}\simeq\rho_{\mathrm{o}}$ for $|r|\gtrsim 2$~fm already at early time.  Between 5 and 10~fm/$c$, the densities reach the box boundaries ($\pm 10$~fm/$c$) and level off toward their asymptotic values ($(1/9)\times (1/20)~\mathrm{fm}^{-1}\approx 5.5\times 10^{-3}$~fm$^{-1}$). At small distance, the dipolar nature of the imaginary potential leads to a reduced equilibration rate, which delays color equilibration as illustrated in the figure. 
We then arrive at the physical picture of a decaying singlet bound state in a growing and equilibrating  background of unbound ${\rm c\bar{c}}$ pairs: At $t=20$~fm/$c$ the distribution at large relative distance reaches a stationary pedestal, while a trace of a bound state stills remains at small relative distance.  

\begin{figure*}[h]
\centering
\includegraphics[width=0.49\linewidth]{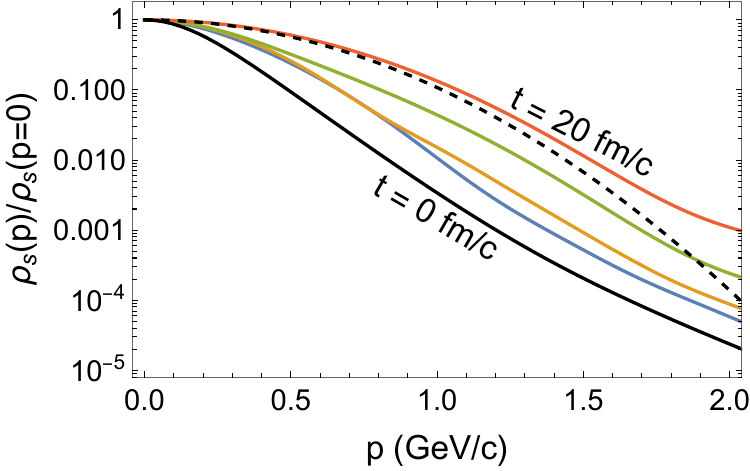}
\includegraphics[width=0.49\linewidth]{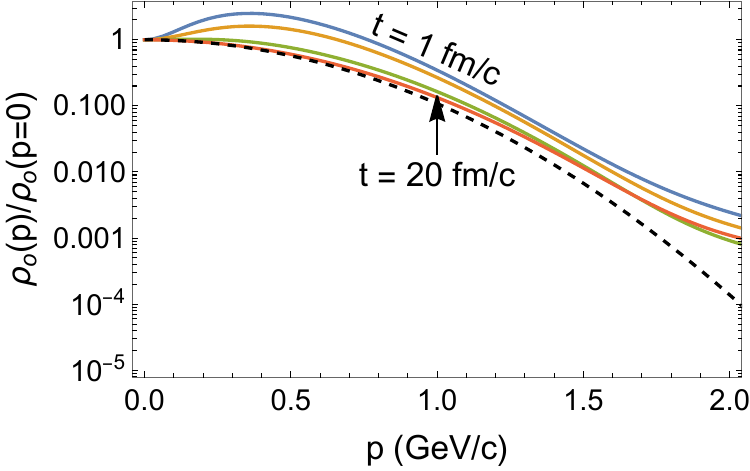}
\caption{Time evolution of the singlet (left) and octet (right) momentum densities for $t=0\mathrm{fm}/c$ (black), 1~fm/$c$ (blue), 2~fm/$c$ (orange), 5~fm/$c$ (green) and 20~fm/$c$ (red). The initial condition (1S in-medium state) and the temperature ($T=300$ MeV) are the same  as in Fig.~\ref{fig:DsDo1Ssingletthermal}. Each curve has been rescaled to unity at $p=0$ for better readability. The classical Maxwell distribution $\exp(-p^2/MT)$ -- with $M=m_c=1.469\,\mathrm{GeV}$ -- is also shown for comparison (dashed). A fair agreement with the actual distribution at $t=$ 20 fm/$c$ holds for $p\lesssim \lambda_{\mathrm{th}}^{-1}\approx 0.7\,\mathrm{GeV}/c$. }
   \label{fig:singletoctetpvstime}
\end{figure*}

Further insight into the equilibration process  can be gained from the momentum distributions $\rho_{\mathrm{s,o}}(p,t)$, defined as 
\begin{equation}
\rho_{\mathrm{s/o}}(p,t)=\frac{1}{2\pi}\int \rmd y\,e^{-i p\,y}\int \rmd r\, 
\mathcal{D}_{\mathrm{s/o}}\left(r+\frac{y}{2},r-\frac{y}{2},t\right), 
\label{eq:rhosofp}
\end{equation}
and displayed in Fig.~\ref{fig:singletoctetpvstime}. One can see there that the momentum densities  evolve gradually toward the thermal distribution $\exp(-p^2/2M_{\mathrm{red}}T)=\exp(-p^2/MT)$, which reflects the statistical population of the continuum momentum states. Note that for $t=20 $ fm/$c$, the thermal (Maxwell-Boltzmann) distribution is well reproduced up to momenta of the order of $\sqrt{MT}\simeq 0.7$ GeV, corresponding to a thermal wavelength $\sim 0.3$ fm. However the tail at large $p$ deviates systematically from the Maxwell-Boltzmann distribution. This behavior is further discussed in Sec.~\ref{sec:3.B2} and App.~\ref{app:equilibrium} where a more detailed analysis of the late-time momentum distribution is given.

\begin{figure}[H]
\centering
\includegraphics[width=0.49\linewidth]{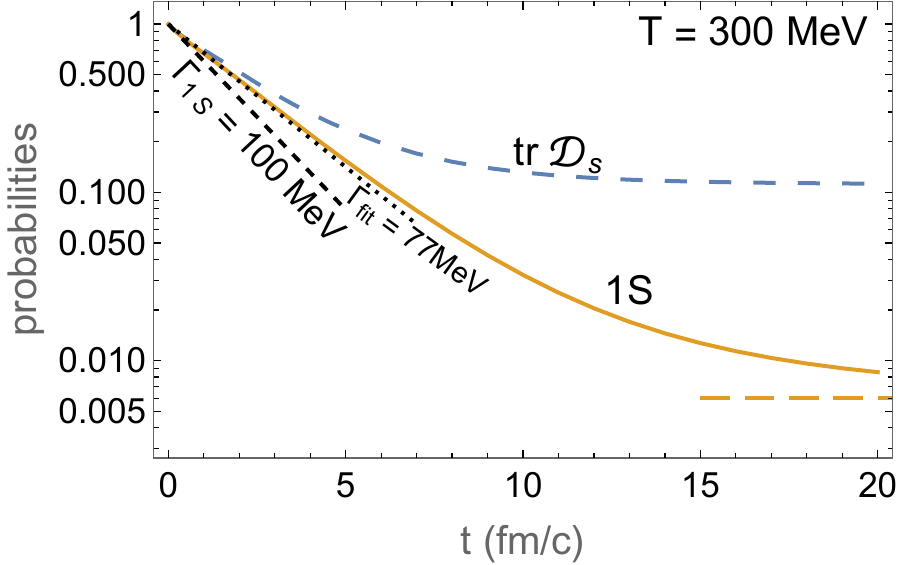}
\includegraphics[width=0.49\linewidth]{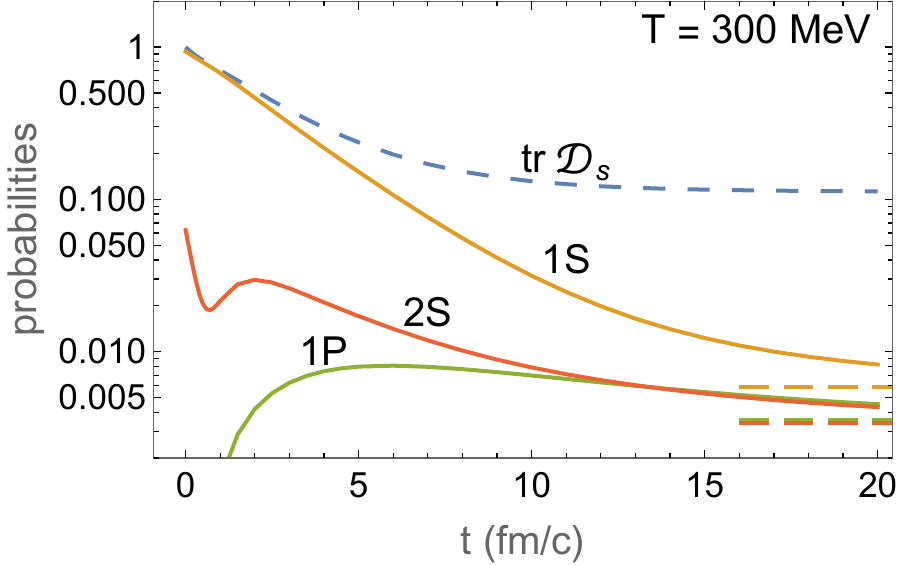}
\caption{Left: Time evolution of the probability $p_{_{1\rm S}}$ of the in-medium 1S state (solid line). The initial condition and temperature ($T=300$ MeV) are the same as in  Fig.~\ref{fig:DsDo1Ssingletthermal}. The dashed curve represents the singlet weight $\mathrm{tr}\,\mathcal{D}_{\mathrm{s}}$ (from Fig.~\ref{fig:singletoctetvstime}). Right: The probabilities of vacuum eigenstates $p_{_{1\mathrm{S}}}$ (orange), $p_{_{1\mathrm{P}}}$ (green) and $p_{_{2\mathrm{S}}}$ (red). On both panels, the horizontal long-dashed lines indicate the respective asymptotic values (see Sec.~\ref{sec:3.B3}).}
\label{fig:Proba1Ssingletthermal300} 
\end{figure}

A different perspective on the time evolution of the density matrix can be obtained by  projecting the singlet density operator $\mathcal{D}_{\mathrm{s}}$ on various charmonium  bound states, that we denote by $\Phi_n$. Recall that we have two types of bound states: in-medium bound states, i.e., eigenstates of the in-medium screened potential, and vacuum bound states. The probability to find the system at time $t$ in one of these bound states is given by 
\beq\label{probablities1}
p_n(t)=\langle \Phi_n|\mathcal{D}(t)| \Phi_n\rangle=\iint \rmd s \rmd s'  \Phi_n^\star(s) {\mathcal D}_{\rm s}(s,s',t) \Phi_n(s') .\eeq
This is of course a conditional probability that depends on the choice of the initial density matrix, the survival probability corresponding to the choice ${\cal D}(0)=\ket{\Phi_n}\bra{\Phi_n}$. The evolution of the survival probability of the in-medium 1S state (the only bound state for $T=300$~MeV) is shown in Fig.~\ref{fig:Proba1Ssingletthermal300} (left panel). 

At early time, the evolution of this in-medium eigenstate is dominated by the imaginary potential, that is, by collisions. These naturally  decrease the survival probability as continuum states are being populated. As already mentioned, at very early time, $t\lesssim 0.5$ fm/$c$,
this initial depletion of the 1S state is well captured by an exponential decay law, with the decay rate $\Gamma_{\!_{\rm 1S}} \approx 100$~MeV. At later times, say up to $t\lesssim 6$ fm/$c$, the survival probability of the initial state still follows an approximate exponential decay law, but with a smaller effective rate of $\approx 77$~MeV, indicated by the dotted curve. We do not have a simple explanation for the emergence of this approximate exponential law, besides the fact that it results from the combined contributions of several operators $\L_i$. We show in App.~\ref{app:L2alone} that the action of $\L_2$ alone predicts a definite deviation from the exponential $\exp(-\Gamma_{\!_{\rm 1S}}t)$ toward an algebraic law (see e.g. Eq.~(\ref{eq:QCDp1SL2alone}) and Fig.~\ref {fig:probvstimeL2}).  Beyond 6~fm/$c$, the trend progressively changes and one starts observing the beginning of an asymptotic regime  where the ${\rm c\bar{c}}$ relative motion reaches some form of equilibrium.  After 20~fm/$c$, about $10\%$ of the singlet weight can be attributed to the surviving  in-medium 1S state, a value which has been checked to scale like the inverse of the box size $L$, provided that $L$ is significantly larger than the potential range (see the discussion in Sec.~\ref{sec:3.B1}, and in particular Fig.~\ref{fig:boxsizedep}). 

On the right panel of Fig.~\ref{fig:Proba1Ssingletthermal300}, we display the projection of the density matrix on the three vacuum eigenstates, i.e., the probabilities $p_{_{1\mathrm{S}}}$, $p_{_{1\mathrm{P}}}$ and $p_{_{2\mathrm{S}}}$, the initial state being as before the in-medium 1S state. The motivation for this analysis is that if the medium were to freeze-out at say, 10 fm/$c$, these projections would yield the actual populations, at that particular time, of the various (vacuum) states. The in-medium 1S  state has initially a projection to the vacuum 1S state nearly equal to unity and a small component (of the order of $5\%$) on the vacuum 2S. Because of the large overlap of the in-medium and vacuum 1S states, the evolution of the vacuum $p_{_{1\mathrm{S}}}$ does not differ significantly from the in-medium $p_{_{1\mathrm{S}}}$   displayed on the left panel. The probability $p_{_{2\mathrm{S}}}$  has a more complex evolution than the $p_{_{1\mathrm{S}}}$, reflecting the effective couplings of the three vacuum eigenstates, in addition to their decay to continuum states.  Such a coupling is obvious for the 1P state, which is not present in the initial state. It is only populated through the chain of transitions (1S singlet $\to$ odd octet $\to$ even octet $\to$ 1P singlet) found at the 3rd order in time-dependent perturbation theory (this explains in particular the slow growth (as $t^3$) of the probability $p_{_{1\mathrm{P}}}$). We note that $p_{_{2\mathrm{S}}}$ and $p_{_{1\mathrm{P}}}$ reach a common value when $t\gtrsim 10$~fm/$c$, and then slowly converge, as well as $p_{_{1\mathrm{S}}}$, to their respective asymptotic values. That $p_{_{2\mathrm{S}}}$ and $p_{_{1\mathrm{P}}}$ reach a common value before reaching their asymptotic limit can  be understood from the fact that the density matrix becomes rapidly diagonal, as shown in Fig.~\ref{fig:DsDo1Ssingletthermal}. If it were completely diagonal, $\mathcal{D}(s,s')=\rho(s)\delta(s-s')$, then the projection on any state would be almost the same, since $\bra{\Psi}\mathcal{D}\ket{\Psi}=\int \rmd r \rho(r) |\Psi(r)|^2$ and $\rho(r)$ is nearly constant, except in a small region around the origin. Another important factor that contributes is the fact that the 1P and 2S states become rapidly delocalized, and are therefore subject to frequent transitions toward octet states. A more detailed discussion of the late time behavior and the corresponding asymptotic state is presented in 
Sec.~\ref{sec:3.B3}.
\begin{figure}[h]
\centering
\includegraphics[width=0.50\linewidth]{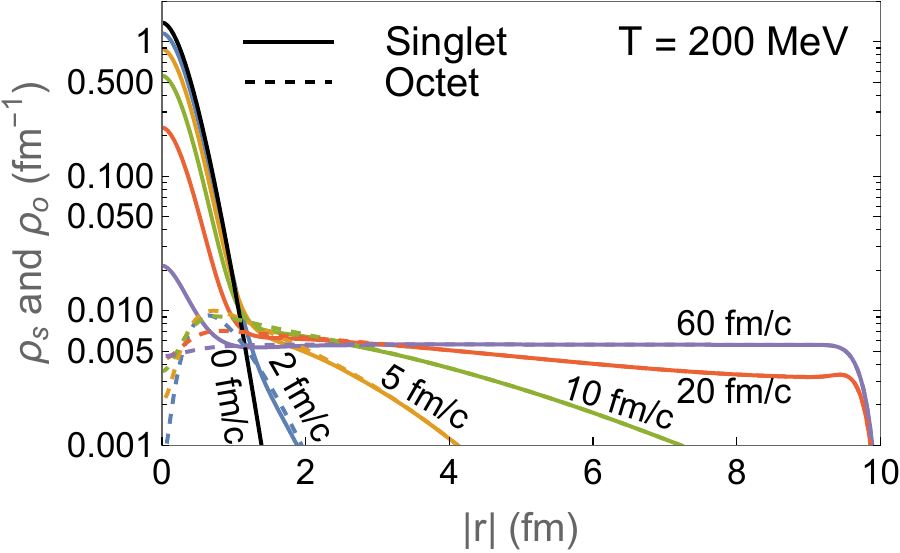}
\includegraphics[width=0.49\linewidth]{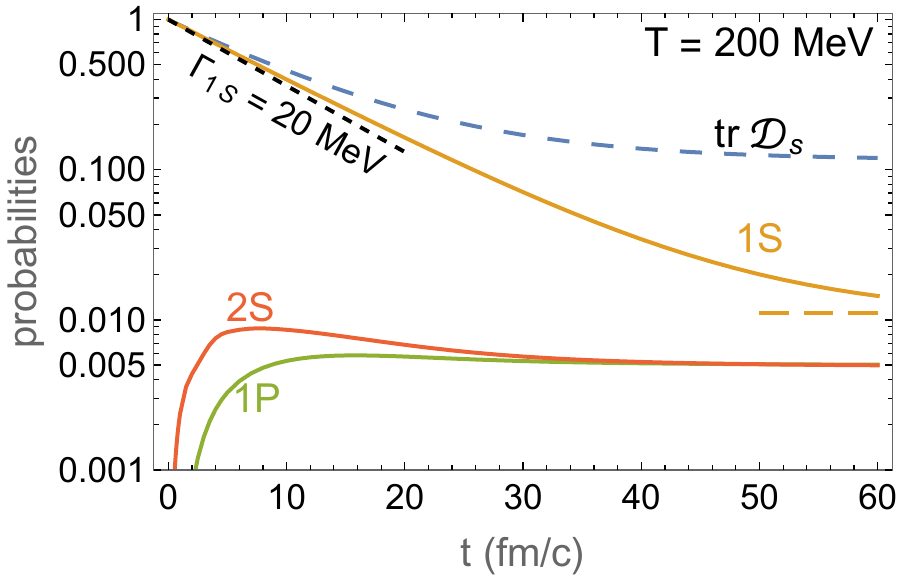}
\caption{Left: Same as Fig.~\ref{fig:singletoctetvstime} (right panel) for $T=200$~MeV, with one extra time ($t=60$~fm/$c$, purple). Right: The probabilities $p_{_{1\mathrm{S}}}$ (orange), $p_{_{1\mathrm{P}}}$ (green) and $p_{_{2\mathrm{S}}}$ (red) corresponding to the in-medium bound states. The horizontal dashed line indicates the asymptotic value of $p_{_{1\mathrm{S}}}$. The asymptotic values of $p_{_{1\mathrm{P}}}$ and $p_{_{2\mathrm{S}}}$ are already reached before 60~fm/$c$. The black dashed line illustrates the exponential decay law with $\Gamma_{\!_{\rm 1S}}$ evaluated from Eq.~(\ref{eq:appendixEGamman}).}
\label{fig:singletoctetvstime200}
\end{figure}

It is finally instructive to contrast the evolution of the in-medium 1S state at temperature $T=300$ MeV with the corresponding evolution taking place at $T=200$ MeV. According to Fig.~\ref{fig:scaleanalysis}, this temperature lies on the borderline of the QBM regime, but as can be seen in Fig.~\ref{fig:singletoctetvstime200}, the results are qualitatively the same as  the ones displayed in Fig.~\ref{fig:singletoctetvstime} for $T=300$~MeV, with a lowering of the transition probabilities (and hence a dilatation of the time scale), and a stronger peak at small $r$ remaining at late time. For $T=200$~MeV, the 1D potential still  supports  three bound  states, with the following binding energies: 0.474~GeV (1S), 0.221~GeV (1P), and 0.103~GeV (2S) (see App. \ref{app:numerics}). The probabilities $p_{_{1\mathrm{S}}}$, $p_{_{1\mathrm{P}}}$ and $p_{_{2\mathrm{S}}}$, corresponding to these in-medium bound states, are displayed on the right panel. One notes that the  decay of $p_{_{1\mathrm{S}}}$ follows more accurately the exponential law than at higher temperature, with the value $\Gamma_{\!_{1\rm S}}=20$ MeV corresponding to $T=200$ MeV. One notes also that $p_{_{1\mathrm{P}}}$ and $p_{_{2\mathrm{S}}}$ reach their common asymptotic value before the end of the evolution, as already observed for $T= 300\,{\rm MeV}$. 

\subsection{1S vacuum singlet initial state}
\label{sec:2.B}
In this subsection, we consider as initial state a  1S vacuum state, and analyze its evolution in terms of the projections of the density matrix on the various vacuum bound states, i.e., in terms of the vacuum  probabilities $p_{_{1\mathrm{S}}}$, $p_{_{1\mathrm{P}}}$ and $p_{_{2\mathrm{S}}}$, for temperatures in the range $T=200-600$ MeV.

\begin{figure}[h]
\centering
\includegraphics[width=0.6\linewidth]{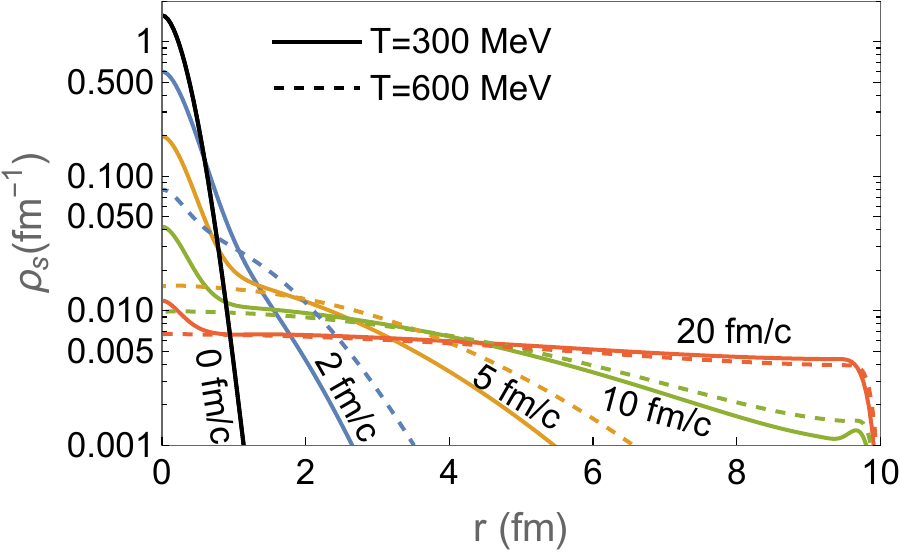}
\caption{Time evolution of the singlet density $\rho_{\rm s}(r)=\bra{r}{\cal D}_{\rm s}\ket{r}$ for an initial 1S vacuum state and two different medium temperatures: $T=300$~MeV (solid curves) and $T=600$~MeV (dashed curves). }
\label{fig:dens1Ssingletvacuum300600}
\end{figure}

We start by considering the evolution of the singlet density at $T=600$~MeV, a temperature at which the in-medium potential does not sustain any bound state. As shown in Fig.~\ref{fig:dens1Ssingletvacuum300600} the qualitative features of the evolution are quite similar to those observed for $T=200$ MeV in Fig.~\ref{fig:singletoctetvstime200} (left panel) and for $T=300$ MeV in Fig.~\ref{fig:singletoctetvstime} (right panel). The variations with the temperature can be understood simply in terms of the collisions: At $T=600$~MeV, the system suffers harder and more frequent collisions, the ${\rm c\bar{c}}$ relative distance increases faster and the central peak disappears earlier, as compared to the $T=300$ and $200$~MeV cases. In fact beyond $t\simeq 5\,{\rm fm}/c$, no peak survives at small $r$ that would signal the presence of a bound state, in line with the fact that the in-medium potential is not binding at $T=600\,{\rm MeV}$.

\begin{figure}[h]
\centering
\includegraphics[width=0.49\linewidth]{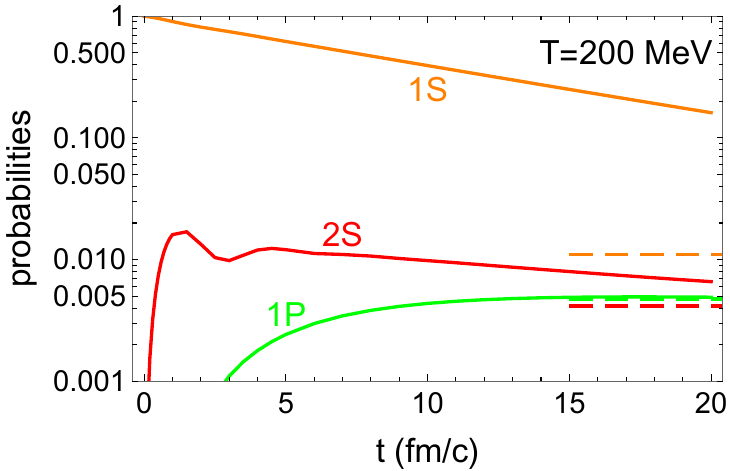}
\includegraphics[width=0.49\linewidth]{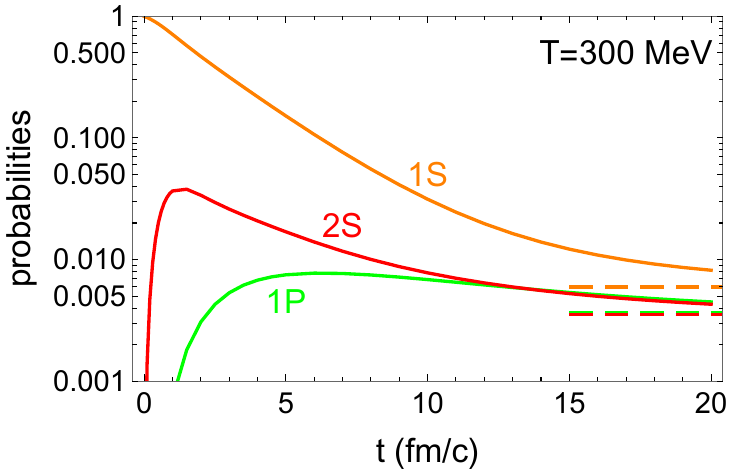}
\includegraphics[width=0.49\linewidth]{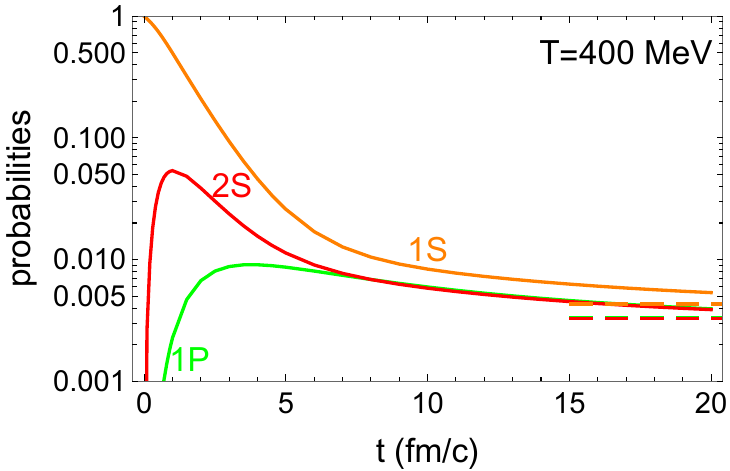}
\includegraphics[width=0.49\linewidth]{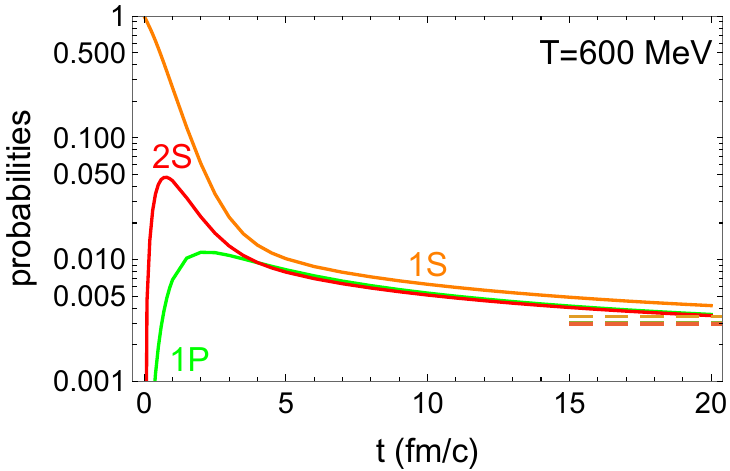}
\caption{Up-left: Time evolution of the probabilities $p_{_{1\mathrm{S}}}$ (orange), $p_{_{1\mathrm{P}}}$ (green) and $p_{_{2\mathrm{S}}}$ (red) of the  vacuum singlet states in a medium at temperature $T = 200$~MeV (solid lines). The initial state is the 1S vacuum state.  The horizontal long-dashed lines represent the asymptotic values. Up-right, low-left, and low-right : same for $T=300$, 400 and 600~MeV.}
\label{fig:Proba1Ssingletvacuum300}
\end{figure}

The time evolution of the vacuum probabilities $p_n$ are  shown in Fig.~\ref{fig:Proba1Ssingletvacuum300} for four different values of the temperature.
At early times, $p_{_{1\mathrm{S}}}$ exhibits an exponential decay, while the growth of $p_{_{1\mathrm{P}}}$ and $p_{_{2\mathrm{S}}}$ reflects the generation of the respective populations via various transitions, as already discussed. After this initial transient stage, $p_{_{1\mathrm{S}}}$  and $p_{_{2\mathrm{S}}}$ appear to decrease with similar rates.  As the temperature increases, the same pattern develops, while the three probabilities become rapidly of comparable magnitudes, eventually reaching a nearly common value at late time, for the reasons invoked earlier about the diagonal structure  of the density matrix and the delocalized nature of the various states. Finally, note that the oscillatory behavior of the 2S weight found at early time for the lowest temperature ($T=200$~MeV) disappears for temperature $\simeq T_{\mathrm{damp}}\approx 300~{\rm MeV}$ and beyond, in agreement with the interpretation of $T_{\mathrm{damp}}$ given in Sec.~II as marking the onset of an overdamped regime. 
\begin{figure}[h]
\centering
\includegraphics[width=0.6\linewidth]{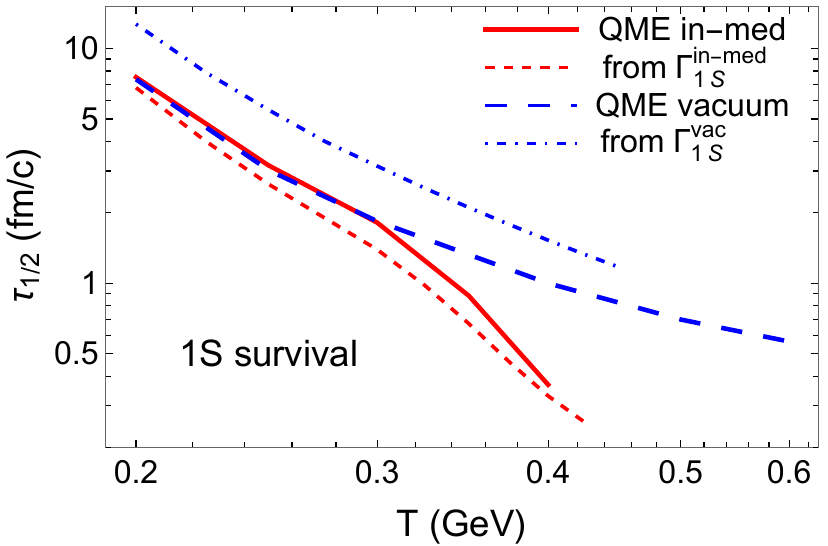}
\caption{Half-life of an initial in-medium (resp. vacuum) 1S-state estimated from Figs.~\ref{fig:Proba1Ssingletthermal300},\ref{fig:singletoctetvstime200} and other evolutions not shown
(resp. Fig.~\ref{fig:Proba1Ssingletvacuum300}), in solid red line (resp. long-dashed blue line), as a function of the temperature and compared to the
estimation made using the decay rate of either the in-medium 1S state (red dashed) or the vacuum 1S state (blue dot-dashed) evaluated in App.~\ref{app:dissocrates} (see Fig.~\ref{fig:decayrate2}).}
\label{fig:tauhalifr}
\end{figure}

The time dependence of the various probabilities in Fig.~\ref{fig:Proba1Ssingletvacuum300} appears to follow, for $T\gtrsim 300$ MeV, a common pattern, with the plots at different temperatures being related by a simple scaling of the time scale with the decay rate. In fact, because the exponential decay with the appropriate in-medium or vacuum decay width as calculated in App.~\ref{app:dissocrates}, holds only at very early time, and can be affected with transient effects, a more accurate  characterization of the decay of the (dominant) 1S state is provided by its half-life $\tau_{1/2}$, that is the time it takes for $p_{_{1\rm S}}$ to drop by a factor 2. The half-life derived from the QME is plotted in Fig.~\ref{fig:tauhalifr}. We found that the plots of Fig.~\ref{fig:Proba1Ssingletvacuum300} that correspond to temperatures $T\gtrsim 300$ MeV approximately  overlap once plotted as a function of the reduced time $t/\tau_{1/2}$. 

\subsection{Initial compact wave packet}
\label{sec:2.C}
As a somewhat more realistic initial condition, we now investigate the case where the initial system is described by a compact wave packet. If one imagines the ${\rm c\bar{c}}$ pairs as being produced in a  gluon fusion process one expects an initial mixture of singlet and (dominantly) octet pairs. We shall examine separately these two color components of the initial density matrix, respectively $\mathcal{D}_{\mathrm{s}}$ and $\mathcal{D}_{\mathrm{o}}$, for which we choose the following simple Gaussian forms\footnote{The octet state is taken here in a ``P wave", as it is meant to result from a $g\to {\rm c}\bar{\rm c}$ splitting. }
\beq\label{eq:compact}
\mathcal{D}_{\mathrm{s}}(s,s')\propto \exp(-\frac{s^2+{s'}^2}{2 \sigma^2})\qquad \mathcal{D}_{\mathrm{o}}(s,s') \propto s\,s' \exp(-\frac{s^2+{s'}^2}{2\sigma_{\rm o}^2}).
\label{eq:DsDoinitial}
\eeq
The parameter $\sigma$ of $\mathcal{D}_{\mathrm{s}}$ is adjusted so that the rms radius of the density $\mathcal{D}_{\mathrm{s}}(r,r)$ is $\frac{\sigma}{\sqrt{2}}\approx 0.12\,\mathrm{fm}$ (i.e. less than one half of the 1S rms radius (see table~\ref{tab:Eandrvacuum1D})), leading to  $p_{_{\rm 1S}}\approx 0.74$ and  $p_{_{\rm 2S}}\approx 0.12$. For $\mathcal{D}_{\mathrm{o}}$, $\sigma_{\rm o}$ is chosen such that the rms of the momentum distribution, has   the same value as for the initial singlet Gaussian state, rms$(p)=0.83$~GeV/$c$, corresponding to $\sigma_{\rm o}\approx 0.29$~fm.\footnote{The specific choice of these values of $\sigma$ and $\sigma_o$ is made so that the two wave packets have initially the same kinetic energy content. }
\begin{figure}[h]
\centering
\includegraphics[width=0.49\linewidth]{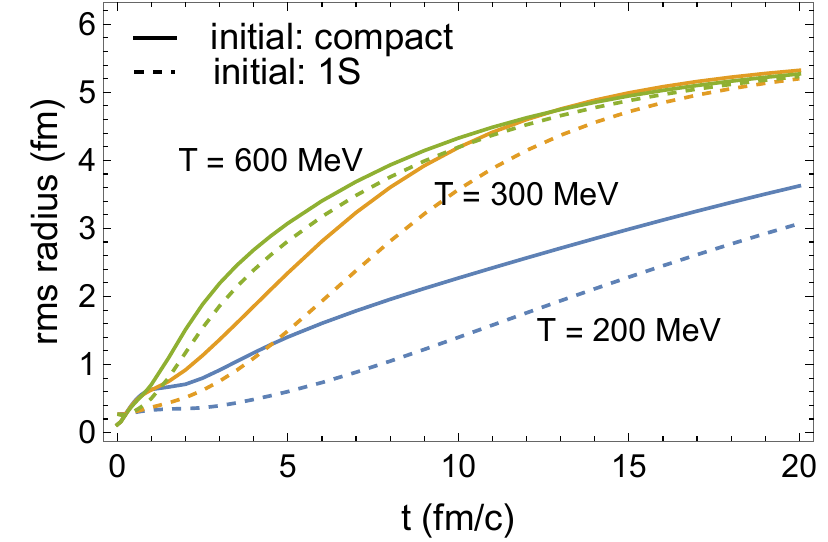}
\includegraphics[width=0.49\linewidth]{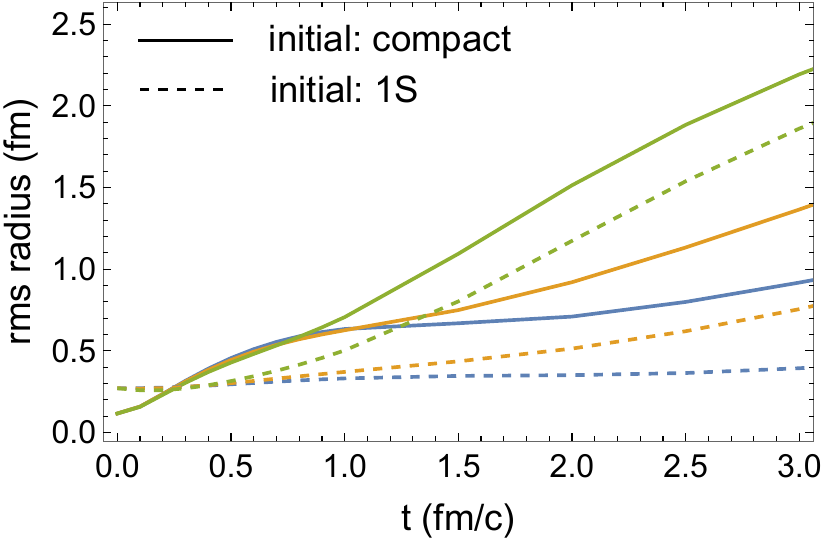}
\caption{Left: Rms radius of $\mathcal{D}_{\rm s}(r,r,t)$ for the 1S initial state (dashed curves) and the compact initial wave packet (solid curves),  for three values of the medium temperature,  $T=200$, 300 and 600~MeV (bottom to top, i.e. blue, orange, green). Right: Same, focusing on the  first 3 fm/$c$ of the evolution. }
\label{fig:radii}
\end{figure}

The time evolution of the rms radius of $\mathcal{D}_{\mathrm{s}}(r,r,t)$ is displayed in Fig.~\ref{fig:radii} and compared with that of the rms of the vacuum 1S initial state. One can see, in particular in the right panel, that the size of the  distribution corresponding to the 1S state does not change much for the first half fm/$c$. This is because the 1S state is, as we have already indicated,  nearly an eigenstate of the in-medium Hamiltonian. When collisions  set in  the system starts to expand, the faster the higher the temperature. When the initial condition corresponds to a compact wave packet one observes an initial rapid growth due its natural spreading induced by the operator $\L_0$. This initial growth, of purely quantum mechanical origin, is essentially independent of the temperature, and lasts for less than  about 1 fm/$c$, at which point the collisions take over and lead to smooth evolution.

\begin{figure}[H]
\centering
\includegraphics[width=.49\linewidth]{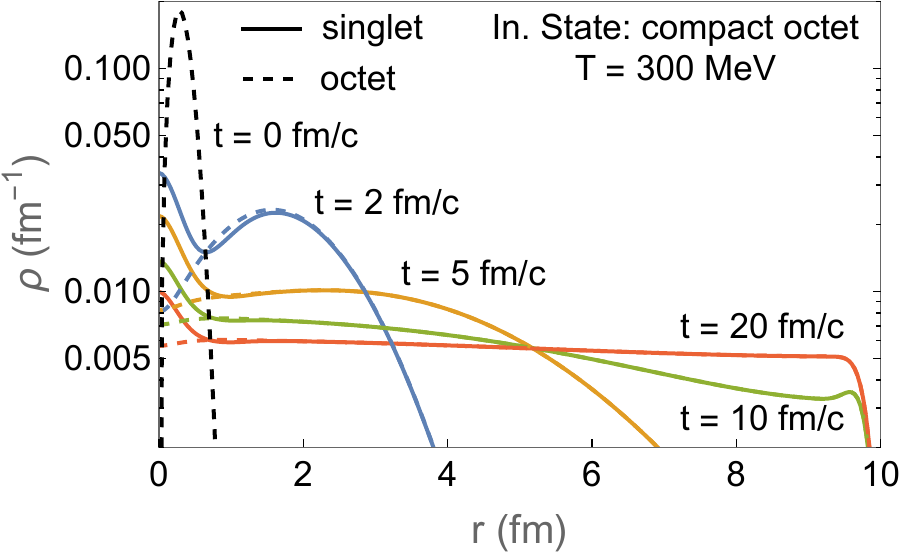}
\includegraphics[width=.48\linewidth]{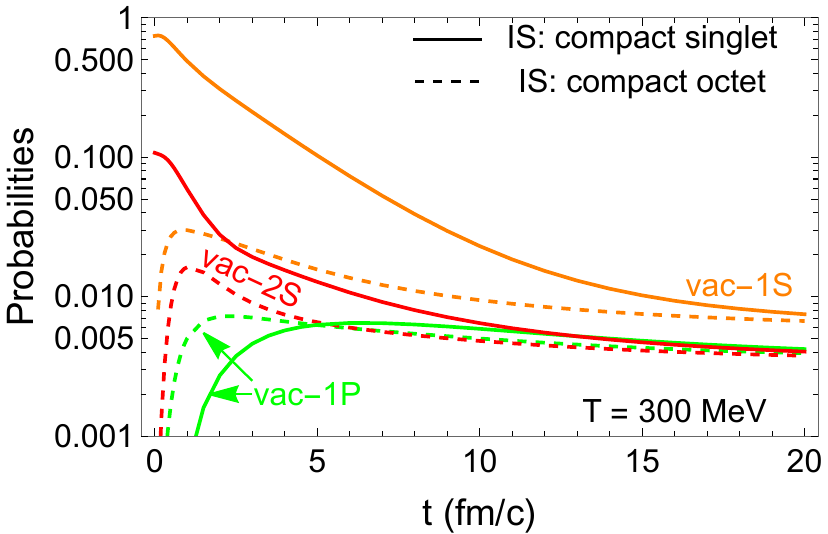}
\caption{Left: Spatial densities $\rho_{\mathrm{s}}(r)$ (solid) and $\rho_{\mathrm{o}}(r)$ (dashed) for various times: 0~fm/$c$ (black), 2~fm/$c$ (blue), 5~fm/$c$ (orange), 10~fm/$c$ (green) and 20~fm/$c$ (red), corresponding to an initial compact octet state. Right: Time evolution of the populations of the first three vacuum eigenstates starting either from a compact P-octet wave packet (dashed curves) or a compact S-singlet  wave packet (solid curves). The medium temperature is $T = 300$~MeV. }
\label{fig:compactoctetevol}
\end{figure}

The role of $\L_0$ in the early stage of the evolution can also be seen in the probabilities $p_n(t)$, as illustrated in the right panel of Fig.~\ref{fig:compactoctetevol}.  The early  transient evolution of the compact initial singlet state  is dominated by quantum mechanics, and transitions between the three states, as revealed for instance by the rapid population of the 1P state originally absent. A distinctive feature of the compact-state evolution seen on the right panel of Fig.~\ref{fig:compactoctetevol}, and absent from Fig.~\ref{fig:Proba1Ssingletvacuum300}, is the early time evolution of $p_{_{\rm 1S}}$ and $p_{_{\rm 2S}}$ which appears to be mostly quadratic in time during the first 0.5 fm/c. Such a behavior is known to originate from the unitary operators $\mathcal{L}_0$ and $\mathcal{L}_1$ (see e.g. \cite{Cugnon:1993ye}). After this transient regime, the probabilities follow a regular pattern analogous to the ones illustrated on Fig.~\ref{fig:Proba1Ssingletvacuum300} and eventually reach their asymptotic values at late time. 

We turn now to the evolution corresponding to the octet  wave packet initial condition. Except for the very early stage, the overall  evolution of the densities is similar to that of a singlet initial state (see Fig.~\ref{fig:DsDo1Ssingletthermal}). In Fig.~\ref{fig:compactoctetevol} (left panel), the singlet and octet densities are shown for $t=0$, 2, 5, 10 and 20~fm/$c$. One can see that, thanks to the dipolar transitions, within less than two fm/$c$ the singlet density develops a peak at small relative distance, while the peak corresponding to the initial octet density disappears. After that, the evolution of the singlet and octet density is similar to that observed in the right panel of Fig.~\ref{fig:singletoctetvstime}. The early time behavior of the densities is also reflected in a fast increase of the probabilities $p_{_{1\mathrm{S}}}$, $p_{_{1\mathrm{P}}}$ and $p_{_{2\mathrm{S}}}$, as can be observed on the right panel of Fig.~\ref{fig:compactoctetevol}.  In particular, one observes  a faster growth of P singlet states in the case of an initial P octet state (behaviors in $t^2$ in time-dependent perturbation theory  vs $t^3$ for initial singlet state). Generically, S (resp. P) states appear to be less (resp. more) populated when starting from a P octet initial state. Note finally that at late time the various probabilities reach their asymptotic values, independently of the initial conditions. 

\subsection{Charmonium in an expanding plasma}
\label{sec:2.E}

In the previous subsections, we have assumed a medium with a fixed temperature. In order to get closer to potential applications in  heavy ion collisions, we now consider an expanding medium with a time-dependent temperature  given by the commonly used (see e.g. \cite{Kajimoto:2017rel,Akamatsu:2021vsh})  Bjorken expansion 
\beq
T(t) = T_{0}\left(\frac{1}{\tau_0 + t}\right)^{1/3}, 
\eeq 
 with $T_{0}=600$ MeV the initial medium temperature, and $\tau_0=1$ fm/$c$.  We limit the evolution to time $t\le 20$ fm/$c$ in order to fulfill the condition $T(t)>T_{_\mathrm{QBM}}$ for all times ($T(20 {\rm fm}/c) \approx 217\,\mathrm{MeV}$). As for the initial state, we consider the two compact states discussed in the previous subsection, namely the singlet and the octet wave packets in Eq.~(\ref{eq:compact}). 

\begin{figure}[H]
\centering
\includegraphics[width=.49\linewidth]{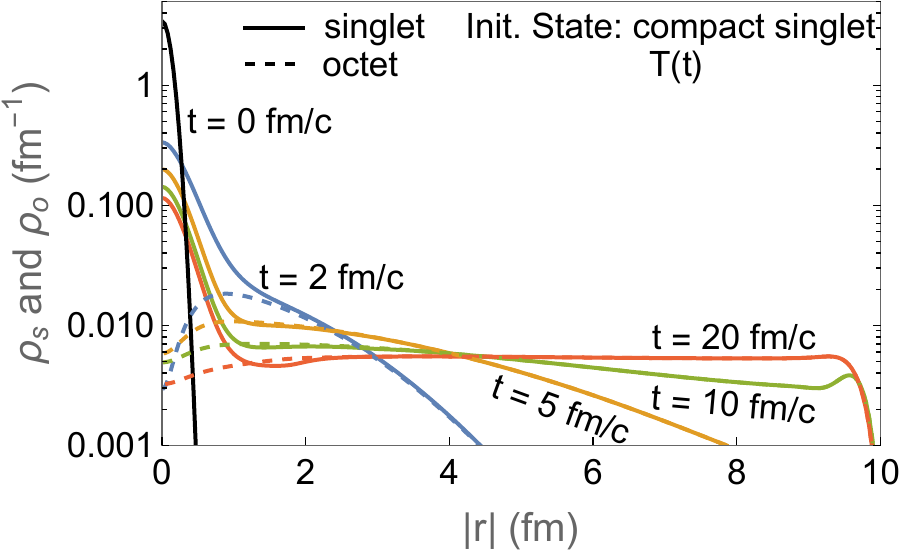}
\includegraphics[width=.49\linewidth]{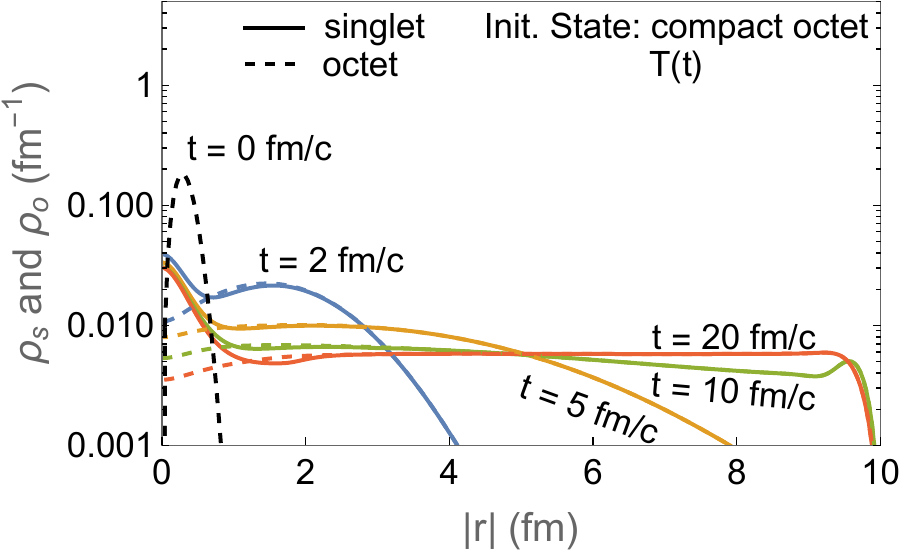}
\caption{Left: Spatial densities for several times -- 0 fm/$c$ (black), 2 fm/$c$ (blue), 5 fm/$c$ (orange), 10 fm/$c$ (green) and 20 fm/$c$ (red) -- as a function of the relative distance for a singlet compact initial state; full (resp. dashed) lines are for the singlet (resp. octet) density. Right: Same for the P octet compact initial state.}
\label{fig:densityTDep}
\end{figure}

The time evolution of the densities $\rho_{\mathrm{s}}(r)$ and $\rho_{\mathrm{o}}(r)$ is shown on Fig.~\ref{fig:densityTDep}. A question of phenomenological interest  is whether the peak near the origin in the singlet density $\rho_{\mathrm{s}}$ survives at the end of the evolution, as it is correlated with significant weights of the vacuum eigenstates. Fig.~\ref{fig:densityTDep} shows that this is clearly the case: The magnitude of the peak at the end of the expansion, is about five times larger than the ``equilibrium'' value reached in a static medium  of constant temperature $T=200\,\mathrm{MeV}$, as observed on Fig.~\ref{fig:singletoctetvstime200}. A fully quantitative analysis is delicate because of effects coming from the finite size of the box used in the calculation, but the main point is that a significant peak survives at the end of the expansion. This suggests that, in the present setting,  the early hot stage of the QGP may not last long enough to eradicate the peak that arises at small $r$ from the presence of a bound state, as would be the case if the medium, instead of expanding, would remain at its initial temperature $T=600$ MeV (see Fig.~\ref{fig:dens1Ssingletvacuum300600}). This conclusion is reinforced by the fact that color has not completely relaxed at the end of the evolution either, as can be inferred from the densities shown in the left panel of Fig.~\ref{fig:densityTDep} for $\lesssim 1$~fm. In fact, after 20~fm/$c$, still 17\% of the ${\rm c\bar{c}}$ pairs are found in singlet states whereas the equilibrium value would be $1/9\simeq 11\%$.

The evolution of the densities obtained when one  starts with the P octet density matrix, resembles that displayed in the left panel of Fig.~\ref{fig:compactoctetevol}. One observes fast transitions toward the singlet channel, building a peak near the origin, as can be seen in the right panel of Fig.~\ref{fig:densityTDep}. Color relaxation is reached within $\approx 2\,\mathrm{fm}/c$ for $r\gtrsim 2$~fm.  The later cooling preserves the peak which appears nearly frozen to a constant magnitude. The final value of $\rho_s(0)\approx 0.03\,\mathrm{fm}^{-1}$ is sizable (compare with Fig.~\ref{fig:compactoctetevol}) and of the same order of magnitude as that of the peak obtained at the end of the evolution of the singlet initial state (left panel). This argues in favor of considering both octet and singlet initial states in  phenomenological applications.

\section{\label{sec:3}Discussion of more technical  issues}
In this section, we return to some of the main features observed in the analysis of the previous section. In particular, we clarify certain technical aspects of the QME that we use, the physical roles of the various superoperators in the evolution of the density matrix, the interplay of collisional decoherence and dissipation in driving the system to a steady state at late time,  and the specific role of the superoperator $\L_4$ in ensuring the positivity of the evolution. We also discuss characteristics of the asymptotic steady state that is reached at the end of the evolution in a static medium with a constant temperature. We analyze the origin of the small observed deviations of the corresponding  momentum distribution from a thermal distribution. We compare more broadly the late-time steady state with thermal equilibrium, and point out interesting differences between abelian and QCD plasma.   

\subsection{The role of various superoperators in the evolution}
\label{sec:3.A}

\subsubsection{Global overview}
\label{sec:3.A1}

\begin{figure}[H]
\centering
\includegraphics[width=.49\linewidth]{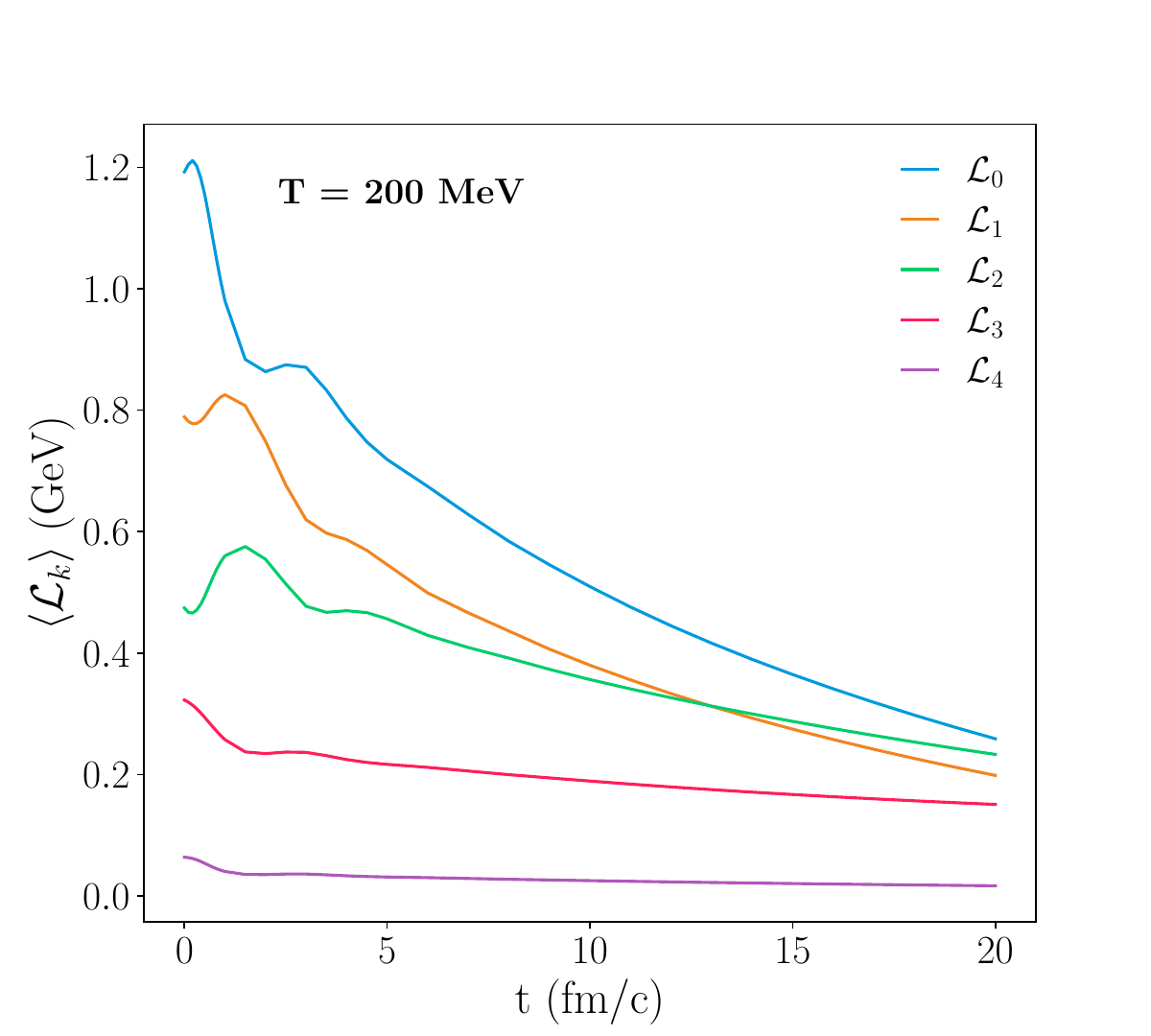}
\includegraphics[width=.49\linewidth]{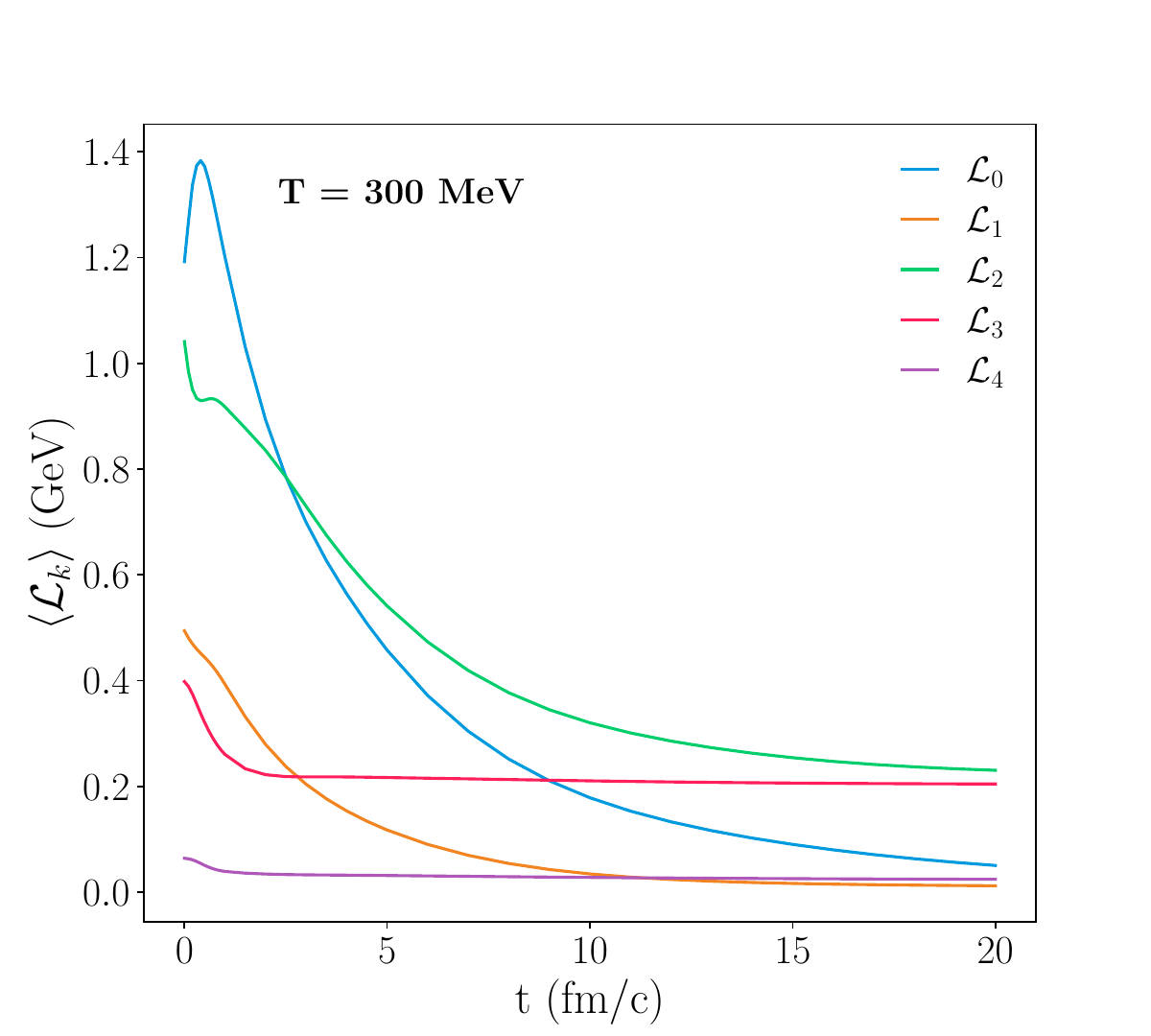}
\includegraphics[width=.49\linewidth]{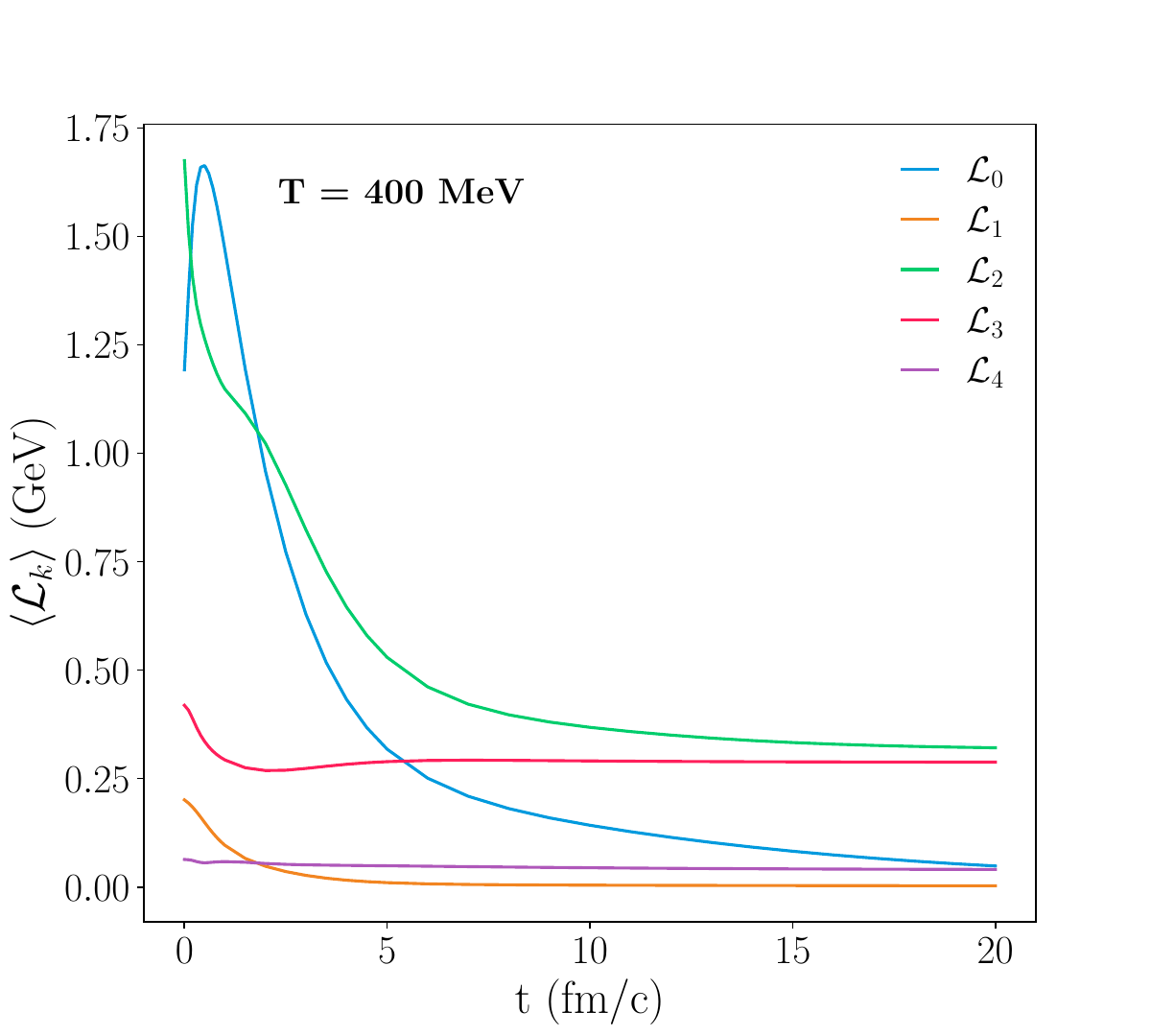}
\includegraphics[width=.49\linewidth]{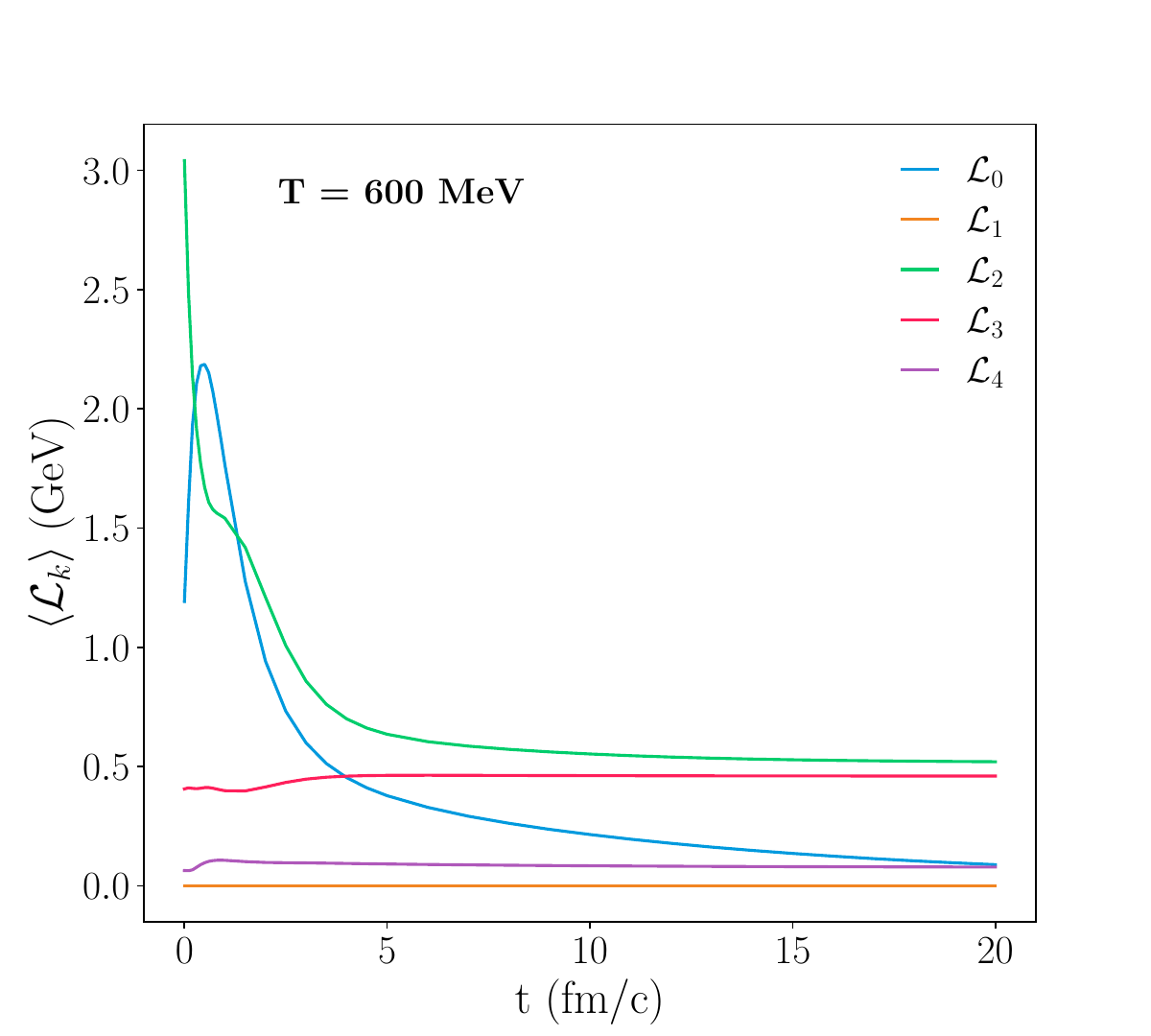}
\caption{Mean values of the $\mathcal{L}_{k}$ superoperators (defined according to Eq.~(\ref{eq:averages})) as a function of time for a medium temperature of 200~MeV (top left panel), 300~MeV (top right panel), 400~MeV (bottom left panel) and 600~MeV (bottom right panel). The initial state is the singlet 1S  vacuum state. }
    \label{fig:MeanValues}
\end{figure}

As can be seen on the formulae given in App.~\ref{app:transition_operators}, the non-unitary superoperators $\mathcal{L}_2$, $\mathcal{L}_3$ and $\mathcal{L}_4$ are multiplied by powers of the thermal wavelength, $\lambda_{\rm th}=1/\sqrt{MT}$, with $\L_2$ independent of $\lambda_{\rm th}$, $\L_3\propto \lambda^2_{\rm th}$ and $\L_4\propto \lambda^4_{\rm th}$. This suggests a natural hierarchy for the magnitudes of these superoperators (see also Eq.~(\ref{eq:gamma3overgamma2})), which has been exploited to some extent in the derivation of Eqs.~(\ref{eq:2.2}) in the QBM regime.  However, the superoperators $\L_3$ and $\L_4$ contain also gradients, of the imaginary potential, and of the density matrix. These have to be taken into account in order to quantify the action of each superoperator on a given density matrix. Thus, in order to gauge (numerically) the relative importance of the action of the  various superoperators $\L_k$ on the density matrix, we define the following averages (for $k=0,\ldots,4$)
\begin{equation}\label{eq:averages}
\left<\mathcal{L}_{k}\right>=
\frac{\sum_{i,i'}\left|\mathcal{L}^{\rm ss}_{k}[\mathcal{D}_{\rm s}]
+\mathcal{L}^{\rm so}_{k}[\mathcal{D}_{\rm o}]\right|_{i,i'}
+\left(N_{c}^{2} - 1\right)\left|\mathcal{L}^{\rm os}_{k}[\mathcal{D}_{\rm s}]
+\mathcal{L}^{\rm oo}_{k}[\mathcal{D}_{\rm o}]\right|_{i,i'}}{
\sum_{i,i'}\left|\mathcal{D}_{\rm s}\right|_{i,i'}
+\left(N_{c}^{2} - 1\right)\left|\mathcal{D}_{\rm o}\right|_{i,i'}},
\end{equation}
where $i$ (resp $i'$) is a discrete index running  over all the values of  $s$ (resp. $s'$) coordinates on the grid of the numerical simulations. The linear combinations $\mathcal{L}^{\rm ss}_{k}[\mathcal{D}_{\rm s}]
+\mathcal{L}^{\rm so}_{k}[\mathcal{D}_{\rm o}]$ and
$\mathcal{L}^{\rm os}_{k}[\mathcal{D}_{\rm s}]
+\mathcal{L}^{\rm oo}_{k}[\mathcal{D}_{\rm o}]$
have been chosen as they respectively represent the contribution from the superoperator $\mathcal{L}_{k}$ 
to the evolution of $\mathcal{D}_{\rm s}$ and $\mathcal{D}_{\rm o}$. The particular weightings $|\dot{\mathcal{D}}_{\rm s}| + \left(N_{c}^{2} - 1\right)|\dot{\mathcal{D}}_{\rm o}|$ and $|\mathcal{D}_{\rm s}| + \left(N_{c}^{2} - 1\right)|\mathcal{D}_{\rm o}|$ are privileged in the numerator and in the denominator as they correspond to those appearing in the conserved trace. The time evolution of the various $\left<\mathcal{L}_{k}\right>$ is illustrated in Fig.~\ref{fig:MeanValues} for medium temperatures  $T=200$, 300, 400, and 600~MeV. This allows us to follow the gradual change of the dynamics  from $T\approx T_{_\mathrm{QBM}}$ to beyond $T_{\rm melt}$.
At moderate temperature ($T=200$~MeV, upper left panel of Fig.~\ref{fig:MeanValues}), the hierarchy between the various superoperators is well recognized. The decay width $\Gamma$ is small compared to the binding energy $E_{_{\rm BS}}$ of the initial state, or equivalently $\tau_{_{\rm R}}> \tau_{_{\rm BS}}$ (from Fig.~\ref{fig:scaleanalysis} one gets approximately $\Gamma\approx 50$~MeV, $E_{_{\rm BS}}\approx 500$~MeV so that $\tau_R\approx 4\,{\rm fm}/c$ and $\tau_{_{\rm BS}}\approx 0.4\,{\rm fm}/c$). In these conditions, the collisions represent a perturbation of the dynamics which is dominated at early time by  $\mathcal{L}_0$ and $\mathcal{L}_1$. 
Such dynamics lead to a slow evolution of all the ${\rm c\bar{c}}$ properties (size, internal energy,\ldots) and of the probabilities to find the system in a given eigenstate. The ${\rm c\bar{c}}$ system evolves then quantum mechanically for several ``periods" $\tau_{_{\rm BS}}$ before the effects of the collisions encoded in $\mathcal{L}_{2}$ and $\mathcal{L}_{3}$ take over after $t\approx 10\,{\rm fm}/c$.

With increasing temperature ($T=300$~MeV, upper right panel of Fig.~\ref{fig:MeanValues}), $\mathcal{L}_2$ plays an increasing role already at early time, and $\langle \L_2\rangle$  becomes comparable in magnitude with $\langle\L_0\rangle$,  while the potential becomes more and more screened and plays a negligible role after a few fm/$c$. As a consequence, the size of the ${\rm c\bar{c}}$ pair grows faster and the time over which $\mathcal{L}_0+\mathcal{L}_1$ governs the evolution shrinks to $t\lesssim 2$~fm/$c$. Note that $T=300$~MeV  corresponds approximately to the temperature $T_{\mathrm{damp}}$ at which the magnitude of the decay width is comparable to the binding energy (or  $\tau_{R} \approx \tau_{_{\rm BS}}\approx 1.3\,{\rm fm}/c$).

For even larger temperature (lower panels), collisional effects dominate right from the start ($\tau_R\ll 1$~fm/$c$) and naturally contribute to enhance the contribution of $\langle \L_0\rangle $ which remains of comparable magnitude as $\langle \L_2\rangle $ during the first few  fm/$c$. In contrast, binding effects induced by $\L_1$ are completely negligible.  The physical picture is clearly that of a pair with negligible internal interaction, whose constituents   separate rapidly and are driven toward thermal equilibrium by stochastic forces. 

Note that in all the plots, but this is more obvious at the highest temperature, the magnitudes of $\langle \L_2\rangle$ and $\langle \L_3\rangle$ reach a common value. This reflects the effect of friction (encoded in $\L_3$) in taming the effect of $\L_2$,  eventually leading to equilibrium with the heat bath where $\langle \L_2\rangle\approx\langle \L_3\rangle$.  As can be inferred from the discussion in App.~\ref{app:L2alone}, $ \L_2 \left[{\cal D}(s,s',t)\right]$ generically decreases with time, as $\exp(-\kappa(s-s')^2 t)$, with $\kappa$ a constant. This particular dependence on $s-s'$ is responsible for collisional decoherence, whose effect on the density matrix $\bra{s}{\cal D}\ket{s'}$ is to decrease the range of the off-diagonal elements from a typical bound state value, $|s-s'| \approx \frac{1}{M\alpha_S}$, to $|s-s'|\approx 0$ where the density matrix reaches a diagonal form, characteristic of a classical behavior. The superoperator $\L_3$ prevents this limit to be reached, restricting instead the off-diagonal elements of $\bra{s}{\cal D}\ket{s'}$ to remain non negligible over a distance of the order of the thermal wavelength when equilibrium is reached and $\langle \L_2\rangle\approx\langle \L_3\rangle$ (see  Fig.~\ref{fig:DsL2L3L4}, and also Fig.~\ref{fig:diagonalDss}).

As for $\langle \mathcal{L}_1\rangle$, its general behavior is easy to understand. As the average size of the ${\rm c\bar{c}}$ pair increases with time, under the effect of collisions, the quark and the antiquark move away from the screened range of the potential and are therefore less affected by the attractive forces present in the singlet channel. One may also observe that the shrinking of the $|s-s'|$ range  of the off-diagonal elements of the density matrix, caused by $\L_2$, contributes to an additional weakening of $\langle\L_1\rangle\propto \langle V(s)-V(s') \rangle$. 

One sees also in  Fig.~\ref{fig:MeanValues} that, after a short transient time where kinetic energy increases because of collisions with the heat bath particles,  $\langle \L_0\rangle$ decreases regularly with time. Since $\L_0\propto y\del_r$, with $y=s-s'$ and $r=(s+s')/2$, this regular decrease can be associated to that of  the average value of $y$, while the gradient $\del_r$ remains small and also decreases as the distance between the c and the $\bar{\rm c}$ increases. Finally, Fig.~\ref{fig:DsL2L3L4} suggests that the effect of $\langle \L_4\rangle$ is marginal. A specific analysis of the role of this operator is presented later in this subsection. 

\subsubsection{More on $\L_0+\L_1$, $\L_2$ and $\L_3$}
\label{sec:3.A2}

\begin{figure}[H]
\centering
\includegraphics[width=0.98\linewidth]{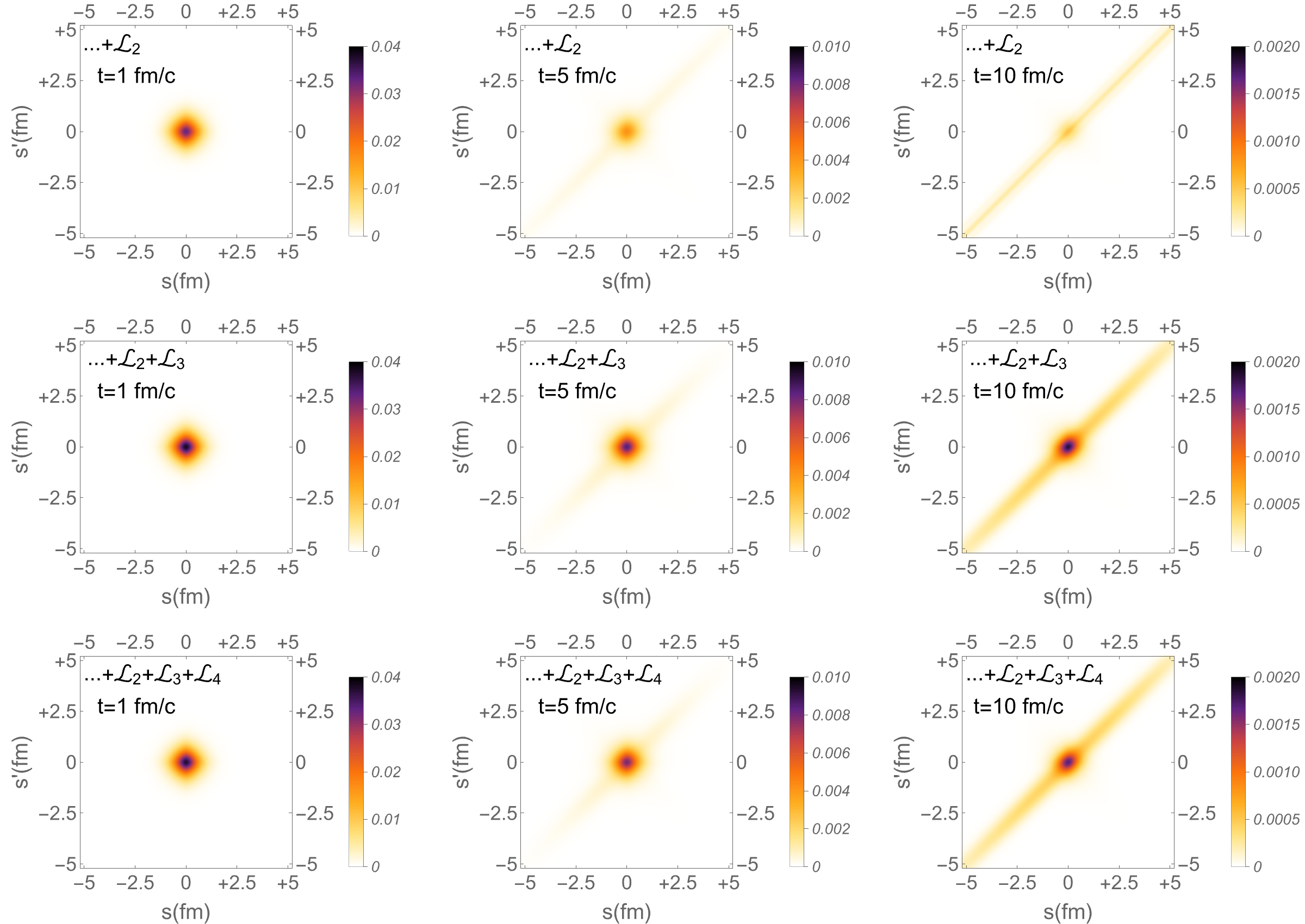}
\caption{Time evolution of the singlet density matrix $\mathcal{D}_{\rm s}(s',s)$, for three different cases: including superoperators up to $\mathcal{L}_{2}$ (top panels), up to $\mathcal{L}_{3}$ (middle panels) and up to $\L_4$ (bottom panels). From left panel to right panel: $t=$ 0.1, 5 and 10 fm/$c$. For better readability, the color scale is different for each time. The initial state is a ${\rm c\bar{c}}$ bound vacuum 1S singlet state, and the medium temperature is $T=300$~MeV. }
\label{fig:DsL2L3L4}
\end{figure}

Fig.~\ref{fig:DsL2L3L4} illustrates the impact of the superoperators $\L_3$ and $\L_4$ on the evolution of the singlet density operator $\mathcal{D}_{s}$. With $\L_2$ alone (in addition to $\L_0+\L_1$), the  density matrix becomes quickly diagonal in coordinate-space, and by the end of the evolution, corresponding to $t=10$~fm/$c$, most of the central peak reflecting the initial presence of a bound state is lost. The dynamics is dominated by collisions without dissipation, leading to collisional decoherence which makes the density matrix diagonal in coordinate space. As already discussed, the effect of $\L_3$ is to tame the effect of $\L_2$, and one can see that indeed in Fig.~\ref{fig:DsL2L3L4}: At 10~fm/$c$ the ``diagonal'' of the density matrix keeps a width of the order of the thermal wavelength ($\lambda_{\rm th}\simeq 0.3$~fm), and the correlation at small $r$ remains clearly visible.

\begin{figure}[h]
\centering
\includegraphics[width=0.49\linewidth]{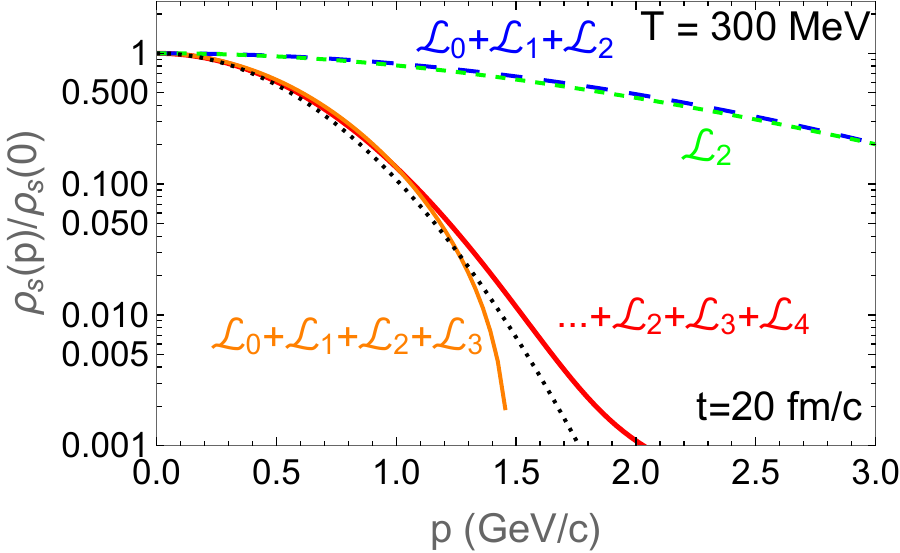}
\caption{Momentum space singlet density obtained after evolving the vacuum 1S state up to 20~fm/$c$, in a medium at temperature $T=300$~MeV. The impact of the superoperators $\L_3$ and $\L_4$ is illustrated by implementing various truncations, as indicated on the figure. The red line represents the evolution obtained with the full superoperator $\L=\L_0+\L_1+\L_2+\L_3+\L_4$. 
The dashed green curve shows the result of the evolution with $\L_2$ alone ($\L\mapsto \L_2$). The black dotted curve represents the Maxwell-Boltzmann distribution $\exp(-\frac{p^2}{M T})$.  } 
\label{fig:effectvariousorders}    
\end{figure}

The role of $\L_2$ is further illustrated in Fig.~\ref{fig:effectvariousorders}, which displays the singlet momentum space density obtained at $t=20$~fm/$c$. It can be seen that the density obtained with $\L_2$ alone and with the combination  $\mathcal{L}_0+\mathcal{L}_1+\mathcal{L}_2$ are pretty similar  and both deviate significantly from the equilibrium Maxwell-Boltzmann distribution, already at small momenta: Overall the distribution is much harder than the equilibrium one, a situation sometimes qualified of ``overheating". The full evolution yields a distribution which agrees with the thermal equilibrium distribution up to momenta of order $\sqrt{MT}$, in agreement with the results displayed in Fig.~\ref{fig:singletoctetpvstime} (for the initial in-medium 1S state). This agreement with the equilibrium distribution is 
to a large extent due to the superoperator $\L_3$, which  brings in  friction, and slows down the diffusive evolution caused by $\L_0+\L_1+\L_2$ (see also  \cite{Miura:2019ssi, Miura:2022arv}).  Finally, adding  $\mathcal{L}_{4}$ does not change the picture in any visible way, except at large momenta where the presence of $\L_4$ is needed to ensure that the distribution remains positive. 

\begin{figure}[h]
\centering
\includegraphics[width=0.49\linewidth]{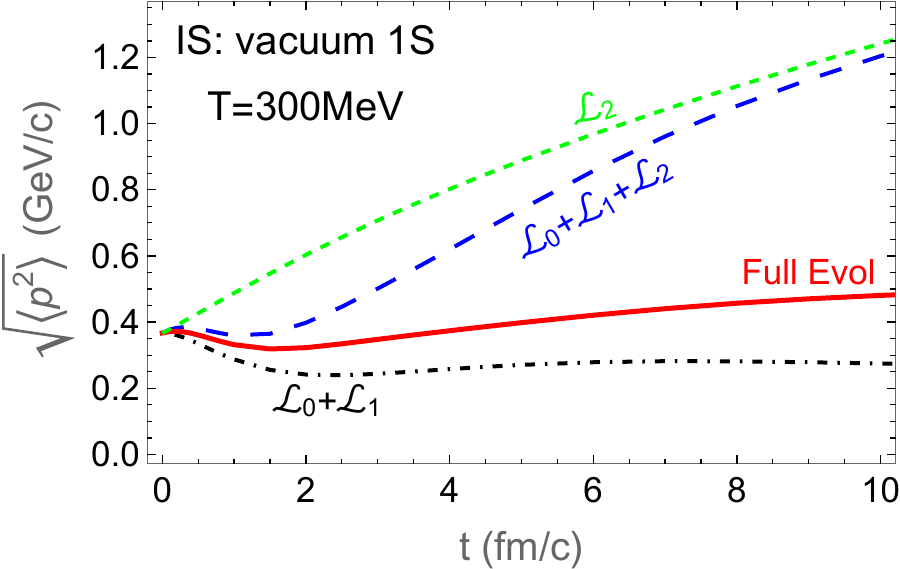 }
\includegraphics[width=0.49\linewidth]{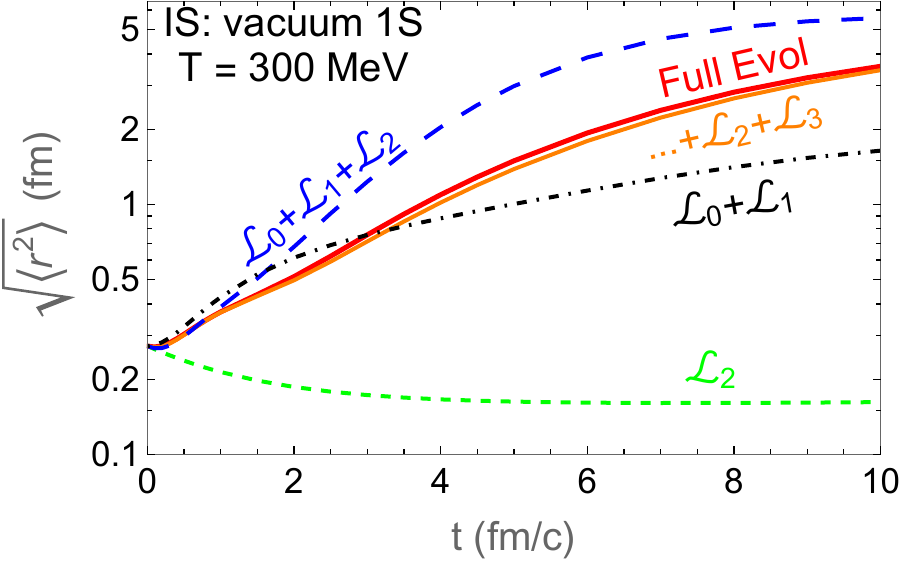}
\caption{Left: Time-evolution of rms(p) of the singlet component for various truncated QME as well as for the full QME (same conventions as for Fig.~\ref{fig:effectvariousorders}). Right: Time evolution of the rms radius of the singlet spatial density for the same situation and the same selections of superoperators as in Fig.~\ref{fig:effectvariousorders}. The saturation for $t\gtrsim 10$~fm/$c$ is due to the finite size of the grid.}
\label{fig:p2oftime}
\end{figure}

The impact of the various superoperators is further analyzed on Fig.~\ref{fig:p2oftime}, illustrating the evolution of $\sqrt{\langle p^2\rangle}$ and $\sqrt{\langle r^2\rangle}$. Starting from the 1S  vacuum state, the unitary part $\mathcal{L}_0+\mathcal{L}_1$ does not modify significantly the momentum distribution and  leads to a moderate spatial expansion. The superoperator $\mathcal{L}_2$ steadily converts heat bath energy into ${\rm c\bar{c}}$ kinetic energy but locally conserves the probability. In fact the evolution with $\mathcal{L}_2$ alone  leads to an early-time reduction of $\langle r^2\rangle$ (see Eq.~(\ref{eq:rmsL2}) and the corresponding discussion in App.~\ref{app:L2alone}). When considering $\mathcal{L}_0$ on top of $\mathcal{L}_2$, this effect is counterbalanced by the enhanced spatial diffusion triggered by collisions.  The combined action of $\mathcal{L}_0$, $\mathcal{L}_1$ and $\mathcal{L}_2$ hence eventually yields a fast enhancement of both $\sqrt{\langle p^2\rangle}$ and $\sqrt{\langle r^2\rangle}$, which is tamed by the  friction brought in by $\L_3$. 

\subsubsection{$\L_4$ and the positivity issue}

\begin{figure}[hbt!]
\centering
\includegraphics[width=.6\linewidth]{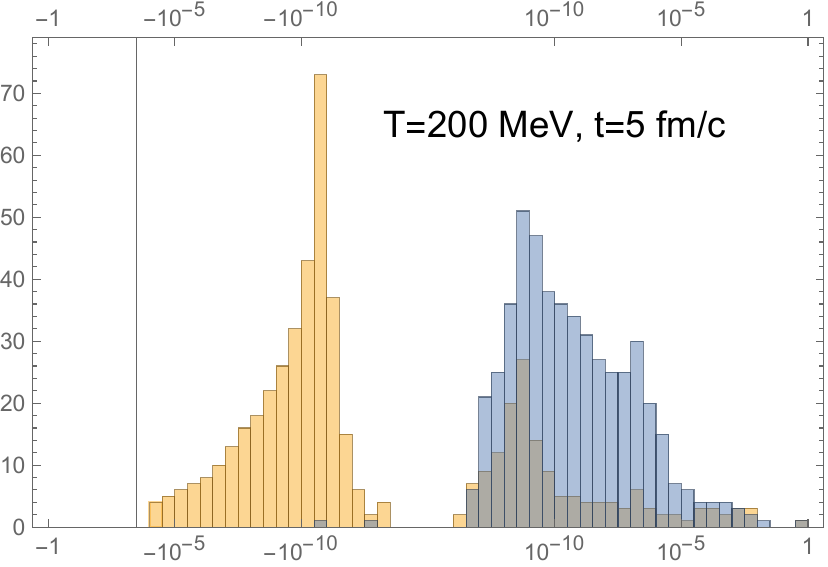}
\caption{Distribution of eigenvalues of the singlet density matrix ${\cal D}_{\rm s}$ after 5~fm/$c$ starting from the 1S vacuum state. The orange (resp. blue) -- light gray (resp. mid-gray) in B\&W printing -- histogram represents the distribution obtained without (resp. with) the $\mathcal{L}_4$ superoperator. The temperature of the medium is $T=200$~MeV.}
\label{fig:eigenvaluesds}
\end{figure}

The role of the superoperator $\L_4$ in insuring the positivity of the evolution equation for the density matrix is discussed in App.~\ref{app:positivity} (see also App.~\ref{app:dissocrates} and in particular the left panel of Fig.~\ref{fig:decayrate1}). In this section, we present a more direct analysis, by considering the spectrum of eigenvalues of the density matrix $\bra{s}{\cal D}\ket{s'}$, after an evolution of 5~fm/$c$ starting from the 1S vacuum state, in a medium of temperature $T=200$~MeV. The low value of the temperature is chosen on purpose, in order to maximize the possible effect of the $\mathcal{L}_4$ superoperator. At $t=0$~fm/$c$, all eigenvalues vanish, except the one corresponding to the initial pure 1S state. As can be seen in  Fig.~\ref{fig:eigenvaluesds}, the evolution without the $\mathcal{L}_4$ term generates negative eigenvalues, a clear signal that the evolution does not preserve the positivity of the density matrix. Note however, that the negative eigenvalues are small, of the order of $10^{-4}$ at most in absolute magnitude, and their accumulated effect on the bulk of the dynamics is not significant, as we have seen before in several occasions. After adding the $\mathcal{L}_4$ term, one finds that the distribution of eigenvalues is shifted so that most eigenvalues end up positive, except for a very limited set of negative ones, whose  magnitude $\sim 10^{-12}$ make them indistinguishable from the numerical noise affecting the evolution. Focusing on the distribution of the largest positive eigenvalues one finds that the impact of $\mathcal{L}_4$ remains quite moderate. Note finally that the impact of the negative eigenvalues is expected to be maximum at the low temperature chosen for the present analysis, which is at the boarder line of the validity of the master equations. At higher temperatures the impact of $\L_4$ on the evolution is found to be quite small, and in most cases negligible.  

\subsubsection{Probabilities and preferred basis}

\begin{figure}[hbt!]
\centering
\includegraphics[width=.6\linewidth]{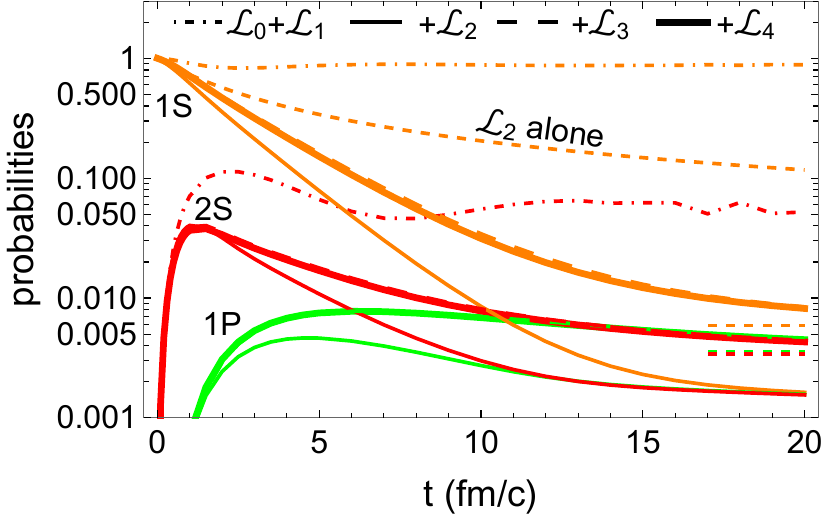}
\caption{Time evolution of the probabilities of the first three vacuum eigenstates (from top to bottom: $p_{_{1\rm S}}$, $p_{_{2\rm S}}$ and $p_{_{1\rm P}}$). These are obtained by evolving the density matrix first  with  $\mathcal{L}_{0}+\mathcal{L}_{1}$ only (dot-dashed lines), then adding $\mathcal{L}_{2}$  (solid lines),  $\mathcal{L}_{3}$ (long-dashed lines) and finally $\mathcal{L}_{4}$ (thick lines); The 1S evolution under $\mathcal{L}_{2}$ alone is also represented, by a dashed orange line. The medium temperature is $T=300$~MeV, and the initial state is the vacuum 1S state. The horizontal dashed lines represent the asymptotic values of the survival probabilities.}
\label{fig:ProbaL2L3L4}
\end{figure}

We turn now to a different way to analyze the results of the calculations, namely an analysis based on the probabilities $p_n=\bra{\Phi_n}{\cal D}_{\rm s} \ket{\Phi_n}$ where $\ket{\Phi_n}$ is one of the vacuum eigenstates (see Eq.~(\ref{probablities1})).
In Fig.~\ref{fig:ProbaL2L3L4}, we display the time evolution of these probabilities and, as we did before for other quantities, we consider the impact of the various superoperators on the evolution.  When only the unitary superoperators $\mathcal{L}_0$ and $\mathcal{L}_1$ are involved, the system exhibits some kind of Rabi oscillations, mostly visible in the plot corresponding to $p_{_{2\rm S}}$. These oscillations find their origin in the fact that the initial state, the vacuum 1S state, is not an eigenstate of the in-medium Hamiltonian. This is also, partially, the cause of the rapid rise of $p_{_{2\rm S}}$ at early time. When  $\mathcal{L}_{2}$ is added, collisions cause a rapid  decrease of the various probabilities  before these  equilibrate to a common value for $t\gtrsim 20$~fm/$c$. If one tries to interpret the corresponding steady state as a thermal equilibrium state, and extract an effective temperature from the momentum density $\rho_{\rm s}(p)$ at $t=20~\mathrm{fm}/c$, one finds a huge temperature, $T\approx 1.9$~GeV (hence the terminology of ``overheating'' alluded to earlier). 
When the $\mathcal{L}_3$ friction term is included, the decrease saturates earlier to larger values, as the ${\rm c\bar{c}}$ system equilibrates at the temperature of the system. Again, one can note that the addition of the $\mathcal{L}_{4}$ terms leads only to tiny  differences when $T > 200$~MeV, which confirms that those terms are, for practical purpose, unimportant when $T \gtrsim T_{_\mathrm{QBM}}$. As a final remark, consider the time evolution of the probability $p_{_{1\rm S}}$  under $\mathcal{L}_2$ alone. This is represented in Fig.~\ref{fig:ProbaL2L3L4} by the orange-dashed curve, where the evaluation is made with the same imaginary potential $W$ as for other curves. While it coincides with the full evolution for $t\lesssim 1\,{\rm fm}/c$, it deviates at later time. We note in particular that combining $\mathcal{L}_2$ with $\mathcal{L}_0+\mathcal{L}_1$ yields a much more rapid suppression  of $p_{_{1\rm S}}$. This can be attributed to the fact that, thanks to the operator $\L_0$,  the ${\rm c\bar{c}}$ pair, while gaining   kinetic energy through the collisions with the plasma constituents can also expand spatially, as already observed on Fig.~\ref{fig:p2oftime}, resulting in a substantial suppression of the 1S probability at intermediate time.\footnote{In App.~\ref{app:L2alone}, the evolution under $\L_2$ alone is studied. In the absence of $\L_0$, diffusion occurs only in momentum space, and leads to broadening of the momentum distribution.} 

\begin{figure}[h]
\centering
\includegraphics[width=0.6\linewidth]{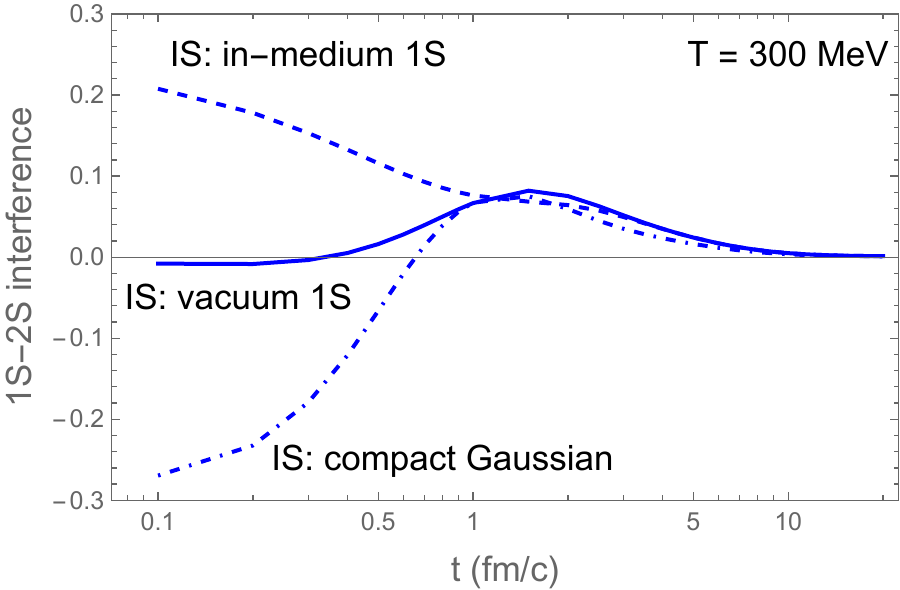}
\caption{The matrix element $\langle {\rm 2S} |\mathcal{D}_{\rm s}| {\rm 1S}\rangle$ as a function of time for various initial conditions. }
\label{fig:interference}
\end{figure}
The analysis of the results in terms of eigenstates of the Hamiltonian complements that carried out earlier in terms of the densities, or more generally in terms of the coordinate space representation of the density matrix. On may  argue that the coordinate  space representation plays here the role of a ``preferred basis'' in the sense where this notion is used in the field of open quantum systems (see e.g. \cite{breuer2007theory}). It is in this basis that we see most clearly the effect of decoherence induced by the collisions and leading the density matrix to a diagonal form. The representation in terms of ``states'', meant to be localized states representing bound states, hides the fact that a dominant aspect of the physics is precisely the dissolution of bound states into continuum states: As can be read on Fig.~\ref{fig:ProbaL2L3L4}, at $t=10$ fm/$c$, $p_{_{1{\rm S}}}\lesssim 5\%$ and $p_{_{2{\rm S}}}\simeq 1\%$, meaning that most of the ``states'' at that time are continuum states. These represent weakly interacting c and $\bar{\rm c}$, and are clearly better described in the coordinate representation. There is another issue. In the coordinate representation, decoherence occurs quickly. This is not the case, for instance, in the basis of vacuum eigenstates. To illustrate the point, consider Fig.~\ref{fig:interference}, where the matrix element $\langle {\rm 2S} |\mathcal{D}_s| {\rm 1S}\rangle$ between the  first two vacuum-eigenstates is displayed as a function of time for $T= 300\,{\rm MeV}$ for the three initial conditions considered in Sec.~\ref{sec:2}. By looking at Fig.~\ref{fig:interference}, one sees that this non-diagonal matrix element evolves on appreciable time scales (and even develops when not present in the initial condition). At late time, this non-diagonal element however vanishes. This reflects the fact that the density matrix approaches a unit matrix at late time, as we have seen already. 

\begin{figure}[H]
\centering
\includegraphics[width=0.6\linewidth]{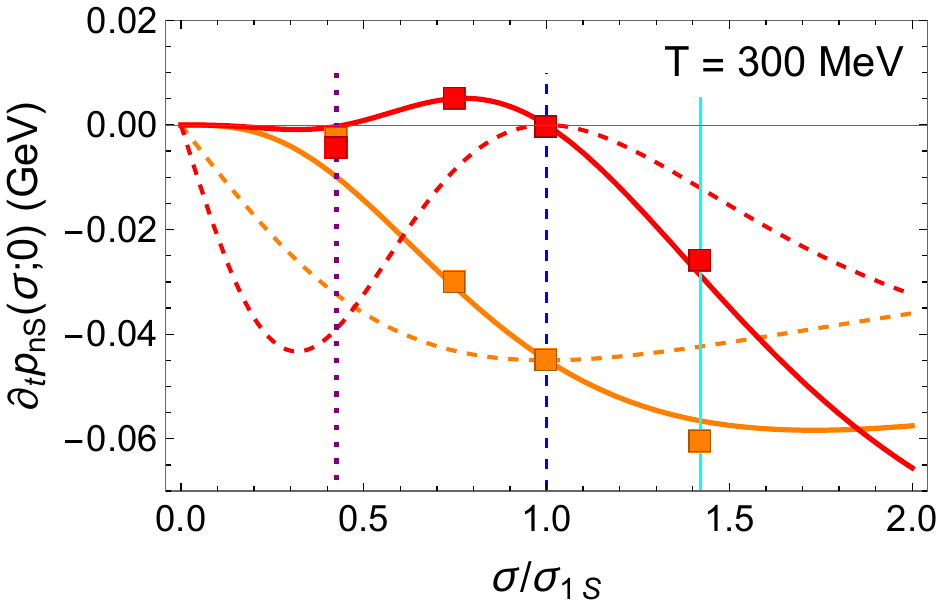}
\caption{Orange (resp. red) solid line: Initial time derivative of the probability to observe a vacuum-1S (resp. 2S) from a initial Gaussian state (\ref{eq:DsDoinitial}) under the evolution of $\mathcal{L}_2$ alone, according to Eq.~(\ref{eq:evolp2SgaussianL2alone}); The $\kappa$ value is tuned to reproduce $\Gamma_{\!_{\rm 1S}}^{\rm vac.}=45\,{\rm MeV}$ at $T=300\,{\rm MeV}$. The 3 vertical lines correspond to the 3 initial states studied in Sec.~III: vacuum 1S ($\sigma=\sigma_{\rm 1S}\equiv 0.387\,{\rm fm}$), in-medium  1S state ($\sigma \equiv$~0.55 fm), and compact Gaussian state ($\sigma =$~0.165 fm), while the square symbols represent the results from the full QME at $T=300\,{\rm MeV}$ for these IS  (plus one extra simulation performed for $\sigma/\sigma_{\rm 1S}=0.75$). The dashed lines represent the usual decay law $\dot{p}_{n{\rm S}}(\sigma,0)=-\Gamma_{n{\rm S}}\;p_{n{\rm S}}(\sigma;0)$, thus simply reflecting the variation of $p_{n{\rm S}}(\sigma;0)$ with $\sigma$.}
\label{fig:2Sslope}
\end{figure}

A further illustration of the interplay between quantum mechanical effects and the collisions is provided by the evolution of the $1S$ and $2S$ vacuum states,  starting from an initial Gaussian wave packet with parameter $\sigma$ (see Eq.~(\ref{eq:compact}) and App.~\ref{app:L2alone}), and a medium temperature $T=300\,{\rm MeV}$. In Fig.~\ref{fig:2Sslope}, we compare the time derivatives of $p_{\rm 1S}$ and $p_{\rm 2S}$ at $t=0$ obtained by solving the full QME  with the values obtained form a calculation with $\mathcal{L}_2$ alone.  The values  $\sigma=0.55\,{\rm fm}$, $0.387\,{\rm fm}$ and $0.165\,{\rm fm}$ are representative of the situations described respectively in Fig.~\ref{fig:Proba1Ssingletthermal300} (right panel) for a initial in-medium 1S, Fig.~\ref{fig:Proba1Ssingletvacuum300} (upper right panel) for the vacuum 1S and Fig.\ref{fig:compactoctetevol} (right panel) for a compact singlet state. Also shown in Fig.~\ref{fig:2Sslope} are the curves corresponding to simple decay laws of the individual states, $\dot{p}_{n{\rm S}}(\sigma,0)=-\Gamma_{n{\rm S}}\;p_{n{\rm S}}(\sigma;0)$ (dashed lines). If the density matrix was of the form ${\cal D}^\sigma(t)=p_{\rm 1S}(t) \ket{1S}\bra{1S}+p_{\rm 2S}(t) \ket{2S}\bra{2S}$,  the probabilities $p_{\rm 1S}$  and $p_{\rm 2S}$ would decay according to these laws. Fig.~\ref{fig:2Sslope} shows that this is clearly not the case. Even in cases where the initial density matrix is nearly in this diagonal form, non-diagonal elements develop in time and alter the evolution of the probabilities in a non trivial way. This is discussed in details in App.~\ref{app:L2alone} in the case where the evolution is driven by $\L_2$ alone. The comparison displayed in Fig.~\ref{fig:2Sslope}  indicates that this approximate evolution with $\L_2$ alone  provides an accurate representation when $\sigma$ is not too different from $\sigma_{\rm 1S}$.  The effect of coherence of the initial Gaussian wave packet is particularly visible for small values of $\sigma$. For $\sigma \lesssim 0.5$, one sees very little evolution, in particular for the 2S. This is because the small size of the (coherent) wave packet hinders the effect of the dipolar transitions. Also, the 2S exhibits a non-uniform pattern, increasing for $\sigma \gtrsim 0.5$ and decreasing for $\sigma>\sigma_{\rm 1S}$. Such a non-uniform behavior can be attributed to the destruction of the coherence in the initial overlap by collisions with the medium constituents.  More details are given in App.~\ref{app:L2alone}.

\subsection{Asymptotic distributions and steady states}
\label{sec:3.B}

We have seen in Sec.~\ref{sec:2} that, at late time, the ${\rm c\bar{c}}$ density matrix  appears to reach a steady state, independent of the initial conditions. This steady state presents features of a thermal equilibrium state,\footnote{To our knowledge, there is no guarantee that the late time behavior of a system described by a Lindblad equation in the QBM regime always corresponds to a thermal equilibrium distribution expressed in terms of the energy levels like the canonical Gibbs-Boltzmann distribution. In fact, this subject is still debated nowadays (see e.g. \cite{Ghosh:2023ppc} and refs. therein). See also \cite{spohn1977algebraic,rivas2012open}.} with  however some (small) deviations (see  Fig.~\ref{fig:singletoctetpvstime}). The goal of this section is to provide some understanding of these deviations  and, more broadly, to better characterize the steady state. To do so, we shall examine the fixed point solutions of the QME, that is the solutions ${\cal D}_{\rm asymp}$ of the equation
\beq 
\L\, {\cal D}_{\rm asymp}=0.
\eeq
It will be interesting to contrast the solution of the full QCD equations with that of their abelian version, which we shall refer to (somewhat abusively) to QED, and which is obtained by setting ${\cal D}_{\rm s }={\cal D}_{\rm o }$ in the QCD equation for ${\cal D}_{\rm s}$, and take ${\cal D}_{\rm s}$ to be proportional to the abelian density matrix, which we shall denote  ${\cal D}_{_{\rm QED}}$. In order to ease the comparison of the numerical results, we keep the color factor $C_F$ in the expression of the singlet potential when solving the equation for ${\cal D}_{_{\rm QED}}$. Note finally that the normalization ${\rm tr}{\cal D}_{\rm s}+(N_c^2-1){\rm tr}{\cal D}_{\rm o}=1$ entails, when ${\cal D}_{\rm s}={\cal D}_{\rm o}$, that ${\rm tr}{\cal D}_{\rm s}=1/N_c^2$. As ${\cal D}_{_{\rm QED}}$ is normalized so that ${\rm tr}{\cal D}_{_{\rm QED}}=1$, one should therefore confront ${\cal D}_{_{\rm QED}}/N_c^2$ and ${\cal D}_{\rm s}$ when comparing QCD and QED. 

\subsubsection{Generic structure of ${\cal D}_{\rm asymp}$}
\label{sec:3.B1}
\begin{figure}[h]
\includegraphics[width=0.49\linewidth]{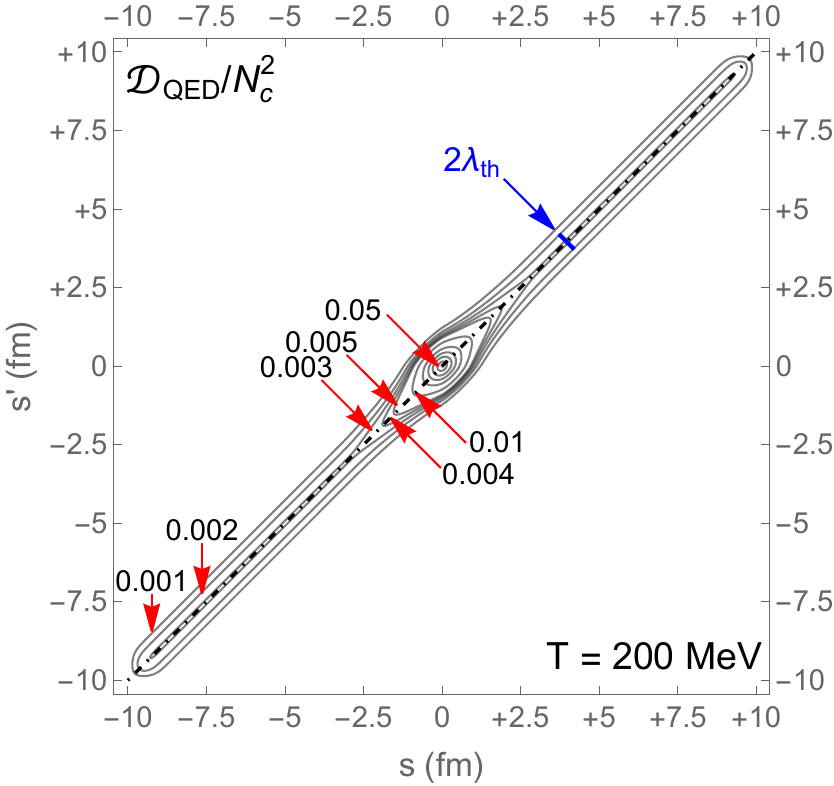}
\includegraphics[width=0.49\linewidth]{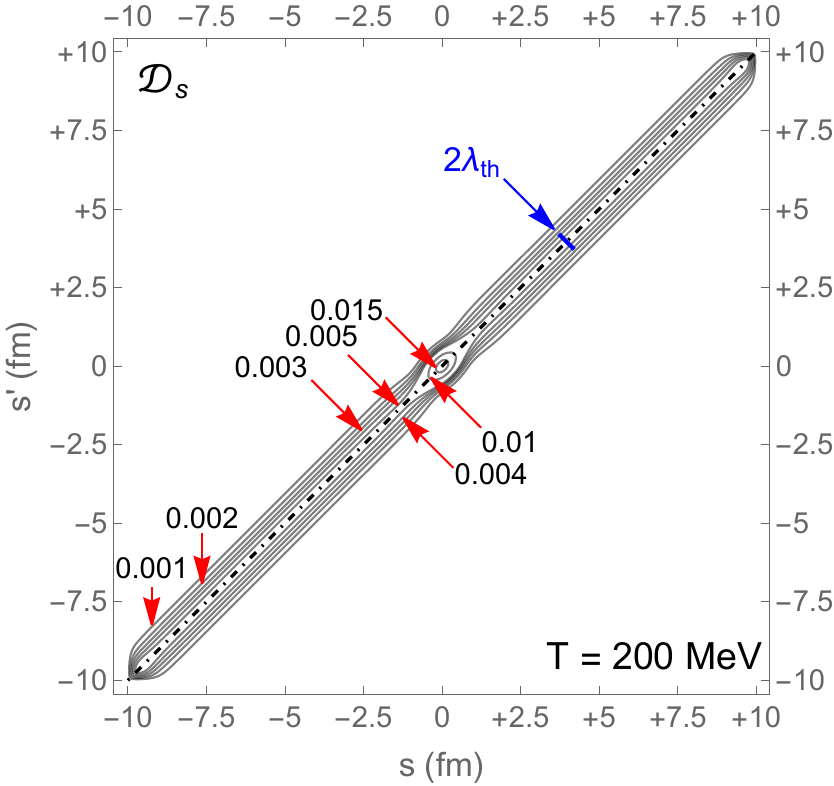}
\caption{Left: Contour plots of ${\cal D}_{_{\rm QED}}(s,s')/N_c^2$ (levels : $0.001, 0.002, \ldots, 0.005, 0.01, 0.02, \ldots, 0.05\,{\rm fm}^{-1}$); the skewed blue line across the main diagonal represents $2\lambda_{\rm th}$ for $T=200\,{\rm MeV}$. Right: Same for ${\cal D}_{\rm s}(s,s')$ (levels : $0.001, 0.002, \ldots, 0.005, 0.01, 0.015\,{\rm fm}^{-1}$).}
\label{fig:diagonalDss}
\end{figure}

A generic feature of the asymptotic solution, common to QED and QCD,  is illustrated in Fig.~\ref{fig:diagonalDss}. The nearly diagonal structure of the density matrix is the result of the collisions of the ${\rm c\bar{c}}$ pair with the surrounding medium constituents. We have associated this phenomenon to collisional decoherence and it is further discussed in App.~\ref{app:L2alone} where the role of the operator $\L_2$ is thoroughly discussed, as well as the important role of the coordinate space basis as the preferred basis in which decoherence is the fastest. The final width is eventually controlled by the thermal wavelength (in both cases of QCD and QED), as indicated in the figure. 

The peak seen near the origin is stronger in the case of QED than in the case of QCD. It reflects the correlations induced by the singlet attractive potential, which leads to an increase of the  density near the origin. 
To quantify this effect, one can consider the Wigner transform of the density matrix 
\begin{equation}
\mathcal{W}_{{\rm s/QED}}(r,p) =\int \rmd y\,e^{-i p\,y}\, 
\mathcal{D}_{\mathrm{s}/\mathrm{QED}}(r+\frac{y}{2},r-\frac{y}{2}). 
\label{eq:wignerdef}
\end{equation}
A semi-classical approximation  (see e.g. \cite{Blaizot:2017ypk}) can provide useful guidance at sufficiently high temperature. This approximation, in its leading order,  yields the following simple form for  the Wigner transform for the asymptotic distribution
\begin{equation}
\mathcal{W}_{_{\mathrm{SC}}}(r,p) \propto \exp\left(-\frac{p^2/M+V(T,r)}{T} \right).
\label{eq:WasymptQED}
\end{equation}
\begin{figure}[h]
\centering
\includegraphics[width=0.49\linewidth]{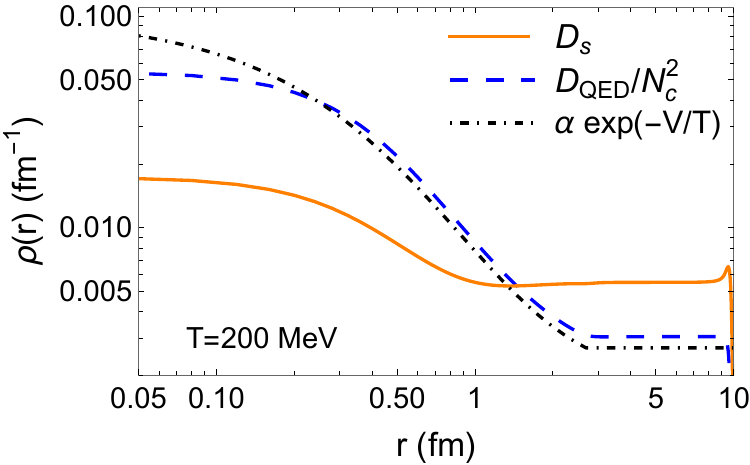} 
\includegraphics[width=0.49\linewidth]{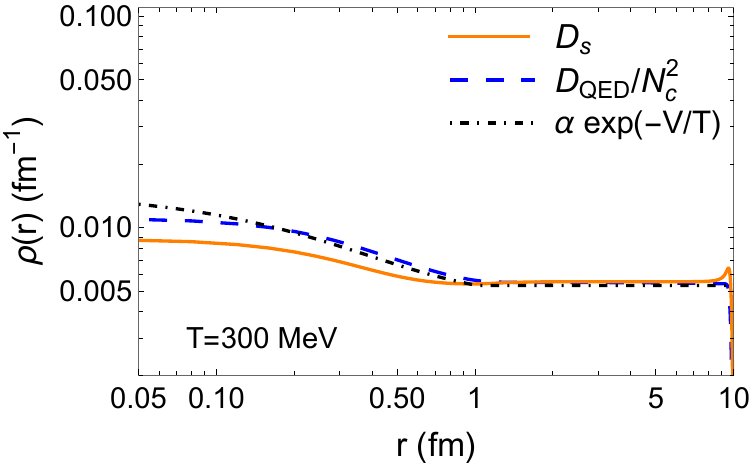}   
\caption{Left: Asymptotic density associated to the steady state of the QM equations for $T=200$~MeV: ${\cal D}_{\rm s}$ (solid curve), ${\cal D}_{_{\rm QED}}$ (dashed blue), and semi-classical density $\propto \exp(-V(T,r)/T)$ (dot-dashed black), properly normalized. Right: The same for $T=300$~MeV.}
\label{fig:rhoQEDT200}
\end{figure}

The corresponding density, $\rho_{_{\mathrm{SC}}}(r)\propto\exp\left(-V(T,r)/{T} \right)$,  is shown in Fig.~\ref{fig:rhoQEDT200} for the two temperatures, $T=200$~MeV and $T=300$~MeV. In the abelian case, the semi-classical expression (\ref{eq:WasymptQED}) indeed compares well with the exact  asymptotic density, except at very small $r$.\footnote{The agreement with the semi-classical expressions holds provided the potential $V$ is bounded from below, which is the case for the one-dimensional potential used in our calculations.} At $T=200$~MeV, there is a significant difference between the abelian and the non-abelian cases. In the QCD case, the singlet to octet transitions, that are particularly efficient at large $r$, bring  the pedestal of continuum states near its equilibrium value ($\simeq (1/9) (1/20)$ fm$^{-1}$). In the QED case, the pedestal is smaller. This can be attributed to the attraction of the potential which maintains a significant fraction of pairs concentrated at small distance. The picture changes somewhat at $T=300$~MeV. At this temperature, collisions are sufficiently strong in QED to balance the effect of the attractive potential, and the pedestal is equilibrated in both cases. Correlatively the magnitude of the peak is significantly reduced. 

\begin{figure}[h]
\includegraphics[width=0.65\linewidth]{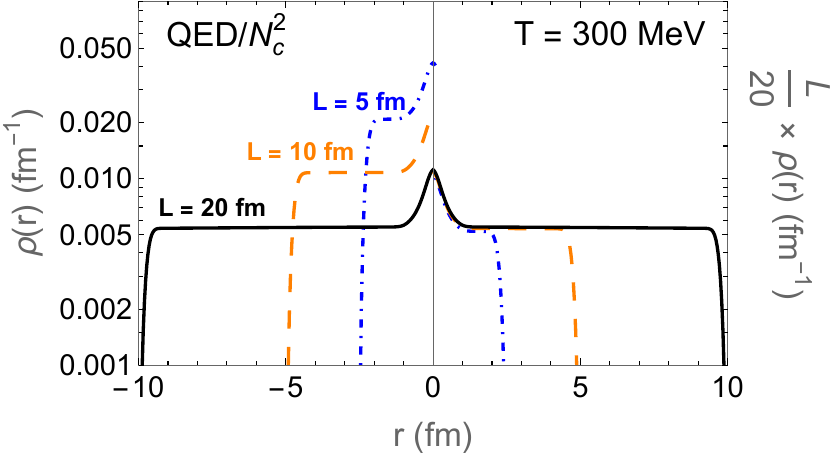}
\caption{The density $\rho(r)$ for various sizes of the box ($L=$ 5 fm, 10 fm and 20 fm); QED case. For positive values of $r$, a multiplicative factor  ${L}/{20}$  has been applied to illustrate the scaling of the central peak with the pedestal.}
\label{fig:boxsizedep}
\end{figure}

The density in the steady state depends on the size $L$ of the box where the system is confined. One expects in particular that in a box of infinite length, the weight of the bound state vanishes. The dependence on $L$ of the density $\rho(r)$ is  illustrated in Fig.~\ref{fig:boxsizedep} in the abelian case. As $L$ increases, the shape of the peak remains essentially unaltered. This is because this peak is dominantly composed of localized bound states that do not depend too much on the box size. We may write $\rho(r)=c_1\,\rme^{-V(r)/T}$, with $c_1$ a normalization constant. That is, the density is essentially constant outside the region of the peak (assuming that there $V(r)=0$). The integral of the density is composed of two contributions: the peak contribution  $\sim c_1 a_0 \rme^{V_0/T}$ -- where $a_0$ is the width of the peak and $-V_0$ is some average value of the potential over the peak -- and the continuum contribution $\sim c_1(L-a_0)$. Now, the normalization is kept fixed as the size of the box increases, $\int\rmd r\, \rho(r)=1\simeq c_1\left[L-a_0+  a_0 \rme^{V_0/T}\right]$. It follows that when $L$ is large enough ($L\gg a_0$), $c_1\sim 1/L$ and the weight of the peak becomes indeed a negligible contribution to the total weight. Note however that the relative magnitude of the peak with respect to the pedestal is mostly determined by the potential $V(r)$ and is essentially independent of $L$. This is indeed illustrated in Fig.~\ref{fig:boxsizedep}.

\subsubsection{Asymptotic momentum distribution}
\label{sec:3.B2}

\begin{figure}[h]
\centering
\includegraphics[width=0.6\linewidth]{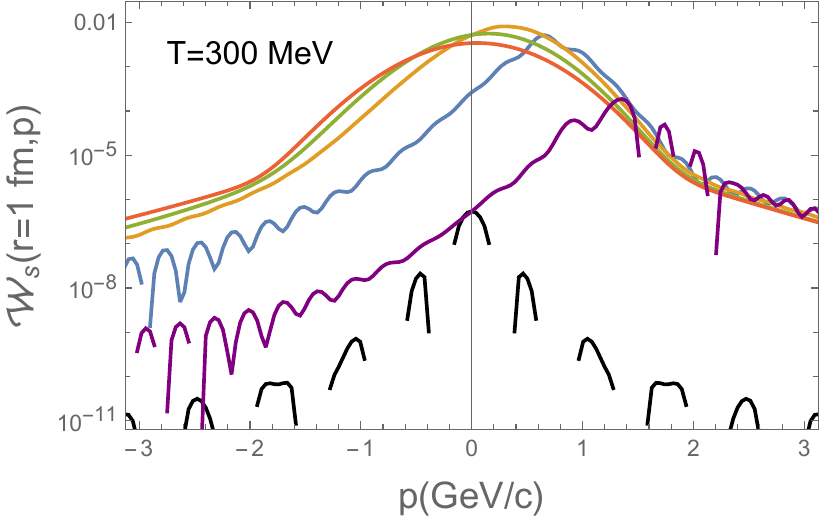}
\caption{The Wigner distribution $\mathcal{W}_{\rm s}(r,p)$ for $r=1$~fm and $T=300$~MeV, showing the emergence of the Maxwell distribution, becoming gradually peaked at $p=0$ as time increases: $t=0$~fm/$c$ (black), 1~fm/$c$ (purple), 2~fm/$c$ (blue), 5~fm/$c$ (orange), 10~fm/$c$ (green), and 20~fm/$c$ (red). For early times, the Wigner distribution presents oscillations reaching negative values that  cannot be displayed in the semi-log plot.}
\label{fig:WignerDistribution}
\end{figure}

\begin{figure}[h]
\centering
\includegraphics[width=0.49\linewidth]{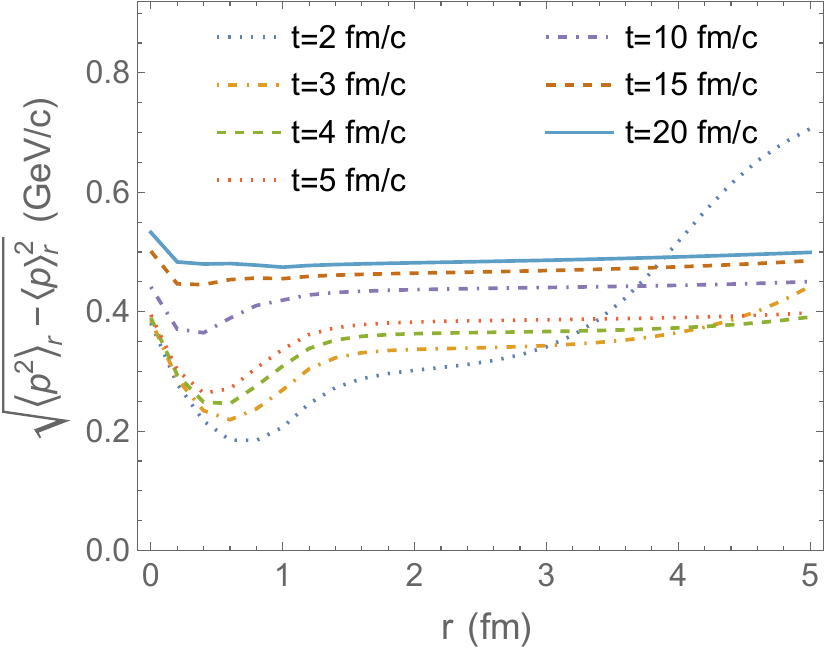}
\includegraphics[width=0.49\linewidth]{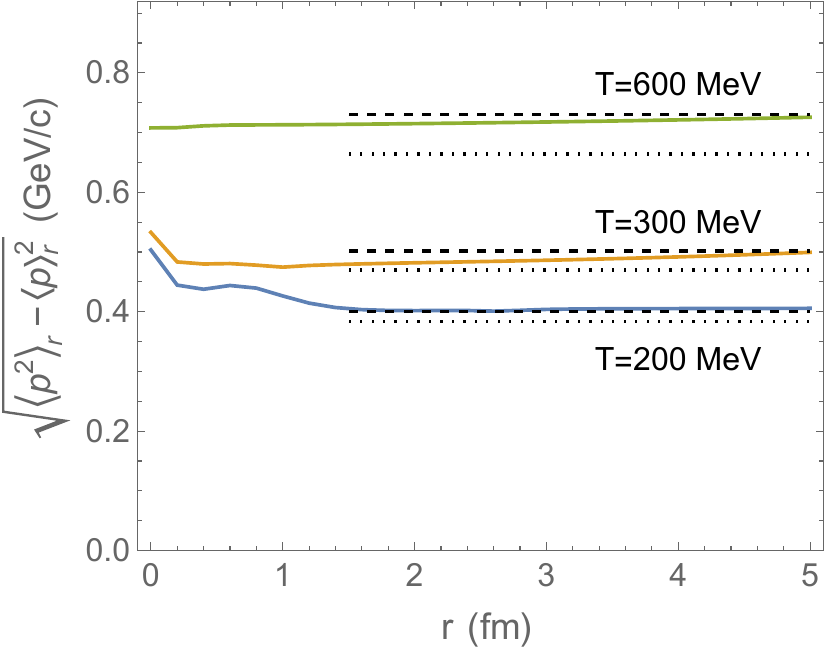}
\caption{Left: Time evolution of the standard deviation of the momentum distribution as a function of $r=(s + s')/{2}$ for a fixed medium temperature $T = 300$~MeV. Right: Same in the large time limit at for three different medium temperature $T = 200$~MeV ($t = 60$~fm/$c$), 300~MeV ($t = 20$~fm/$c$) and 600~MeV ($t = 20$~fm/$c$). The black dotted lines correspond to the classical result $\sqrt{MT/2}$ while the black dashed lines correspond to the corrected value $\sqrt{MT/(2+\gamma)}$ (see text for details). }
\label{fig:MomentumTDep}
\end{figure}

At large relative distance, where the potential becomes essentially constant, the Wigner transform predicts, in the semi-classical approximation  (\ref{eq:WasymptQED}), that the momentum distribution is a Maxwell distribution. This is indeed the qualitative behavior that one observes in Fig.~\ref{fig:WignerDistribution}, even for distances as small as $r=1$ fm from the origin. 
The standard deviation $\sqrt{\langle p^{2}\rangle_r - \langle p \rangle_r^{2}}$  is shown as a function of $r$ in the left panel of Fig.~\ref{fig:MomentumTDep}. One sees that  the classical expectation $p_{\mathrm{th}}\simeq\sqrt{{M T}/{2}}\approx 0.5\,\mathrm{GeV}/c$ is reached at late times, for $r\gtrsim 2$~fm. The relaxation is somewhat slower for the region $r\lesssim 2$~fm/$c$ which remains under the influence of the potential for a longer time. 
The influence of the temperature on the asymptotic state is illustrated on the right panel of Fig.~\ref{fig:MomentumTDep}. Focusing on the large values of $r$ ($\gtrsim 2$~fm/$c$), one observes small but definite deviations with respect to the Maxwell-Boltzmann distribution, which do not disappear at the highest temperature considered. Their origin is clarified in App.~\ref{app:equilibrium} where it is shown that the variance of $p$ at large relative distance is given by 
\begin{equation}
\mathrm{var}(p)= \frac{M T}{2+\gamma}
    \quad\text{with}\quad \gamma = \frac{\tilde{W}^{(4)}(0)}{16 MT\,\tilde{W}''(0)}.
\end{equation}
The deviations from the expected value $\frac{M T}{2}$ are encoded in the coefficient $\gamma$ which involves 2nd and 4th derivatives of the imaginary potential at the origin.\footnote{This coefficient $\gamma$ should not be confused with the mass-shift introduced in some QME approaches.} These derivatives depend on the prescription adopted to control the singular behavior of $W(r)$ near $r=0$.\footnote{For well-behaved $W$, one would expect $\frac{\tilde{W}^{(4)}(0)}{\tilde{W}''(0)} \sim m_D^2$, resulting in a subleading value of $\gamma$ in the QBM regime.} For the choice made in App.~\ref{app:equilibrium}, we have $\frac{\tilde{W}^{(4)}(0)}{\tilde{W}''(0)} \sim -\frac{3\,q_{\mathrm{max}}^2}{20\ln(q_{\mathrm{max}}/m_D)}$, with $q_{\mathrm{max}} \gtrsim m_D$ (see Eq.~(\ref{eq:defphireg})).\footnote{The correction $\gamma$ depends on a choice of regularization and  is cut-off dependent. In~\cite{Blaizot:2015hya}, a fixed cutoff with a large value,  $q_{\mathrm{max}}= 4$~GeV, is adopted. As a consequence, $\gamma$ decreases $\propto \frac{q_{\mathrm{max}}^2}{MT\ln(q_{\mathrm{max}}/(CT)})$ and the Maxwell-Boltzmann distribution is restored at ``large" temperatures $\sim \frac{q_{\mathrm{max}}^2}{M}$. The drawback of such a prescription is that it leads to large deviations for $T\simeq T_{_\mathrm{QBM}}$, up to $50\%$ for c-quarks (depending on the precise value of $m_D$). In the present work, the definition adopted for the regularized $\tilde{W}$ is equivalent to take $q_{\mathrm{max}}= \sqrt{M T}$, which yields a milder  temperature dependence.} The dashed lines in Fig.~\ref{fig:MomentumTDep} show that the resulting estimate of $\gamma$ is in good agreement with the numerical simulations.

Besides the deviations just noted for ${\rm var}(p)$, one should also mention that the equilibrium distribution exhibits large deviations from the Maxwell-Boltzmann distribution when $p\gg \sqrt{MT}$,  as seen e.g. on Fig.~\ref{fig:effectvariousorders}, and Fig.~\ref{fig:probofeigen} below (see also the discussion after Eq.~(\ref{eq:equilL4MS}) in App.~\ref{app:equilibrium}). The asymptotic solution in the semi-classical approximation is, to within a normalizing factor, $\mathcal{D}_s(y) =\tilde{y}^{\eta} K_\eta(\sqrt{\eta} \tilde{y})$, where $\eta = \frac{1}{2}+\frac{1}{|\gamma|}$ and $\tilde{y}=\sqrt{M T}\,y$. The Wigner transform is $\propto (p^2+\eta)^{-(\frac{1}{2}+\eta)}$. When $\gamma \to 0$, one recovers the correct MB distribution, but for finite $\gamma$,  the decay is algebraic, in line with that of the Wigner transform of the QME, as illustrated for instance in Fig.~\ref{fig:WignerDistribution}.

We thus conclude that the QME used in the present paper lead to steady state momentum distributions that deviate slightly from thermal distributions. Analogous  deviations have been pointed out recently by several authors~\cite{PhysRevB.102.115109,Tupkary:2021bcd}, who have also suggested methods to cure them. We shall not explore these here, since these deviations remain quite small in the present setting,  and they have no impact on the main physics results presented in this paper.

\subsubsection{Analysis in terms of in-medium eigenstates}
\label{sec:3.B3}
\begin{figure}[h]
\includegraphics[width=\linewidth]{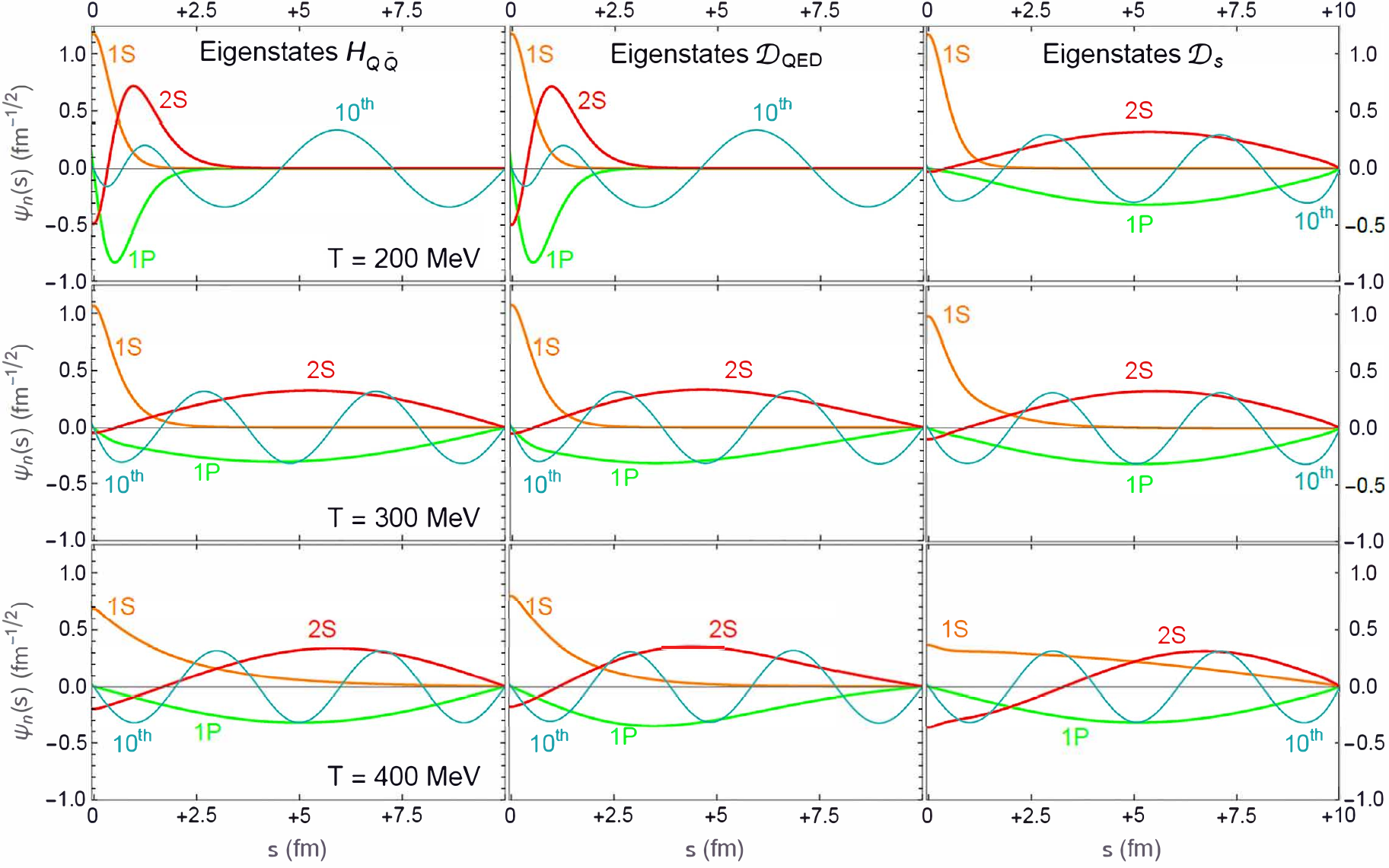}   
\caption{Left panels: Eigenstates (the first 3 as well as the 10th one) of the unitary ${\rm c\bar{c}}$ Hamiltonian (with the $T$-dependent potential); central and right panels: Eigenstates of the asymptotic ${\cal D}_{_{\rm QED}}$ density matrix in the abelian case (central) and of the asymptotic $\mathcal{D}_{\rm s}$ for QCD (right), for $T=200\,{\rm Mev}$ (top), $T=300\,{\rm Mev}$ (middle) and $T=400\,{\rm Mev}$ (bottom). The eigenstates of ${\cal D}_{_{\rm QED}}$ and of $\mathcal{D}_{\rm s}$ are generically complex-valued. For the selected temperatures, eigenstates of $\mathcal{D}_{\rm s}$ have a small imaginary part which has not been represented. See text for comments.}
\label{fig:spectrumD}
\end{figure}

We now turn to the analysis of  $\mathcal{D}_{\mathrm{\rm asymp}}$ in terms of its projections $p_n=\langle n |\mathcal{D}_{\mathrm{\rm asymp}}|n\rangle$ on the eigenstates $|n\rangle$ of the in-medium Hamiltonian. The spectrum of the first three eigenstates of the in-medium Hamiltonian are displayed in the left panel of Fig.~\ref{fig:spectrumD} for three values of the temperature. At $T=200$ MeV, one can identify three localized states that correspond to the three bound states. At the higher temperatures, $T=300, 400$ MeV, only the 1S state remains localized, and its size is seen to increase with temperature. The 2S and 1P states are completely delocalized already at $T=300$ MeV. And so are the higher eigenstates, which are continuum states spread over all the available volume of the simulation box. For comparison, we have also plotted the 10th eigenstate which displays a regular plane-wave behavior, with a slight distortion near the origin which gradually disappears as the temperature increases. 

The Hamiltonian eigenstates are compared to the eigenstates of the abelian (middle column) and non-abelian (right column) density matrices. A striking similarity with the Hamiltonian eigenstates can be observed in the abelian case, suggesting that, in this case, the asymptotic density matrix is diagonal in the Hamiltonian eigenbasis. 
This is not so in the QCD case, as can already be observed at $T=200$ MeV: While the 1S looks quite similar to the 1S eigenstate of the Hamiltonian, the 2S and 1P states are not localized states, but are completely delocalized; they are best viewed as continuum states. We note also that the spreading of the 1S state is much more rapid with increasing temperature than in the abelian case, while at $T=400$~MeV, an apparent shift of the 1S and the 2S  weights to higher values of $s$ is observed.

\begin{figure}[h]
\centering
\includegraphics[width=0.49\linewidth]{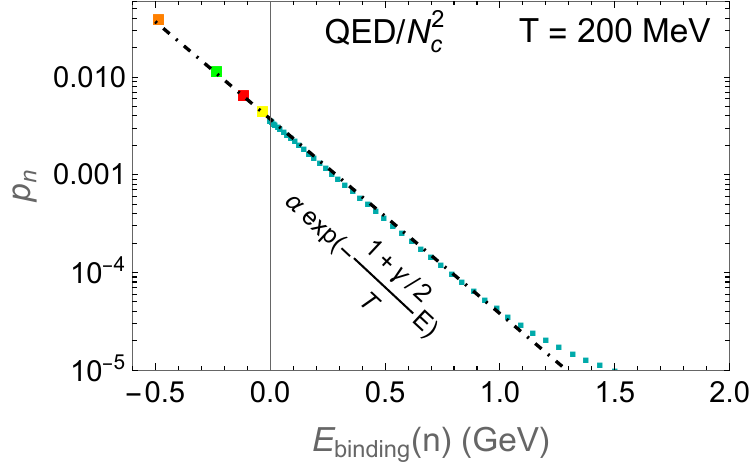}    
\includegraphics[width=0.49\linewidth]{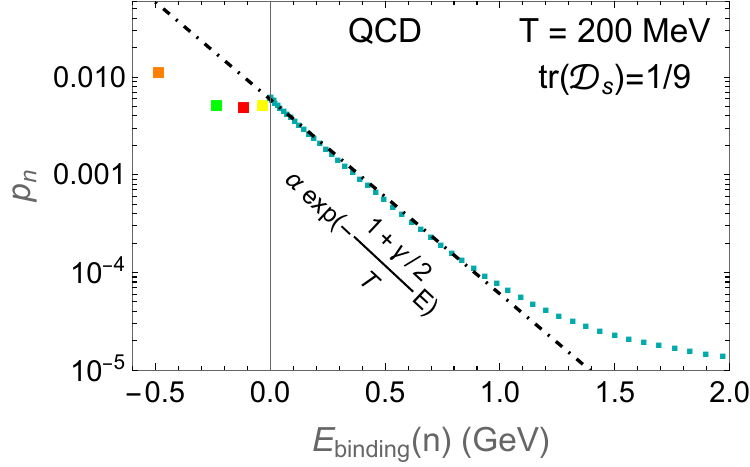}\\
\includegraphics[width=0.49\linewidth]{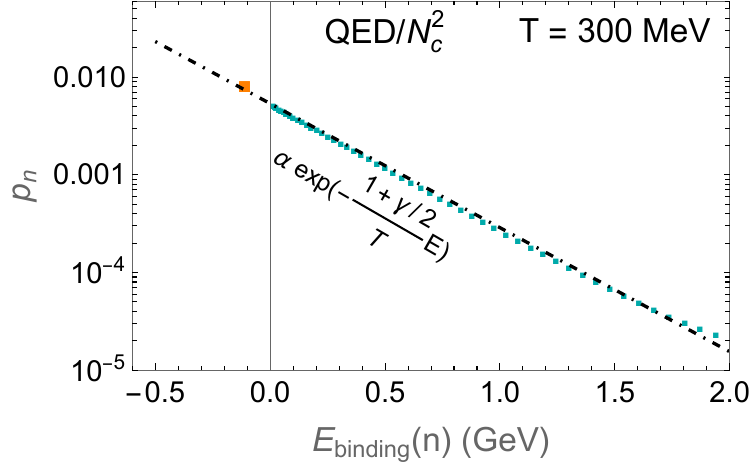}     \includegraphics[width=0.49\linewidth]{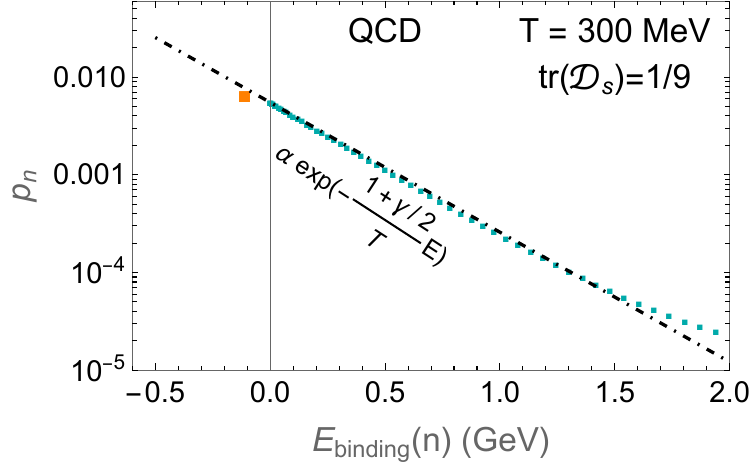}
\caption{Probability to find in-medium eigenstates of the screened Hamiltonian in the asymptotic density operator $\mathcal{D}$ for either the abelian case (left) or QCD (right), for both $T=200\,{\rm MeV}$ (top) and $T=300\,{\rm MeV}$ (bottom). }
\label{fig:probofeigen}
\end{figure}

These features have a strong impact on the interpretation, in terms of Boltzmann weights, of the probabilities $p_n$ to find a given in-medium eigenstate in the asymptotic density matrix.
If $\mathcal{D}_{\mathrm{\rm asymp}}$ was close to a thermal equilibrium distribution, one would expect $p_n$ to be proportional to the Boltzmann factor,  $p_n \propto\exp(-E_n/T)/Z$. This is indeed what one observes in the abelian case, for all the states with $E\lesssim 1$ GeV  including the bound states, as illustrated in the left panel of Fig.~\ref{fig:probofeigen}. The QCD case presents a similar pattern, except for the lowest eigenstates at $T=200$ MeV which significantly depart from the general trend. A similar behavior can be observed in  the results of ~\cite{Miura:2019ssi,Miura:2022arv}, although a detailed comparison is difficult, given the differences in the potentials and in the numerical techniques employed to solve the QME. We note that deviations are observed only at the lowest temperature $T=200$ MeV, and the 1S bound state exhibits thermal behavior at $T=300$~MeV. We do not have a clear interpretation of this observation. Clearly $T=200$ MeV is at the borderline of the QBM regime, and this could be the signal that some physical effects are missed by the present QME \cite{Akamatsu:2020ypb,Blaizot:2021xqa}. However, one may also note that at $T=200$ MeV,  the eigenstates of $H_{Q\bar Q}$ differ significantly for those of the density matrix ${\cal D}_{\rm s}$, in contrast to the QED case (see Fig.~\ref{fig:probofeigen}). Clearly this issue deserves further investigation. 

\begin{figure}[h]
\centering
\includegraphics[width=0.49 \linewidth]{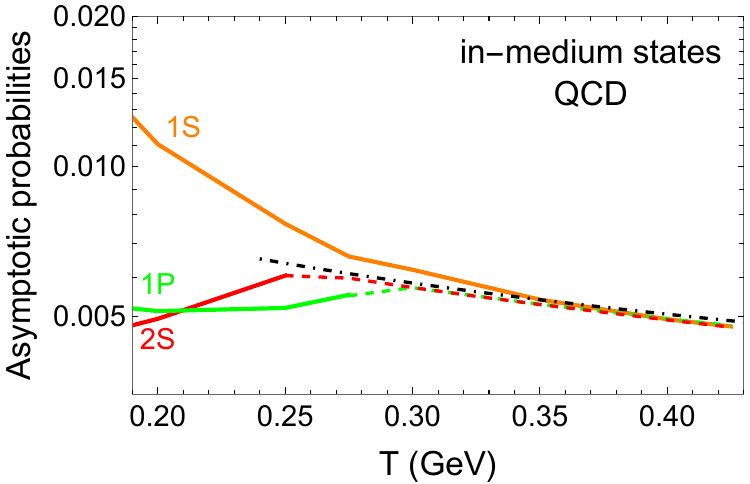}
\includegraphics[width=0.49 \linewidth]{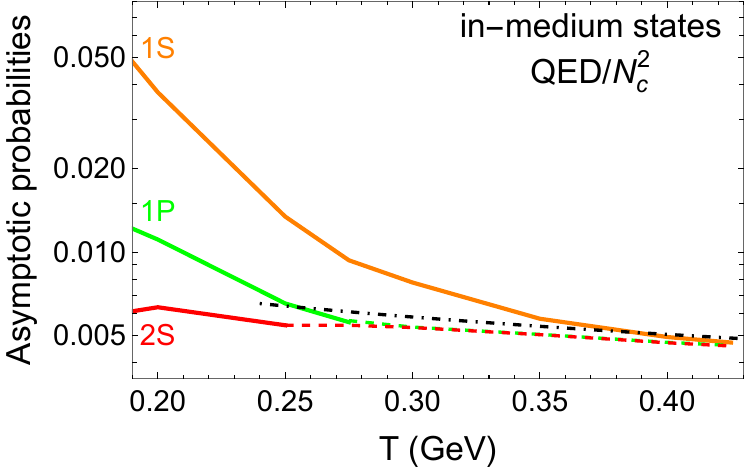}
\caption{Left: The probabilities $p_{_{1\mathrm{S}}}$ (orange), $p_{_{1\mathrm{P}}}$ (green) and $p_{_{2\mathrm{S}}}$ (red) to observe asymptotically  the ${\rm c\bar{c}}$ pair in the corresponding in-medium singlet state, as a function of the medium  temperature $T$ (solid lines); for each state, the transition from a solid to a dashed curve corresponds the melting temperature. The (black) dashed-dotted line is the approximation (\ref{eq:pnapproxSC}). Right: The same in the abelian approximation. } 
\label{fig:probstatio}
\end{figure}

The variation with temperature of the in-medium probabilities $p_n$ is  illustrated in Fig.~\ref{fig:probstatio},  for both the QCD and the abelian cases. Also shown in this figure is the analytic continuation of the probabilities (dashed lines) when these states cease to be bound. One expects that when the temperature is high enough all states are fully delocalized in the box. Then a simple calculation allows us to get the value of $p_n$. We have indeed
\beq
p_n=\int\rmd s\rmd s' \psi_n^*(s) {\cal D}_{\rm s}(s,s') \psi_n(s'),
\eeq 
where $\psi_n(s)$ is the wave function of one of the delocalized state. 
When the temperature is large enough, we expect the density matrix to become diagonal quickly, to within the thermal wavelength $\lambda_{\rm th}$ which is much smaller that the typical range of variation of $\psi_n(s)$, which, for the states considered, is of the order of a fraction of the size of the box, i.e. $\lambda_{\rm th}\ll L$. For instance at $T=300 $ MeV,  $\lambda_{\rm th}\simeq 0.3$~fm, while $L=20$~fm. Equivalently the typical momenta that control the dependence of ${\cal D}(s,s')$ on $|s-s'|$ are large compared to those in $\psi_n$ (the typical momentum components of $\psi_n$  are of the order of a few times $\pi/L$).  Under these conditions, we can approximate
\beq
p_n&=&\int\rmd r\rmd y\, \psi_n^*(r+y/2)\psi_n(r-y/2) {\cal D}_{\rm s}(r,y) \nn 
&\approx& \int\rmd r\rmd y\,\psi_n^*(r)\psi_n(r){\cal D}_{\rm s}(r,y).
\eeq 
For ${\cal D}_{\rm s}(r,y)$ we can use the following factorized expression, which reproduces the Maxwell-Boltzmann momentum distribution with $\langle p^2\rangle=MT/2$, 
\beq 
{\cal D}_{\rm s}(r,y)=\rho_{\rm s}(r) \rme^{-\frac{y^2}{4\lambda_{\rm th}^2}}
\eeq 
Noticing that in the considered regime the density is approximately constant $\rho_{\rm s}(r)\simeq (1/N_c^2)(1/L)$, and assuming $\phi_n(r)$ to be normalized to unity, one gets
\beq\label{eq:pnapproxSC}
p_n\approx \frac{2\sqrt{\pi}}{N_c^2}\frac{\lambda_{\rm th}}{L} .
\eeq 
As can be seen in Fig.~\ref{fig:probstatio}, this simple formula accounts quite well for the values of the probabilities at temperatures $T\gtrsim 300\,{\rm  MeV}\approx T_{\rm damp}$. In fact the same agreement is found for the abelian case, but at higher temperature. This may be attributed to the fact that, in QCD, color singlet-octet  transitions accelerate the delocalization of states, as is visible for instance in Fig.~\ref{fig:spectrumD}.

\section{\label{sec:conclu}Conclusion}

We have solved numerically the non-abelian quantum master equations derived in Ref.~\cite{Blaizot:2017ypk} in the quantum Brownian regime. These can be put in the form of a Lindblad equation. This was used to study  the dynamics of a heavy charm-anticharm pair in a Quark-Gluon Plasma in thermal equilibrium. The operators entering these equations were simplified in order to cope with various artificial divergences induced by successive derivatives of the complex potential. The resulting ``minimal'' set of operators conserves the basic property of a Lindblad equation, in particular the positivity of the evolution. The specific role of the  operator $\L_4$ in guaranteeing this property was emphasized.  The equations were solved in a one-dimensional setting, exploiting the  complex potential specifically developed for one-dimensional studies in \cite{Katz:2022fpb}. This potential is adjusted so as to yield numbers for physical quantities that are comparable to those of a three dimensional potential. This allowed us, along the paper, to make remarks of phenomenological interest, although a detailed phenomenological study is beyond the scope of this paper. 

The evolution of the reduced density matrix of the ${\rm c\bar{c}}$ pair was studied for different initial conditions of the pair (size and color), and various temperatures of the quark-gluon plasma. The case of an expanding plasma was also briefly analyzed. A generic feature of the evolution is the fact that collisions lead to a rapid squeezing of the coordinate space density matrix along its diagonal, a characteristic effect of decoherence making part of the semi-classical dynamics. It also underlines the role of coordinate space as a ``preferred basis'' for the analysis of the dynamics.  The color dynamics, captured by coupled equations that connect the singlet and octet sectors of the reduced density matrix, involve a subtle coupling between coordinate space and color: this is because of the coordinate space dependence of the color-dipolar interactions that control the overall dynamics. Also, the continuum states play an important role. They are quickly populated, and it is through them that color equilibrates the fastest. In contrast, since dipolar interactions vanish at vanishing relative distances, color equilibrates on a much longer time scale in the region occupied by bound states, which enhances the survival probability of the bound states. Two subregimes were identified in the quantum Brownian regime. For $T\gtrsim T_{\rm damp}$ (overdamped subregime), one observes the dominance of the collisional decoherence over the binding effects, while for $T\lesssim T_{\rm damp}$ the vacuum bound states gain in significance as the temperature decreases. Much of the qualitative features of the overdamped regime can be understood from the dominant role in this regime of a specific operator of the QME, the operator $\L_2$. 

We have shown that at late time the density matrix evolves toward a steady state that is close to a thermal equilibrium state if the temperature is high enough. Because of dissipation the typical range of the non-diagonal matrix elements of the reduced density matrix ends up being of the order of the thermal wavelength. The equilibrium state is that of a bound pair in equilibrium with continuum states that occupy the box where the simulation is done. The momentum distribution for the continuum states is close to a Maxwell-Boltzmann distribution with a small deviation that depends on the details of the complex potential. For the lowest temperature available in our calculations, one finds that the populations of the bound states deviate significantly from the Boltzmann factors.  This is to be contrasted with the abelian situation where the populations of the various states are nearly perfectly given by the Boltzmann weights. This feature, and more broadly the nature of the steady state, deserve further study.

The analysis in terms of  projections on charmonium eigenstates as a function of time provided further insight. By projecting on vacuum eigenstates one gets access to the probability to find a particular charmonimum state in the plasma as a function of time, a quantity which is often calculated in phenomenological studies. The projections onto the in-medium screened Hamiltonian provides a complementary view of the system properties. This is complicated by the fact that such bound states quickly disappear in favor of continuum states as the temperature increases.  The 1S state, however, survives for a longer time, and its evolution dominates that of the density matrix. Thus the evolution of the latter can, to a large extent, be understood in terms of the evolution of the 1S state. In fact,  the very  first instants of the evolution depend much on the initial state, and can exhibit quantum mixing before such effects start to be hindered by collisions. The later  stages of the evolution, on the other hand, have a universal character, driven essentially by the generic evolution of the coordinate space density matrix which, when analyzed in terms of projections on vacuum eigenstates, yields indeed a similar evolution of both the 1S and the 2S. 

There are several aspects of the present work that are worth investigating further. One of them concerns the nature of the steady state, as mentioned above. Another one is related to the semi-classical dynamics and its interplay with color dynamics. As pointed out in \cite{Blaizot:2017ypk} this interplay  hinders the straightforward application of semi-classical approximations that have proven to be successful in the abelian case (see e.g. \cite{Blaizot:2015hya}).  Having at our disposal complete solutions of the QME constitutes a benchmark that could be very helpful for deepening our understanding of this issue. At the lowest temperature that we have explored, we have seen that the eigenstates of the asymptotic QCD density matrix differ from those of the in-medium Hamiltonian, in contrast to what happens in the analogous abelian situation.  This points possibly to missing ingredients in the QME, which would be needed for instance to cope with the energy gaps between various bound states when these bound states are playing a prominent role in the dynamics \cite{Blaizot:2018oev}. Ways to improve the present QME in order to address this issue are known and implementing them in the present context would represent an interesting extension of the present study. Finally   the tools that have been developed for the present study may constitute a useful playground for a number of more phenomenologically oriented studies.

\begin{acknowledgments}

We wish to thank Aoumeur Daddi Hammou and Joerg Aichelin for valuable discussions. The authors thank the Région Pays de la Loire and Subatech for support. S.D.~is supported by the Centre national de la recherche scientifique (CNRS) and Région Pays de la Loire and acknowledges the support of Narodowe Centrum Nauki under grant no. 2019/34/E/ST2/00186. R.K.~was under contract No.~2015-08473. PBG is supported by the European Union’s Horizon 2020 research and innovation program under grant agreement No 824093 (STRONG-2020).
\end{acknowledgments}

\appendix
\section{Positivity and structure of the QME}
\label{app:positivity}
In this appendix, we sketch the proof that, once the term $\L_4$ is included, the equations  used in this paper preserve the positivity of the density matrix.\footnote{Although the proof holds for any dimension, we focus on the 1D case when it comes to specific expressions.} To that aim, we first note that the  Liouville superoperator $\mathcal{L}$ of Eq.~(\ref{eq:2.1}) can be written explicitly in terms of sub-operators that represent the various color transitions:
\begin{align}\label{eq:A.2}
    \frac{\rmd}{\rmd t}\mathcal{D}_{\rm s} &= \mathcal{L}^{ss}\mathcal{D}_{\rm s} + \mathcal{L}^{\rm {so}}\mathcal{D}_{\rm o} \nonumber \\ 
    \frac{\rmd}{\rmd t}\mathcal{D}_{\rm o} &= \mathcal{L}^{\rm {os}}\mathcal{D}_{\rm s} + \mathcal{L}^{\rm {oo}}\mathcal{D}_{\rm o}.
\end{align}
In turn, each of these sub-operators can be decomposed in terms of the $\L_i$'s defined in Eq.~(\ref{eq:2.2}). 
We shall be concerned in this appendix only with the $\L_i$'s which are proportional to the imaginary part $W$ of the complex potential and its spatial derivatives, namely $\L_2$, $\L_3$ and $\L_4$. These are the operators that are involved in the positivity constraint that we want to study ($\L_0$ and $\L_1$ are unitary operators which preserve positivity). Recall that the operators $\L_2$, $\L_3$ and $\L_4$  have ordered dependence on the mass of the heavy quark, 
$\mathcal{L}_2\propto \mathcal{O}(M^0)$, $\mathcal{L}_3\propto \mathcal{O}(M^{-1})$ and $\mathcal{L}_4\propto \mathcal{O}(M^{-2})$. The same ordering holds for the color transition operators introduced above. As an illustration consider the operator $\mathcal{L}^{\rm so}$ which realizes transitions from  octet to singlet states. Using the convention of Ref.~\cite{Blaizot:2017ypk}, we write this operator  as follows:
\begin{equation}\label{eq:calLso}
\mathcal{L}^{\rm so} = C_F \left(\mathcal{L}_{2b}+ \mathcal{L}_{3b} + \mathcal{L}_{4b}\right),
\end{equation}
where the superoperators of type $\mathcal{L}_b$ correspond to one-gluon exchange diagrams in which the gluon propagator connects quark lines that are located on both sides of the Schwinger-Keldysch contour.  Referring to  Eq.~(\ref{eq:2.2}), and to the explicit expressions listed in App.~\ref{app:transition_operators}, one sees that $\mathcal{L}_{2b}$ just involves $W$ while   $\mathcal{L}_{3b}$ and $\mathcal{L}_{4b}$ contain both derivatives of $W$ and derivatives acting on $\mathcal{D}_{\rm o}$. The derivative content of the various operators is summarized in  table~\ref{table:structure}.
\begin{table}[H]
\begin{center}
\begin{tabular}{ |c|c|c|c| } 
 \hline
& $\mathcal{D}_{\rm o}$ & $\partial\mathcal{D}_{\rm o}$  & $\partial^2\mathcal{D}_{\rm o}$\\
\hline
$\vphantom{\sum^{N^N}}\mathcal{L}_{2b}$ & $\mathcal{L}_{2b,{\rm cst}}$: $\times W$ & 0 & 0 \\[2mm] 
$\mathcal{L}_{3b}$ &$\mathcal{L}_{3b,{\rm cst}}$:  $\times W^{(2)}$ & $\mathcal{L}_{3b,{\rm lin}}$: $\times W^{(1)}$ & 0\\[2mm] 
$\mathcal{L}_{4b}$  &$\mathcal{L}_{4b,{\rm cst}}$:  $\times W^{(4)}$ & $\mathcal{L}_{4b,{\rm lin}}$: $\times W^{(3)}$ & $\mathcal{L}_{4b,{\rm quad}}$: $\times W^{(2)}$\\[2mm] 
 \hline
\end{tabular}
\end{center}
\caption{Structure of the various superoperators $\mathcal{L}_{2b}$, $\mathcal{L}_{3b}$ and $\mathcal{L}_{4b}$ in terms of their spatial derivative content,  generically written as $\mathcal{D}_{\rm o}$, $\partial\mathcal{D}_{\rm o}$  and $\partial^2\mathcal{D}_{\rm o}$. $W^{(i)}$ denotes the $i$th derivative of $W$. The subscripts ``cst'', ``lin'' and  ``quad'' refer to the number of derivatives acting on $\mathcal{D}_{\rm o}$, respectively 0, 1, and 2.  }
\label{table:structure}
\end{table}

For the physical interpretation of the various superoperators,\footnote{See also \cite{Akamatsu:2020ypb}, Sec.~3.2.1.} we note that, in the large mass limit, the sole effect of the collisions is to shift the momentum of the heavy quarks and to rotate their  colors. The dominant superoperator in this limit is $\L_2$. Finite-mass corrections, which involve in particular recoil corrections that account for the energy conservation in individual collisions, yield the superoperators $\L_3$ and $\L_4$. The superoperator $\L_3$ contributes a term $\mathcal{L}_{3b,{\rm cst}}$ which has a structure similar to  $\mathcal{L}_{2b,{\rm cst}}$, namely it is independent of the heavy quark momenta, $\mathcal{L}_{3b,{\rm cst}}\sim \frac{q^2}{8MT}$; There is also in $\L_3$ a contribution that depends explicitly on the heavy quark momentum ${\vec p}_Q$,  $\mathcal{L}_{3b,{\rm lin}}\sim -\frac{\vec{q}\cdot\vec{p}_Q}{4MT}$, which leads to  dissipation. Similar considerations apply to the superoperator $\L_4$. 

To proceed, it is convenient to express $W(r)$ in terms of the following spectral representation
\begin{equation}
W(r)=\int_{-\infty}^{+\infty} \rmd q \,\rho(q)(1-\cos qr )=
\int_{-\infty}^{+\infty} \rmd q\, \rho(q)(1-e^{-i qr} ),
\label{spectraldec}
\end{equation}
where the spectral density $\rho (q)$ is an even function of $q$. Its explicit form is actually not needed for the argument of this appendix, aside from the fact that it is a positive definite function of the (one-dimensional) momentum $q$.\footnote{In the case of the HTL imaginary potential of Eq.~(\ref{eq:WHTL}),  $\rho(q)=\frac{\alpha_S T}{2 m_D} 
\left[\frac{\pi}{2}- \tan^{-1}\left(\frac{|q|}{m_D}\right)-\frac{m_D |q|}{m_D^2+q^2}\right]$, which is indeed $>0$.} 

The positivity property is in principle guaranteed by the Lindblad structure of the master equations (\ref{eq:2.2}). However, it is instructive to examine concretely how this is realized in the present case. To that aim, we shall look at the transition rates and verify that they are all positive  definite. We shall do that by focusing on the set of superoperators  (\ref{eq:calLso}) which describe  the octet  to singlet transitions. In the basis that, at time $t$, diagonalizes the density matrix, the octet and singlet density operators at time $t$ take the form $\mathcal{D}_{\rm o}=\sum_l p_{{\rm o},l} |l\rangle \langle l|$,  and $\mathcal{D}_{\rm s}=\sum_m p_{{\rm s},m} \ket{m}\bra{m}$.  One can then show that the time variation of the singlet population  $p_{{\rm s},n}(t)=\langle n |\mathcal{D}_{\rm s}(t)|n\rangle$ in this specific basis is given by the following rate equation 
\beq
\frac{\rmd {p}_{{\rm s},n}}{\rmd t}=\sum_l t_{nl}\, p_{{\rm o},l} - \Gamma_n \,p_{s,n}
\quad\text{with}\quad 
\Gamma_n=\sum_l t_{ln},
\label{eq:kineticequationappA}
\eeq
where the gain (resp. loss) term stems from the $\mathcal{L}^{\rm so}$ (resp. $\mathcal{L}^{\rm ss}$) superoperator.\footnote{The detailed calculations leading to these results can be found in \cite{Delormethesis}.} For the sake of illustration, we show how the transition coefficients $t_{nl}$ relate to  $\mathcal{L}_{\rm so}$ in the case of the superoperator $\mathcal{L}_2=\mathcal{L}_{2b}=2 \left(W(\frac{s+s'}{2})-W(\frac{s-s'}{2}) \right)$ (see App.~\ref{app:transition_operators}). Using the spectral decomposition (\ref{spectraldec}) of $W$, one obtains
\begin{equation}
\mathcal{L}_{2b}=2 \int {\rm d}q \rho(q) \left(\rme^{-iq\frac{s-s'}{2}} - \rme^{-iq\frac{s+s'}{2}} \right)= 4\int {\rm d}q \rho(q)
\sin(\frac{qs}{2}) \sin(\frac{qs'}{2}).
\end{equation}
The contribution from $C_F \mathcal{L}_{2b} \mathcal{D}_{\rm o}$ to 
$\dot{p}_{{\rm s},n}$ thus reads
\begin{equation}
\frac{\rmd {p}_{{\rm s},n}}{\rmd t} = C_F \langle n | \mathcal{L}_{2b} \mathcal{D}_{\rm o} |n\rangle = C_F \sum_l \int {\rmd}q \rho(q) \left|2\int {\rmd}s\,\psi^\star_{{\rm s},n}(s)  \sin(\frac{qs}{2}) \psi_{{\rm o},l}(s)\right|^2\,    p_{{\rm o},l}.
\label{kinetiso}
\end{equation}
On recognizes in Eq.~(\ref{kinetiso}) the form of the gain term in Eq.~(\ref{eq:kineticequationappA}), provided the transition coefficient $t_{nl}$ is defined as $t_{nl}=C_F \int {\rm d}q \rho(q) |T_{n,l}(q)|^2$, where $T_{n,l}(q)$ is  the following transition amplitude 
\[
T_{n,l}(q) = 2 \int \rmd s\, \psi_{{\rm s},n}^\star(s) \sin\left(\frac{q\,s}{2} \right) \psi_{{\rm o},l}(s).
\]
For the full $\mathcal{L}_{\rm so}$ superoperator, the expression of $t_{nl} $ reads
 \begin{equation} 
t_{nl} = C_F \int \rmd q\, \rho(q) \left|(1-(q \lambda_{\rm th})^2/8) T_{n,l}(q)+  T_{n,l}^{(3)}(q)\right|^2
\label{eq:transcoeff}
\end{equation}
where $\lambda_{\rm th}=1/\sqrt{MT}$, while a new transition amplitude
$T_{n,l}^{(3)}(q)$ needs to be introduced at the $\mathcal{L}_{3b}$ level:
\[
T_{n,l}^{(3)}(q) = \frac{q \lambda_{\rm th}^2}{2} \int \rmd s\, \psi_{s,n}^\star(s) \cos\left(\frac{q\,s}{2}\right)\,\psi'_{o,l}(s).
\]
Clearly, the expression (\ref{eq:transcoeff}) is always positive iff $\rho(q)>0$ for all $q$. 

In Eq.~(\ref{eq:transcoeff}), six terms contribute to the integrand, which can be mapped to those enumerated in table \ref{table:structure}. The leading fluctuations term for an infinite mass heavy quark (recoilless limit) corresponds to  $|T_{n,l}(q)|^2$ originating from $\mathcal{L}_{2b}$, as shown above. The cross terms, $\left(T_{n,l}^{(3)}(q)T_{n,l}^\star(q)+{\rm cc}\right)$ and $-(q \lambda_{\rm th})^2/4\, |T_{n,l}(q)|^2$,  correspond respectively to $\mathcal{L}_{3b,{\rm lin}}$ and $\mathcal{L}_{3b,{\rm cst}}$. The last 3 terms $|T_{n,l}^{(3)}(q)|^2$, $-(q \lambda_{\rm th})^2/8\,(T_{n,l}^{(3)}(q)  T_{n,l}^\star(q) + \rm{cc})$ and  $\frac{(q\lambda_{\rm th})^4}{64} |T_{n,l}(q)|^2$,  correspond respectively to the higher order superoperators $\mathcal{L}_{4b,{\rm quad}}$, $\mathcal{L}_{4b,{\rm lin}}$, and $\mathcal{L}_{4b,{\rm cst}}$.

A very similar decomposition applies for the $\mathcal{L}^{\rm os}$, $\mathcal{L}^{\rm ss}$, and $\mathcal{L}^{\rm oo}$ blocks, including superoperators of type $\L_a$ (in particular $\mathcal{L}^{\rm ss}$ will generate the loss terms for the singlet state probabilities in Eq.~(\ref{eq:kineticequationappA})). 
This demonstrates the equivalence between the action of the superoperator $\mathcal{L}_2+\mathcal{L}_3+\mathcal{L}_4$ and the  gain-loss kinetic rate equation (\ref{eq:kineticequationappA}) which preserves positivity.  

\section{Implementation of the higher-order operator and UV regularization}
\label{app:L4}

While the superoperator $\mathcal{L}_3$  involves at most second derivatives of $W(r)$ with respect to $r$, the superoperator $\mathcal{L}_{4}$  involves third and fourth order derivatives at finite values of $r$, as well as a fourth derivative at the origin for $\mathcal{L}_{4a}$. Because of the specific analytic form of $W(r)$, such derivatives entail  possible UV divergences if the integrals entering the definition of $W$ are not properly regulated. 

To appreciate the difficulty, let us recall that the one-dimensional imaginary potential that we use derives from the three-dimensional  HTL expression given in Eqs.~(\ref{eq:VHTL3D}) and (\ref{eq:VHTL3DPhi}), which we recall here for convenience: 
\begin{equation}
W(r,T) = \alpha_S T\phi(m_D r),  
\label{eq:WHTL}
\end{equation}
where $m_{D}$ is the Debye mass, $r$ is the distance between the quark and the antiquark, and
\begin{equation}
    \phi(\tilde{r})=2\int_{0}^{\infty}\mathrm{d}z\;\frac{z}{\left(z^2+1\right)^2}\left(1-\frac{\sin(z \tilde{r})}{z\tilde{r}}\right),
\label{eq:defphi}   
\end{equation}
with $\tilde{r}=m_D r$ and $z=q/m_D$.
The derivatives of $\phi(\tilde{r})$ at the origin are divergent, starting from the second one which suffers from a logarithmic UV divergence.\footnote{At finite $r$ all derivatives are well defined, although starting from the fourth one, they cannot be evaluated by derivation under the integral sign.}
Since the range of $q$ values that lead to such divergences is in fact outside the domain of validity of the HTL approximation, a standard procedure consists in regulating the integrals by imposing some cutoff \cite{DeBoni:2017ocl}. In our work, we have followed an alternate method that has the advantage of simplifying the structure of the  equations. This method is based on the observation that the transition probability (\ref{eq:transcoeff}) suffers from the same UV divergences. To cure the divergences in Eq.~(\ref{eq:transcoeff}), one proceeds as follows: One assumes that higher contributions in the gradient expansion would generate a prefactor $(1-\frac{(q\lambda_{\rm th})^2}{8})$ in front of $T_{n,l}^{(3)}(q)$ and one then substitutes this common prefactor by  $\left(1+ \frac{(q\lambda_{\rm th})^2}{8}\right)^{-1}$, so that 
\[
\rho(q) \left|\left(1-\frac{(q\lambda_{\rm th})^2}{8}\right) T_{n,l}(q)+  T_{n,l}^{(3)}(q)\right|^2 \longrightarrow 
\frac{\rho(q)}{\left(1+ \frac{(q\lambda_{\rm th})^2}{8}\right)^2} 
\left|T_{n,l}(q)+  T_{n,l}^{(3)}(q)\right|^2,
\]
which is UV-convergent and still positive. Up to $\mathcal{O}((q\lambda_{\rm th})^{4})$, the new expression is similar to the former one, apart from different numerical coefficients for the terms $\mathcal{L}_{4b,{\rm cst}}$ and $\mathcal{L}_{4b,{\rm lin}}$ which are anyhow expected to be subdominant at high temperature.
This procedure leads to a regularized potential $\tilde{W}$:
\begin{equation}
\tilde{W}(r)=\int \rmd q \,\frac{\rho(q)}{\left(1+ \frac{(q \lambda_{\rm th})^2}{8}\right)^2} (1-\rme^{-i q r})=\int \rmd q \tilde \rho(q)(1-e^{-i q r}), 
\label{regulW}
\end{equation}
where the modified spectral density $\tilde{\rho}(q)={\rho(q)}{\left(1+ \frac{(q \lambda_{\rm th})^2}{8}\right)^{-2}}$ contains all the corrections discussed above. The corresponding regularized transition rates then read 
\begin{equation} 
\tilde{t}_{nl} = C_F \int \rmd  q \tilde{\rho}(q) \left|T_{n,l}(q)+  T_{n,l}^{(3)}(q)\right|^2.
\label{eq:transcoeffreg}
\end{equation}
Note that the substitution of the original $t_{nl}$ by the regularized $\tilde t_{nl}$ is equivalent to selecting a reduced number of terms in the original $\mathcal{L}_{b}$ superoperator ($\mathcal{L}_{2b}$, $\mathcal{L}_{3b,{\rm lin}}$ and $\mathcal{L}_{3b,{\rm quad}}$, together referred to the ``minimal set" of superoperators warranting positivity), and expressing them in terms of  $\tilde{W}$ (see table \ref{table:structure2}).
\begin{table}[H]
\begin{center}
\begin{tabular}{ |c|c|c|c| } 
\hline
&$\mathcal{D}_{\rm o}$ & $\partial\mathcal{D}_{\rm o}$  & $\partial^2\mathcal{D}_{\rm o}$\\
\hline
$\vphantom{\sum^{N^N}} \mathcal{L}_{2b}$ & $\mathcal{L}_{2b,{\rm cst}}$: $\times \tilde{W}$ & 0 & 0 \\[2mm] 
$\mathcal{L}_{3b}$ & 0 & $\mathcal{L}_{3b,{\rm lin}}$: $\times \tilde{W}^{(1)}$ & 0\\[2mm] 
$\mathcal{L}_{4b}$  & 0 & 0 & $\mathcal{L}_{4b,{\rm quad}}$: $\times \tilde{W}^{(2)}$\\[2mm] 
 \hline
\end{tabular}
\end{center}
\caption{Same as table \ref{table:structure} for the ``minimal set" of superoperators associated to the regularization of the imaginary potential defined by 
Eq.~(\ref{regulW}). }
\label{table:structure2}
\end{table}
Let us emphasize that the present construction of he minimal set allows to recover the original $\mathcal{L}_{3b}$ term while displaying a lighter structure.\footnote{The simplification brought by the use of the minimal set allows a physical illustration  of the role of $\mathcal{L}^{(4)}$. In a semi-classical expansion, it yields in particular a contribution $\propto \tilde{W}^{4}(0) y^2 \partial^2 \mathcal{D}$, or in terms of the Wigner transform $\mathcal{W}(r=\frac{s+s'}{2},p)$, a contribution 
$\propto \partial_p^2 \left( p^2 \mathcal{W} \right)$. This represents a correction $\propto \mathcal{O}(p^2)$ to the diffusion coefficient appearing in the Fokker-Planck equation obtained from $\mathcal{L}_2$ and $\mathcal{L}_3$ \cite{Blaizot:2017ypk}.} We also note that  the definition (\ref{regulW}) amounts essentially to take as UV regulator $\Lambda=\lambda_{\rm th}^{-1}$ in the original expression of the potential $W$ (see the discussion in App.~\ref{app:equilibrium}).

Up to now, we have discussed the $\mathcal{L}_{b}$ type of superoperators, present in the $\mathcal{L}^{\rm so}$ and  $\mathcal{L}^{\rm os}$ blocks, as well as in $\L^{\rm oo}$. We now discuss the case of the $\mathcal{L}_{a}$ type of superoperators, present in the $\mathcal{L}^{\rm ss}$ and $\mathcal{L}^{\rm oo}$ blocks and obtained by considering exclusively diagrams for which the gluon propagator connects quark lines located on a single side of the Schwinger-Keldysch contour. The regularization discussed above, $W\to \tilde{W}$, also leads to a simplification of the $\mathcal{L}_a$ superoperators. The $\mathcal{L}_{2a}$ contribution to the ${\rm ss}$ block is readily defined as (see App.~\ref{app:transition_operators})
\begin{equation}
\mathcal{L}^{\rm ss}_{2a} = C_F(2 \tilde{W}(0)-\tilde{W}(s) - \tilde{W}(s')).
\label{eq:Lss2a}
\end{equation}
This leads to the following contribution to $\Gamma_n$, the dissociation of a singlet  state $|n\rangle$:
\begin{equation}
\Gamma_{n,2}= 2 C_F \langle n | \tilde{W}(s)- \tilde{W}(0) | n \rangle,
\quad \text{with} \quad
\tilde{\Gamma}(s) =\tilde{W}(s)- \tilde{W}(0).
\label{eq:gamman2}
\end{equation}
In order to construct $\mathcal{L}_{3a}$ consistently within the minimal set prescription we start from the transition probabilities $t_{nl}$ defined in App.~\ref{app:positivity} and rely on detailed balance ($t_{ln}=t_{nl}$)
\footnote{Because of the specific structure of the $\L_{3a}$ superoperator, the procedure put forward for the $\L_b$ superoperators can not be blindly applied. This is the reason for the strategy used here.}
to derive the decay rates $\Gamma_n$ governing the loss term in the rate  equation~(\ref{eq:kineticequationappA}). This allows us to calculate the various contributions of $\Gamma_n$  by selecting the corresponding terms in the expression of $t_{nl}$ and ultimately express these contributions as ``$\mathcal{L}_a$-type'' superoperators acting on the density operator. For instance, one explicitly recovers the expression (\ref{eq:gamman2}) by just retaining the $|T_{nl}|^2$ contributions in $t_{nl}$. Following this method, one defines 
\begin{equation}
\Gamma_{n,3} = C_F \sum_l \int \rmd q\, \tilde{\rho}(q) 
\left(T^\star_{ln}(q)  T^{(3)}_{ln} ({q})  + cc \right)
\end{equation}
where
\[
T_{ln}({q}) = 2 \int \rmd s'\, \psi_l^\star({s}') \sin\left(\frac{{q s'}}{2} \right) \psi_n(s')
\quad\text{and}\quad
T_{ln}^{(3)}({q}) = \frac{\lambda_{\rm th}^2}{2}  \int \rmd s\, \psi_l^\star({s}) \cos\left(\frac{{q s}}{2}\right){q} \psi'_n({s}).
\]
Injecting these transition elements into the expression of $\Gamma_{n,3}$, one obtains a threefold integral. However using the summation on $l$ and the completeness relation leads to a factor $\delta({s}-{s}')$ which is used to simplify the final result: 
\begin{eqnarray}
\Gamma_{n,3} &=& C_F \lambda_{\rm th}^2
\int \rmd q\, \tilde{\rho}(q) 
\int \rmd s \,\rmd s'\, \delta(s-s')  \left[\sin\left(\frac{{q s'}}{2} \right)  \cos\left(\frac{{q s}}{2}\right)q\partial_s +
\right.
\nonumber\\
 && \hspace{1cm} \qquad\qquad\qquad\qquad\qquad\qquad\left. 
\sin\left(\frac{{q s}}{2} \right)  \cos\left(\frac{q s'}{2}\right) q \partial_{s'} 
\right]
\psi_n^\star(s') \psi_n(s) 
\nonumber\\
&=&
\frac{C_F \lambda_{\rm th}^2}{2}
\int \rmd q   \int \rmd s\,\tilde{\rho}(q) \sin(qs)
\,q\partial_s |\psi_n(s)|^2.
\end{eqnarray}
Expressing $\sin(q s)\,q$ as $-\partial_s \cos(qs)$, one finds  
\begin{eqnarray}
\Gamma_{n,3}&=& - \frac{C_F \lambda_{\rm th}^2}{2} \int 
\rmd s\, \partial_s \left(
\int \rmd q \tilde{\rho}(q) \cos(q s)\right) \nabla |\psi_n({s})|^2
=\frac{C_F \lambda_{\rm th}}{2} \int \rmd s\,
\tilde{W}'(s ) \partial_s|\psi_n({s})|^2
\nonumber\\ 
&=& 
\frac{C_F \lambda_{\rm th}}{2} 
\left( \int \rmd s \,\psi^\star_n({s})
\tilde{W}'({s}) \partial_s \psi_n({s}) + cc\right),
\label{eq:gamman3}
\end{eqnarray}
corresponding to the following (1D\footnote{For higher dimensions, the expression reads $\mathcal{L}^{\rm ss}_{3a} = -
\frac{C_F \lambda_{\rm th}^2}{2} \left(
\nabla \tilde{W}(\vec{s}) \cdot  \nabla_s  +  \nabla \tilde{W}(\vec{s}\,') \cdot  \nabla_{s'}\right)$; Details of the derivation can be found in \cite{Delormethesis}, Sec.~4.4.2.}) contribution to $\mathcal{L}^{\rm ss}_{3a}$
\footnote{Note that this expression is neither the $\mathcal{L}_{3a}$ nor the $\mathcal{L'}_{3a}$ obtained in~\cite{Blaizot:2017ypk}.}
\beq
\mathcal{L}^{\rm ss}_{3a} = -
\frac{C_F \lambda_{\rm th}^2}{2} \left(
\tilde{W}'({s})   \partial_s  +  \tilde{W}'({s}')  \partial_{s'}\right).
\label{eq:Lss3a}
\eeq
Following the same method, one obtains
\beq
\Gamma_{n,4}&=&\frac{C_F\lambda_{\rm th}^4}{4} 
\int \rmd q\, \tilde{\rho}(q) \int \rmd s \,\rmd s'
\,\delta(s-s') \cos(\frac{q s}{2})
\cos(\frac{q s'}{2})
\,q^2\partial_s \psi_n(s) \,
\partial_{s'} \psi_n^\star({s}') 
\nonumber\\
&=&
\frac{C_F \lambda_{\rm th}^4}{4} 
\int \rmd s\,
\int \rmd q\, \tilde{\rho}(q) \,q^2  \cos^2(\frac{q s}{2})
\,\psi'_n(s) \,{\psi_n'}^\star({s}) 
\nonumber\\
&=&  
\frac{C_F  \lambda_{\rm th}^4}{4} 
\int \rmd s \,
\left(\tilde{W}''({s})+ \tilde{W}''({0})\right )  {\psi'_n}^\star({s}) \psi'_n({s}),
\label{eq:gamman4}
\eeq
while the associated $\mathcal{L}^{\rm ss}_{4a}$ superoperator is\footnote{In higher dimensions, this is $\mathcal{L}^{\rm ss}_{4a} = 
\frac{C_F  \lambda_{\rm th}^4}{16} \left[ \partial_i \partial_j \tilde{W}(0) \left(\partial_{s,i} \partial_{s,j} + s\leftrightarrow  s'\right)+
\left(\partial_{s,i} \left(\partial_{i} \partial_j \tilde{W}(s) \partial_{s,j} \right) +
s\leftrightarrow  s' \right)\right].$} 
\beq
\mathcal{L}_{{\rm ss},4a} = 
\frac{C_F  \lambda_{\rm th}^4}{16} \left[ \tilde{W}''(0) \left(\partial_{s} \partial_{s} + s\leftrightarrow  s'\right)+
\left(\partial_{s} \left(\tilde{W}''(s) \partial_{s} \right) +
s\leftrightarrow  s'\right)\right].
\label{eq:Lss4a}
\eeq
\section{Color transition operators}
\label{app:transition_operators}
This appendix provides the complete list of the color transition superoperators $\mathcal{L}^{\rm ss}$, $\mathcal{L}^{\rm so}$, $\mathcal{L}^{\rm os}$ and $\mathcal{L}^{\rm oo}$ corresponding to the ``minimal set" discussed in App.~\ref{app:L4}. Details on their derivation can be found in \cite{Delormethesis}. The superoperators below are written in the coordinate representation and are supposed to act on the color components $\bra{s}{\cal D}_{\rm s}\ket{s'}$ and $\bra{s}{\cal D}_{\rm o}\ket{s'}$ of the density matrix. The color factors appropriate to the various channels are properly included. Note that the non-diagonal superoperators $\mathcal{L}^{\rm so}$ and $\mathcal{L}^{\rm os}$ (superoperators of type $\L_b$ in the terminology of Ref.~\cite{Blaizot:2017ypk}) are proportional to the imaginary potential $W$ and its derivatives. The diagonal superoperators  $\mathcal{L}^{\rm ss}$ and $\mathcal{L}^{\rm oo}$ (where $\mathcal{L}^{\rm ss}$ is of type $\L_a$, while both $\L_a$ and $\L_b$ types are present in  $\mathcal{L}^{\rm oo}$ ) contain in addition kinetic energy contributions as well as the real part $V$ of the potential. In each superoperator, the terms proportional to $W$ contribute to $\L_2$, the terms proportional to $\hbar^2 c^2/MT$ to $\L_3$ and the terms in $\hbar^4 c^4/M^2T^2$ to $\L_4$. Note the presence in $\L_4$ of terms proportional to $p_{\rm tot}^2$, where $p_{\rm tot}$ is the center of mass momentum. These terms survive the ``elimination'' of the center of mass, and they limit  the application of the formalism to small $p_{\rm tot}$ \cite{Blaizot:2018oev}.
\begin{align}\label{eq:A.3}
   \mathcal{L}^{\rm ss} &= \frac{i\hbar^{2}c^{2}}{M}\left[\partial^{2}_{s} - \partial^{2}_{s'}\right] -iC_{F}\left[V(s) - V(s')\right] \nonumber\\
   &+ C_{F}\left[2\tilde{W}(0) - \tilde{W}(s) - \tilde{W}(s')\right] \nonumber\\
   &-\frac{\hbar^{2}c^{2}C_{F}}{2MT}\left[\tilde{W}'(s)\partial_{s} + \tilde{W}'(s')\partial_{s'}\right] \nonumber\\
   &+\frac{\hbar^{4}c^{4}C_{F}}{64M^{2}T^{2}}\left[4\tilde{W}''(0)\left(\partial^{2}_{s} + \partial^{2}_{s'}\right) - 2 p_{\rm tot}^{2}\tilde{W}''(0)\right] \nonumber\\
   &+\frac{\hbar^{4}c^{4}C_{F}}{64M^{2}T^{2}}\left[4\tilde{W}''(s)\partial^{2}_{s} + 4\tilde{W}''(s')\partial^{2}_{s'} + 4\tilde{W}'''(s)\partial_{s}\right] \nonumber\\
   &+\frac{\hbar^{4}c^{4}C_{F}}{64M^{2}T^{2}}\left[ 4\tilde{W}'''(s')\partial_{s'} + p_{\rm tot}^{2}\tilde{W}''(s) + p_{\rm tot}^{2}\tilde{W}''(s')\right] \nonumber\\
   &\nonumber\\
   \mathcal{L}^{\rm so} &= C_{F}\left[-2\tilde{W}\left(\frac{s - s'}{2}\right) + 2\tilde{W}\left(\frac{s + s'}{2}\right)\right]\nonumber\\
   &+\frac{\hbar^{2}c^{2}C_{F}}{2MT}\left[-\tilde{W}'(\frac{s - s'}{2})\left(\partial_{s} - \partial_{s'}\right) + \tilde{W}'(\frac{s + s'}{2})\left(\partial_{s} + \partial_{s'}\right)\right] \nonumber\\
   & + \frac{\hbar^{4}c^{4}C_{F}}{8M^{2}T^{2}}\left[\tilde{W}''\left(\frac{s - s'}{2}\right)\partial_{s}\partial_{s'} + \tilde{W}''\left(\frac{s + s'}{2}\right)\partial_{s}\partial_{s'}\right] \nonumber\\
   &+ \frac{\hbar^{4}c^{4}C_{F}}{32M^{2}T^{2}}\left[p_{\rm tot}^{2}\tilde{W}''\left(\frac{s - s'}{2}\right) - p_{\rm tot}^{2}\tilde{W}''\left(\frac{s + s'}{2}\right)\right] \nonumber\\
    &\nonumber\\
   \mathcal{L}^{\rm os} &= \frac{1}{2N_{c}}\left[-2\tilde{W}\left(\frac{s - s'}{2}\right) + 2\tilde{W}\left(\frac{s + s'}{2}\right)\right] \nonumber\\
   &+\frac{1}{2N_{c}}\frac{\hbar^{2}c^{2}}{2MT}\left[-\tilde{W}'(\frac{s - s'}{2})\left(\partial_{s} - \partial_{s'}\right) + \tilde{W}'(\frac{s + s'}{2})\left(\partial_{s} + \partial_{s'}\right)\right] \nonumber\\
   & + \frac{1}{2N_{c}}\frac{\hbar^{4}c^{4}}{8M^{2}T^{2}}\left[\tilde{W}''\left(\frac{s - s'}{2}\right)\partial_{s}\partial_{s'} + \tilde{W}''\left(\frac{s + s'}{2}\right)\partial_{s}\partial_{s'}\right] \nonumber\\
   &+ \frac{1}{2N_{c}}\frac{\hbar^{4}c^{4}}{32M^{2}T^{2}}\left[p_{\rm tot}^{2}\tilde{W}''\left(\frac{s - s'}{2}\right) - p_{\rm tot}^{2}\tilde{W}''\left(\frac{s + s'}{2}\right)\right] \nonumber
   \end{align}
   \newline\newline
   \begin{align}
    \mathcal{L}^{\rm oo} &= \frac{i\hbar^{2}c^{2}}{M}\left[\partial^{2}_{s} - \partial^{2}_{s'}\right] + i\frac{1}{2N_{c}}\left[V(s) - V(s')\right] \nonumber\\
     &+2C_{F} \tilde{W}(0) +\frac{1}{2N_{c}}\left[\tilde{W}(s) + \tilde{W}(s)\right] \nonumber\\
     &-\frac{N_{c}^{2}-2}{N_{c}}\tilde{W}\left(\frac{s - s'}{2}\right) -\frac{2}{N_{c}}\tilde{W}\left(\frac{s + s'}{2}\right) \nonumber\\
     &+\frac{1}{2N_{c}}\frac{\hbar^{2}c^{2}}{2MT}\left[\tilde{W}'(s)\partial_{s} + \tilde{W}'(s')\partial_{s'}\right] \nonumber\\
     &- \frac{N_{c}^{2}-2}{2N_{c}}\frac{\hbar^{2}c^{2}}{2MT}\tilde{W}'(\frac{s - s'}{2})\left(\partial_{s}-\partial_{s'}\right) \nonumber\\
     & -\frac{1}{N_{c}}\frac{\hbar^{2}c^{2}}{2MT}\tilde{W}'(\frac{s + s'}{2})\left(\partial_{s}+\partial_{s'}\right) \nonumber\\
     &+ \frac{\hbar^{4}c^{4}C_{F}}{64M^{2}T^{2}}\left[4\tilde{W}''(0)\left(\partial^{2}_{s} - \partial^{2}_{s'}\right) - 2p_{\rm tot}^{2}\tilde{W}''(0)\right] \nonumber\\
    &-\frac{1}{2N_{c}}\frac{\hbar^{4}c^{4}}{64M^{2}T^{2}}\left[4\tilde{W}''(s)\partial^{2}_{s} + 4\tilde{W}''(s')\partial^{2}_{s'} + 4\tilde{W}'''(s)\partial_{s}\right]\nonumber\\
   &-\frac{1}{2N_{c}}\frac{\hbar^{4}c^{4}}{64M^{2}T^{2}}\left[ 4\tilde{W}'''(s')\partial_{s'} + p_{\rm tot}^{2}\tilde{W}''(s) + p_{\rm tot}^{2}\tilde{W}''(s')\right]\nonumber\\
   &+ \frac{N_{c}^{2}-2}{2N_{c}}\frac{\hbar^{4}c^{4}C_{F}}{64M^{2}T^{2}}\left[8\tilde{W}''\left(\frac{s - s'}{2}\right)\partial_{s}\partial_{s'} +2p_{\rm tot}^{2}\tilde{W}''\left(\frac{s - s'}{2}\right)\right] \nonumber\\
   & - \frac{1}{N_{c}}\frac{\hbar^{4}c^{4}C_{F}}{64M^{2}T^{2}}\left[8\tilde{W}''\left(\frac{s + s'}{2}\right)\partial_{s}\partial_{s'} -2p_{\rm tot}^{2}\tilde{W}''\left(\frac{s + s'}{2}\right)\right].
\end{align}

\section{Details on the numerical implementation}
\label{app:numerics}
The equations are solved numerically for the charmonium system (with the charm mass $m_{c} = 1.469$~GeV/$c^{2}$), 
using the Crank-Nicolson scheme. Both the $s$ and $s'$ variables (which correspond to the relative distance between the quark and antiquark) are discretized along $N = 501$ points, from $-10$~fm to $+10$~fm (box size $L=20\,{\rm fm}$), with a spatial step $\Delta s = 0.04$~fm. We choose a time step $\Delta t = 0.01$~fm/$c$ (lower values of $\Delta t$ were found to have no effect on the results). The boundary conditions are such that ${\cal D}(s,s')=0 $  when either $s$ or $s'$ is on the boundary of the box, that is when $s=\pm 10$ fm or $s'=\pm 10$ fm.

\begin{figure}[H]
\centering
\includegraphics[width=0.48\linewidth]{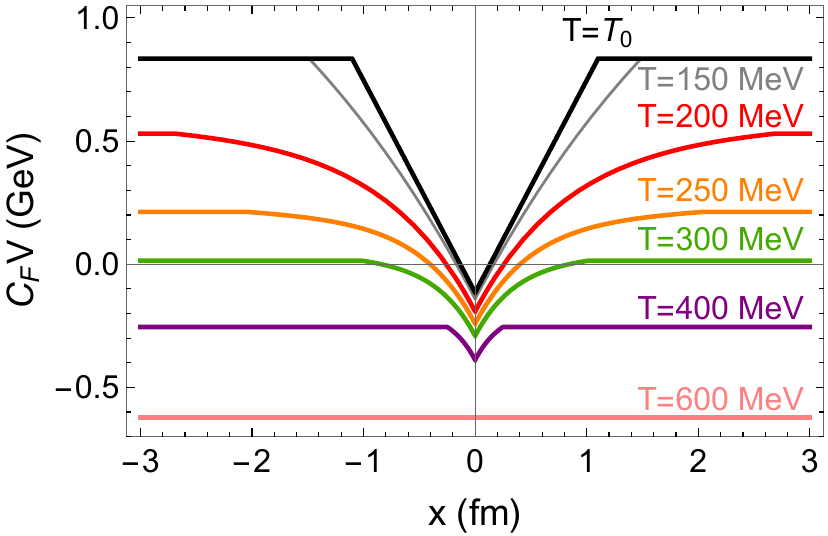}
\includegraphics[width=0.48\linewidth]{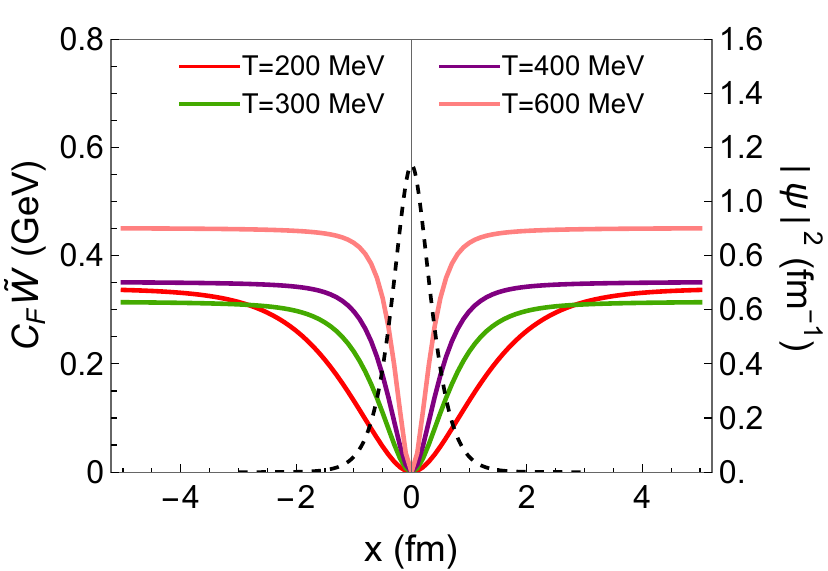}
\caption{Real (left panel) and regularized imaginary (right panel) parts of the one-dimensional potential as a function of the medium temperature $T$ (in GeV) and distance $x$ between the quark and the antiquark of the charmonium  \cite{Katz:2022fpb}. The color factor $C_F$ is that appropriate for the singlet channel. The dashed line in the right panel represents the probability density corresponding to the 1S charmonium state.}
\label{fig:Pot}
\end{figure}

We use a complex potential (shown in Fig.~\ref{fig:Pot}) that was specifically developed for one-dimensional studies \cite{Katz:2022fpb}. The real and imaginary parts of the complex potential were tuned to reproduce at best the temperature-dependent mass spectra and decay widths of quarkonium states obtained with a three-dimensional potential model inspired by lattice QCD results \cite{Lafferty:2019jpr}.\footnote{Note that in \cite{Katz:2022fpb},  as well as in \cite{Lafferty:2019jpr}, the $C_F$ color factor associated to the singlet state was absorbed in the definition of the complex potential. In the present work, we make the color factors explicit, as they differ between the singlet and octet parts of the $\mathcal{L}$ superoperator, as recalled in App.~\ref{app:transition_operators}; this holds both for the real part $V$ and the imaginary part $W$ of the potential.}

The real potential admits three eigenstates (referred to as vacuum eigenstates) which are  obtained by solving the Schrödinger equation for the relative ${\rm c\bar{c}}$ motion. The ground state is referred to as 1S, while the first (odd) and second (even) excited states are referred to respectively as 1P and 2S.

The flattening of the 1D potential at large distance is adjusted to match the corresponding behavior of the 3D potential  and to mimic the temperature dependence of the charm pole-mass, as well as the effect of the ``string breaking" mechanism.

Table \ref{tab:Eandrvacuum1D} shows that the properties of the S states of the charmonium as calculated from a 3D-potential are well accounted for by the 1D potential. Note that the 3D and the 1D calculations both assume that temperature effects are irrelevant below a temperature $T_0$. Thus, the vacuum properties referred to in this paper correspond in fact as well to those calculated at this temperature $T_0$.  For the P-state, the quantitative agreement with the 3D model is not as good, which is probably due to the absence of centrifugal energy in the 1D setup  (see \cite{Katz:2022fpb}). 
\begin{table}[h!]
\begin{center}
    \begin{tabular}{|C{3.5cm}||C{1.5cm}|C{1.5cm}|C{1.5cm}|}
    \hline
At $T=T_0$ & $J/\psi$(1S) & $\chi_c$(1P) & $\psi$(2S)  \\
    \hline
    \hline
$m^{\rm PDG}$ \cite{Zyla:2020zbs} & 3.0969 & 3.525 & 3.6861  \\
    \hline
$m^{\rm EV}_{\rm 3D}$ & 3.0964 & 3.509 & 3.6642  \\
    \hline 
$m^{\rm EV}_{\rm 1D}$ & 3.0981 & 3.4532 & 3.6858  \\ 
    \hline 
$|m^{\rm PDG}-m^{\rm EV}_{\rm 1D}|$  & 0.0012 & 0.072 & 0.0003  \\
    \hline
$E_{\rm binding}$   & 0.675 & 0.320 & 0.087 \\
    \hline  \hline
${\rm 3D} \,\,\, \sqrt{\langle r^2 \rangle}$ & 0.431 & 0.689 & 0.943   \\ 
    \hline  
$ {\rm 1D} \,\,\, \sqrt{\langle x^2 \rangle}$ & 0.271 & 0.543 & 0.856 \\ 
    \hline
    \end{tabular}
\label{tab:Eandrvacuum1D} 
\caption {
\small Masses (in GeV) and root mean square radii (in fm) for the ``vacuum'' S states (at $T_0 \approx 0.126025$ GeV). $m^{\rm PDG}$ is the experimental mass given by the particle data group (for the 1P, it corresponds to a weighted average between $\chi_{c0}$, $\chi_{c1}$, and $\chi_{c2}$), $m^{\rm EV}_{\rm 3D}$ (resp. $\sqrt{\langle r^2 \rangle}$) is the mass (resp. root mean square radius) calculated from the expectation values of the Hamiltonian (the square radius operator) with the three-dimensional potential suggested in \cite{Lafferty:2019jpr} (no spin-orbit term). $m^{\rm EV}_{\rm 1D}$ and $\sqrt{\langle x^2 \rangle}$ are the corresponding values obtained with the one-dimensional potential.}
\end{center}
\end {table}

\begin{figure}[H]
\centering
\includegraphics[width=0.6\linewidth]{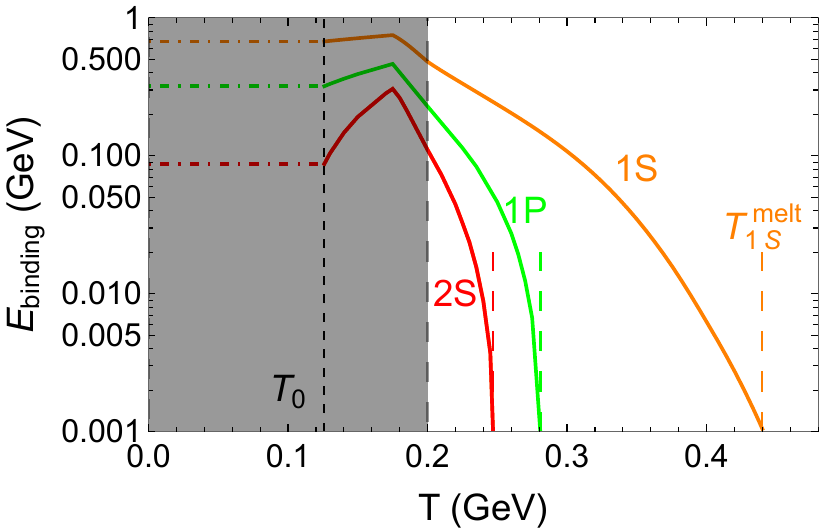}
\caption{Binding energy as a function of the temperature for the 3 vacuum bound states 1S, 1P and 2S, considered to be constant up to $T_0\approx 126\,{\rm MeV}$. Vertical long-dashed lines indicate the melting temperatures, corresponding, for practical purpose, to $E_{\rm binding}=1\,{\rm MeV}$. Note that the unnatural behavior of the binding energies for $T<200$~MeV (darkened area) is an artifact of the specific potential used in this paper, and should not be regarded as physical. All our numerical simulations are done for $T\gtrsim 200$ MeV. }
\label{fig:binding}
\end{figure}

Besides, one has $2 m_c+V_{\rm SB}\approx 2 m(D^+)$ -- where $V_{\rm SB} \approx 0.8352~\mathrm{GeV}$ corresponds to the saturation for the string breaking 1D-potential -- so that the liberation threshold for a ${\rm c\bar{c}}$ pair can roughly be seen as the creation of a $D^++ D^-$ pair for $T=T_0$. The binding energies at larger temperatures can be read directly from Fig.~\ref{fig:binding}.
With increasing temperature,  the potential well gets shallower (see Fig.~\ref{fig:Pot}): The melting temperatures of the 1S, 1P and 2S charmonium state are defined when $E_{\rm binding}<1\,{\rm MeV}$, respectively $T_{\rm melt}\approx 440\,{\rm MeV}$, $\approx 280\,{\rm MeV}$ and $\approx 250\,{\rm MeV}$.\\ 

The imaginary part $W$ of the potential is taken to be:
\begin{equation}
W(x,T) = \alpha W_{C}(|x|,T)+ \beta  W_S(|x|,T),
\label{eq:imaginary}
\end{equation}
with $\alpha = 1.7$, $\beta = 0.8$ for the charmonium system, and where $W_{C}$ and $W_{S}$ are the coulombic and string-like parts, defined as:
\begin{align}
&W_{C}(x,T)= \frac{\tilde{\alpha}_s T}{C_F}\phi(m_{D}x), \quad  \phi(r) =  2\int_{0}^{\infty}\mathrm{d}z\;\frac{z}{\left(z^2+1\right)^2}\left(1-\frac{\sin(zr)}{zr}\right) \nonumber\\
&W_{S}(x,T) =  \frac{\sigma T}{C_F m_D^2}\chi(m_{D}x), \quad \chi(r) = 2\int_0^{\infty}\mathrm{d}p\;\frac{2-2\cos(pr)-pr\sin(pr)}{\sqrt{p^2+\Delta_D^2}\left(p^2+1\right)^2},
\end{align}
where $\tilde{\alpha}_{S} = 0.513 \pm 0.0024$, $\sigma=(0.412\,{\rm GeV})^2$, and $\Delta_{D} \approx 3.0369$.
The $\alpha$ and $\beta$ coefficients were chosen to reproduce at best the decay widths calculated with the three-dimensional potential. At small distance, the imaginary potential exhibits an harmonic behavior\footnote{Up to some logarithmic prefactor.} which can be linked to the heavy quark diffusion coefficient. At large distance, the potential saturates toward twice the reaction rate of individual c quarks \cite{Beraudo:2007ky}.

The imaginary potential given at Eq.~(\ref{eq:imaginary}) suffers from the same UV divergences as the HTL imaginary potential recalled in App.~\ref{app:L4} and is regularized as discussed in that appendix. The right panel of Fig.~\ref{fig:Pot} illustrates the ensuing result, as well as the probability $\rho_{_{\rm 1S}}$ for the vacuum 1S state, which allows to appreciate the action of $W$ on the ground state.

\section{Charmonia dissociation rates}
\label{app:dissocrates}
In this appendix, we evaluate the various transition rates that are referred to in the main text. Let $|n\rangle$ be  a normalized singlet state, and let us assume that the initial density matrix ${\cal D}_{\rm s}(0)$  is the projector on this particular state, i.e. ${\cal D}_{\rm s}(0)=\ket{n}\bra{n}$. The variation with time of the probability $p_n=\bra{n}{\cal D}_{\rm s}(t)\ket{n}$ is given  in Eq.~(\ref{eq:kineticequationappA}),  from which we get, taking into account that $p_{\rm o}(t=0)=0$,
\begin{equation}\label{eq:dampingratedef}
\left.\frac{\rmd p_n}{\rmd t}\right|_{t=0}= \bra{n} \mathcal{L}^{\rm ss}_a{\cal D}(0)\ket{n}=-\Gamma_{n},
\end{equation}
which defines the dissociation rate $\Gamma_{n}$. 
In this equation,  $\mathcal{L}^{\rm ss}_a$ is the $\mathcal{L}_a$ part of the Lindblad superoperator $\mathcal{L}^{\rm ss}$ that involves the imaginary potential $W$, viz.
\[ \mathcal{L}^{\rm ss}_a=\mathcal{L}_{2a}+ \mathcal{L}_{3a}+
\mathcal{L}_{4a}.
\]
We call  $\Gamma_{n,2}$, $\Gamma_{n,3}$ and $\Gamma_{n,4}$ the corresponding contributions to $\Gamma_n$. These are estimated in App.~\ref{app:L4}, see Eqs.~(\ref{eq:gamman2}), (\ref{eq:gamman3}), and (\ref{eq:gamman4}).  Their sum can be written as
\begin{equation}
\Gamma_n = C_F \int {\rm d}q \,\tilde{\rho}(q) 
\int {\rm d}s \left| \left( \sin\left(\frac{{q}s}{2}\right) + \frac{\lambda_{\rm th}^2}{4}
\cos\left(\frac{qs}{2}
\right) q  \partial_s \right) \psi_n(s)  \right|^2.
\label{eq:appendixEGamman}
\end{equation}
It is clearly positive. Incidentally, one can take advantage of this explicit expression to evaluate the hierarchy between the 3 contributions, starting by the $\Gamma_{n,3}/\Gamma_{n,2}$ ratio.
As compared to the evaluation of  $\Gamma_{n,2}$ -- the contribution involving  the $\sin(\cdots)^2$ term in the integrand --,   the calculation of $\Gamma_{n,3}$ -- the cross term -- generates: a) one extra derivative of the wave function, b) one extra $q$ factor which is equivalent -- integrating by parts -- to another derivative of the wave function, and c) an extra $\lambda_{\rm th}^2$ factor. Parametrically, each extra derivative of the wave function $\psi'_n$ generates a factor inversely proportional to the rms radius $\sqrt{\langle r^2\rangle_n}$ of the state.
Hence, the ratio 
$\Gamma_{n,3}/\Gamma_{n,2}$ scales like
\begin{equation}
\frac{\Gamma_{n,3}}{\Gamma_{n,2}} \sim  \frac{\lambda_{\rm th}^2}{\langle r^2\rangle_n}.
\label{eq:gamma3overgamma2}
\end{equation}
Since $\lambda_{\rm th}$ decreases with increasing temperature, while $\langle r^2\rangle_n$ increases with $T$ and $n$, these ratios are decreasing functions of $T$ and $n$. They are moreover expected to be less than unity in the QBM regime.\footnote{For the 1S state, one has indeed 
$\frac{\lambda_{\rm th}^2}{\langle r^2\rangle_{\rm 1S}}\sim \lambda_{\rm th}^2 \langle p^2\rangle_{\rm 1S}\sim \frac{E_{\rm kin}^{\rm 1S}}{T}\lesssim 1$ when $T$ becomes larger than the typical energies associated to the bound state.} 
A similar  analysis results in $\Gamma_{n,4}/\Gamma_{n,2} \propto \left( \Gamma_{n,3}/\Gamma_{n,2}\right)^2$. For all temperatures lying in the QBM regime, one thus expects $\Gamma_{n,3}$ to be small and $\Gamma_{n,4}$ subleading.\\ 

\begin{figure}[h]
    \centering
\includegraphics[width=.49\linewidth]{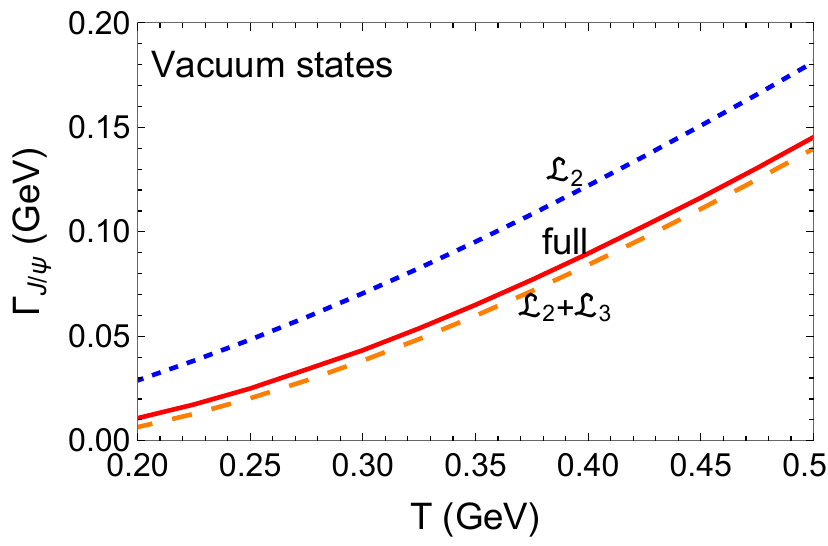}
\includegraphics[width=.49\linewidth]{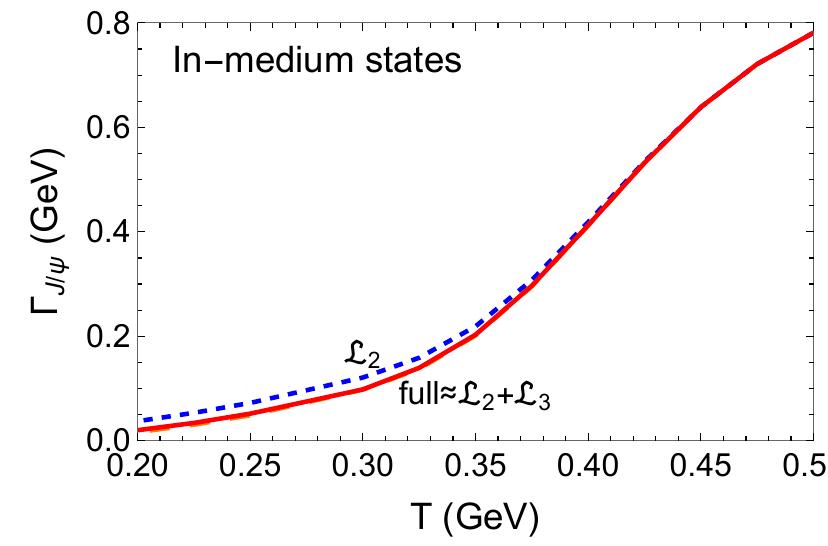}
\caption{Left: Decay rates of the vacuum 1S state (aka $J/\psi$) as a function of the temperature. The respective contributions of $\mathcal{L}_{2a}$ (blue, dashed), $\mathcal{L}_{2a}+\mathcal{L}_{3a}$ (orange, long-dashed) as well as the full result using $\mathcal{L}_{2a}+\mathcal{L}_{3a}+\mathcal{L}_{4a}$ (red, full) are shown. The calculations are done using the regularized imaginary potential of Eq.~(\ref{eq:imaginary}), following the regularization procedure outlined in App.~\ref{app:L4}. Right: Same calculation for the lowest  in-medium 1S state of the $T$-dependent potential.}
\label{fig:decayrate1}
\end{figure}

\begin{figure}[h]
    \centering
\includegraphics[width=.49\linewidth]{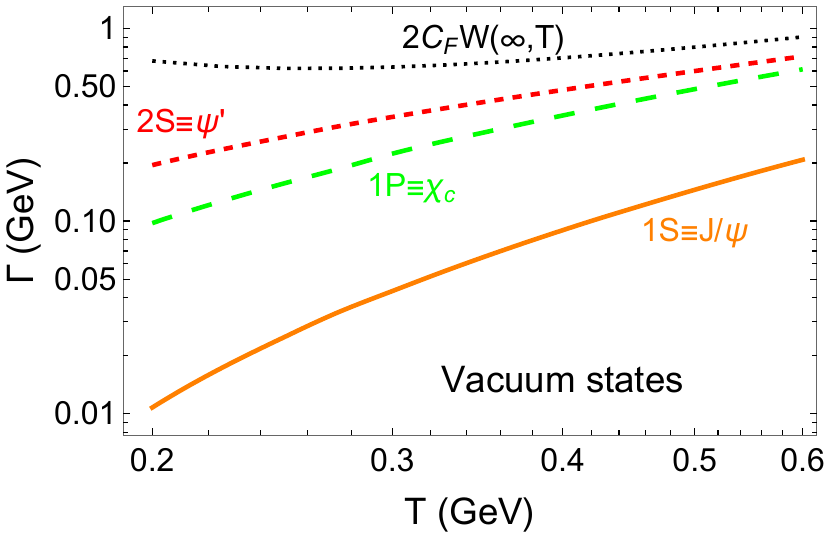}
\includegraphics[width=.49\linewidth]{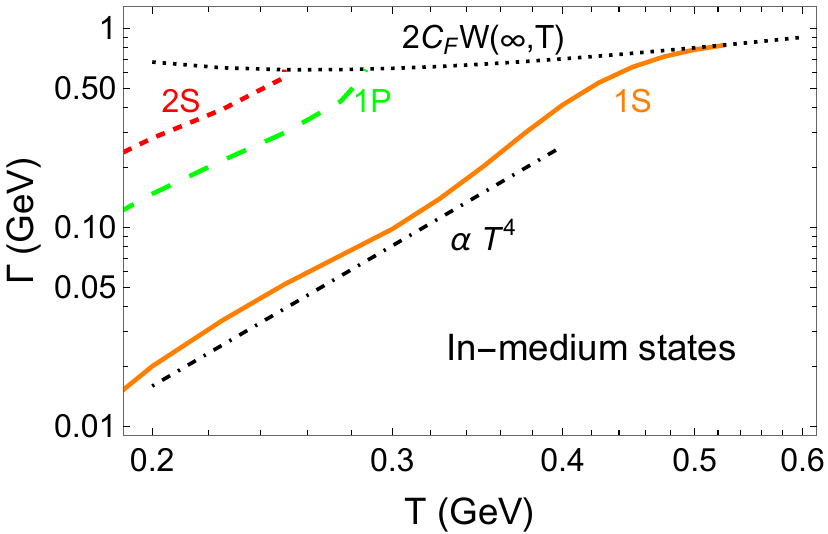}
\caption{Left : Full decay rate ($\mathcal{L}_{2a}+\mathcal{L}_{3a}+\mathcal{L}_{4a}$) for 1S (plain, orange), 1P (long-dashed, green) and 2S (dashed, red) states of the vacuum potential; same method as Fig.~\ref{fig:decayrate1}. On the top, the limiting value $2 C_F W(\infty,T)$ is represented by a black dotted line. Right: Same calculation with in-medium states.}
\label{fig:decayrate2}
\end{figure}

\begin{figure}[H]
    \centering
\includegraphics[width=.5\linewidth]{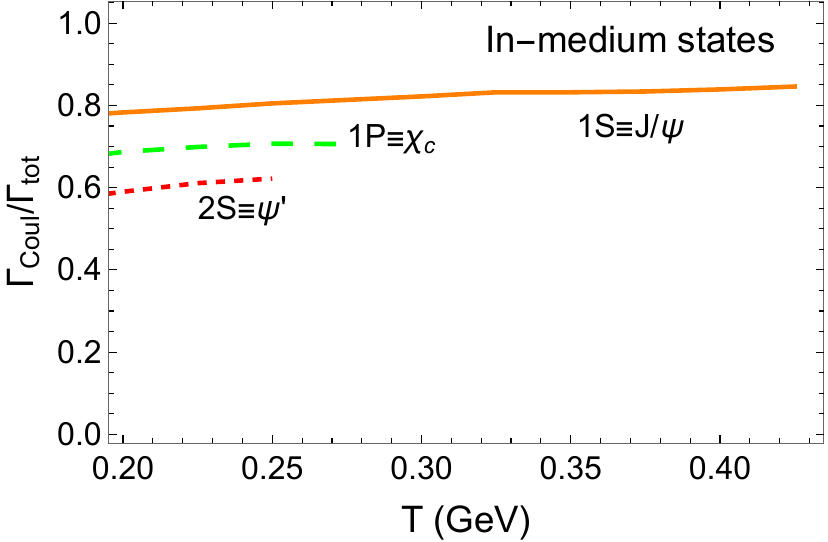}
    \caption{Fraction of the decay rate due to the Coulomb-like part of the imaginary potential (same conventions as for Fig.~\ref{fig:decayrate2}).}
    \label{fig:decayrate3}
\end{figure}

The calculations of the rates of the various  singlet states are displayed as a function of the medium temperature in Figs.~\ref{fig:decayrate1}, \ref{fig:decayrate2} and \ref{fig:decayrate3}.
Fig.~\ref{fig:decayrate1} shows the contributions of the individual superoperators to both the  vacuum (left) and the in-medium (right)  1S ($J/\psi$) state.
In the case of the vacuum 1S state, one sees that the $\mathcal{L}_{3,a}$ term generates a significant reduction as compared to the dominant $\mathcal{L}_{2,a}$ term. The $\mathcal{L}_{4,a}$ term is
completely subdominant for $T\gtrsim 0.25~{\rm GeV}$. For low temperature ($T\lesssim 0.2~{\rm GeV}$, not displayed), it is needed for preserving the positivity of the decay rate, although one may note that the positivity violation occurs only at  the lower edge of the QBM regime. The in-medium eigenstate has  a larger  decay rate. This is related to the fact that it is less bound, and correlatively it has a larger spatial spread than the vacuum states.  

On the left panel of Fig.~\ref{fig:decayrate2}, we compare the decay rates of the three vacuum-eigenstates of the 1D potential and find as anticipated that the deepest bound state has the smallest decay rate and is thus is the most robust against dissociation.  With increasing temperature, $W$ reaches its asymptotic value for smaller and smaller distance (see Fig.~\ref{fig:Pot}, right),  and all decay rates converge toward $2 C_FW(\infty,T)\propto T$ which corresponds to twice the damping rate of an individual c quark.

On the right panel of Fig.~\ref{fig:decayrate2}, the decay rates of the in-medium states are illustrated. As these states are broader than the corresponding vacuum states, their decay rate are significantly larger -- as already noticed on Fig.~\ref{fig:decayrate1} -- and even saturate at finite $T$ to $2 C_FW(\infty,T)$ when the bound state ceases to exist. For $T\gtrsim 200\,{\rm MeV}$, one observes an approximate scaling $\Gamma \propto T^4$. This increase with $T$ slightly exceeds the law (\ref{eq:taursmallT}) due to the aforementioned growth of the states with $T$.

Finally, Fig.~\ref{fig:decayrate3} illustrates the 
 fraction of the decay width that stems from the Coulomb-like part of the imaginary potential~(\ref{eq:imaginary}). As the string-like part behaves as $x^4$ at small distance $x$, it is less important for 1S state than for the 2S or the 1P. But even at high temperature this strong contribution remains significant for all the states. Note that while this contribution is quantitatively significant, as shown in Fig.~\ref{fig:decayrate3}, we have not identified any qualitative impact on the physical quantities that we have evaluated.    

\section{Evolution under $\mathcal{L}_2$ alone}
\label{app:L2alone}
In this appendix, we analyze the evolution of the system under the sole action of the superoperator $\mathcal{L}_2$, which dominates the dynamics at large $T$ ($T\gtrsim T_{\rm damp}$). This will give us the opportunity to comment on the role of coordinate space as a preferred basis (in the sense where this terminology is used in the theory of open quantum systems \cite{breuer2007theory}): The superoperator $\L_2$ is a local operator in the coordinate space representation, which results in a rapid ``diagonalization" of the density matrix as time increases; This illustrates the effect of collisional decoherence in inducing a partial ``classicalization'' of the system. The color degree of freedom remains quantum though, and this is manifest in the fact that the singlet and octet components of the density matrix remain coupled in their equations of motion. 

The explicit equations for $\mathcal{D}_{\rm s}$ and $\mathcal{D}_{\rm o}$ read  (with $\dot {\cal D}=\rmd {\cal D}/\del t$)
\begin{equation}\label{eq:DsDoeqs}
\left(\begin{array}{c} \dot{\mathcal{D}}_{\rm s} \\
\dot{\mathcal{D}}_{\rm o}
\end{array} \right)= \left(
\begin{array}{cc}
C_F \left(2 W(0) - W_c  \right) & - C_F W^- \\
-\frac{1}{2N_c} W^-& 2 C_F W(0) + \frac{1}{2 N_c} W_c
-\left(\frac{N_c^2-2}{2 Nc}W_a + \frac{W_b}{N_c}\right) 
\end{array} \right)\cdot
\left(\begin{array}{c} \mathcal{D}_{\rm s} \\
{\mathcal{D}}_{\rm o}
\end{array} \right),
\end{equation}
where\footnote{We use here the notation of Ref.~\cite{Blaizot:2017ypk}.} $W_a=2 W(\frac{s-s'}{2})$, $W_b=2 W(\frac{s+s'}{2})$, $W^-=W_a-W_b$, and  $W_c= W(s)+W(s')$. Introducing $\Gamma(r)=W(r)-W(0)$, as well as the analogously defined quantities $\Gamma_{a}$, $\Gamma_{b}$ and  $\Gamma^-=\Gamma_{a}-\Gamma_{b}$, we can rewrite Eqs.~(\ref{eq:DsDoeqs}) as 
\begin{equation}
\left(\begin{array}{c} \dot{\mathcal{D}}_{\rm s} \\
\dot{\mathcal{D}}_{\rm o}
\end{array} \right)= \left(
\begin{array}{cc}
-C_F \left(\Gamma(s)+\Gamma(s') \right) & 2 C_F \left(\Gamma(\frac{s+s'}{2})-\Gamma(\frac{s-s'}{2})\right) \\
\frac{1}{N_c} \left(\Gamma(\frac{s+s'}{2})-\Gamma(\frac{s-s'}{2})\right)&  -\frac{1}{2 N_c} \left(\Gamma(s)+\Gamma(s') \right) +
\mathcal{L}_{o\leftrightarrow o} 
\end{array} \right)\cdot
\left(\begin{array}{c} \mathcal{D}_{\rm s} \\
{\mathcal{D}}_{\rm o}
\end{array} \right),
\label{eq:evolF3}
\end{equation}
where we have introduced the $\mathcal{L}_{o\leftrightarrow o}$ superoperator:
$$\mathcal{L}_{\rm o\leftrightarrow \rm o} =
\frac{1}{N_c} \left(\Gamma(s)+\Gamma(s') \right)
-\left(\frac{N_c^2-2}{2 Nc}\Gamma_a + \frac{1}{N_c}\Gamma_b\right),
$$
which leads to transitions exclusively happening within the octet component. It can be verified that  $\mathcal{L}_{\rm o\leftrightarrow \rm o}(s'=s)=0$, so that  $\mathcal{L}_{\rm o\leftrightarrow \rm o}$ does not modify the octet weight (${\rm tr}{\cal D}_{\rm o}$) nor the singlet one (${\rm tr}{\cal D}_{\rm s}$).

Eqs.~(\ref{eq:evolF3}) do not only preserve the trace of the density matrix, but also the total density of heavy quark pairs, $\rho_{\rm s}(r,t)+(N_c-1)^2\rho_{\rm o}(r,t)$, where $\rho_{\rm s}(r,t)=\mathcal{D}_{\rm s}(r,r,t)$ and $\rho_{\rm o}(r,t)=\mathcal{D}_{\rm o}(r,r,t)$. It can indeed be easily verified that this total density remains constant in time, for any $r$. To see that, let us set $r=s=s'$ in Eqs.~(\ref{eq:evolF3}), and note that $\Gamma(0)=0$. Then  these equations simplify into
\begin{equation}
\left(\begin{array}{c} \dot{\rho}_{\rm s} \\
\dot{\rho}_{\rm o}
\end{array} \right)= \Gamma(r) \left(
\begin{array}{cc}
-2 C_F  & 2 C_F  \\
\frac{1}{N_c} &  -\frac{1}{  N_c} 
\end{array} \right)\cdot
\left(\begin{array}{c} {\rho}_{\rm s} \\
{\rho}_{\rm o}
\end{array} \right).
\label{eq:evolF3b}
\end{equation}
The property mentioned above then follows easily. We derive also the following equations 
\beq
\frac{\rmd{\rho}_8}{\rmd t}=-N_c \Gamma(r) {\rho}_8,\qquad \rho_8(r,t)={\cal D}_8(r,r,t), \qquad {\cal D}_8\equiv \frac{1}{N_c^2}\left({\cal D}_{\rm s}- {\cal D}_{\rm o}\right),
\eeq
which indicate that the system is driven to local color equilibrium ($\rho_{\rm s}=\rho_{\rm o}$) at a rate $N_c\Gamma(r)$ which depends on the (fixed) size $r$ of the $Q\bar Q$ pair.

To proceed further, we approximate the imaginary part of the potential $W$ by a harmonic expansion,\footnote{This approximation is a priori valid only when $x\lesssim m_D^{-1}$. We use it here for all values of $x$ in order to get simple analytical expressions.} $W(x)\approx W(0)+\frac{\kappa}{2}x^2$. 
The densities are then given by
\beq
\rho_{\rm s}(r,t)-\rho_{\rm o}(r,t)=\rme^{-N_c r^2\tau} \rho_0(r),\quad \rho_{\rm s}(r,t)=\left[ \frac{N_c^2-1}{N_c^2} \rme^{-N_c r^2\tau}+\frac{1}{N_c^2} \right] \rho_0(r),
\label{eq:densitiesos}
\eeq
where $\rho_0(r)=\rho_{\rm s}(r,t=0)$ is the initial singlet density ($\rho_{\rm o}(r,t=0)=0$), and where we have introduced the reduced time $\tau={\kappa t}/{2}$.  At late time, the densities tend to a common value that is independent of time,  $\rho_{\rm s}^{\rm asymp}(r)= \rho_{\rm o}^{\rm asymp}(r)=\frac{1}{N_c^2} \rho_0(r)$. This corresponds to a maximum entropy state for the color degrees of freedom: All the color states are equally populated, with the densities varying with $r$  according to the initial density profile $\rho_0(r)$. A similar result holds if the initial density is an octet density, i.e., $\rho_{\rm o}(r,t=0)=\rho_0(r)$, and $\rho_{\rm s}(r,t=0)=0$. One finds $\rho_{\rm o}(r,t)=\frac{1}{N_c^2} \left[ N_c^2-1+\rme^{-N_c r^2 \tau}  \right]\rho_0(r)$.

From the densities $\rho_{\rm s}(r,t)$ and $\rho_{\rm o}(r,t)$, one can calculate the evolution in time of the singlet and octet weight, respectively  ${\rm tr}{\cal D}_{\rm s}(t)$ and ${\rm tr}{\cal D}_{\rm o}(t)$. In order to get explicit expressions, we choose  as initial (singlet) density the normalized Gaussian  
\beq\label{eq:gaussianwp2}
\rho_0(r)=\frac{1}{\sqrt{\pi \sigma^2}} \rme^{-\r^2/\sigma^2} .
\eeq
We then obtain
\beq
{\rm tr}{\cal D}_{\rm s}(t)= \frac{N_c^2-1}{N_c^2} \frac{1}{\sqrt{1+\sigma^2 N_c \tau} }+\frac{1}{N_c^2}.
\label{eq:trDsL2alonegaussian}
\eeq
This is a much slower decay than the exponential relaxation of the density toward its equilibrium value, as indicated in Eq.~(\ref{eq:densitiesos}). To understand in more details how this behavior emerges, let us expand ${\rm tr}{\cal D}_{\rm s}(\tau)$ in
 leading order in $\tau$. The leading order contributions to ${\cal D}$ read
 \begin{equation}\label{eq:Dfirstordertau}
\mathcal{D}^{(1)}_{\rm s}(\tau)=\left[1-
C_F (s^2+{s'}^2) \tau \right] 
\mathcal{D}_{\rm s}(0)
\quad\text{and} \quad
\mathcal{D}^{(1)}_{\rm o}(\tau)=
\frac{1}{N_c} s s' \tau \; \mathcal{D}_{\rm s}(0).
\end{equation}
From these equations, or from the direct expansion of (\ref{eq:trDsL2alonegaussian}), it follows that 
\beq\label{eq:trDs0}
{\rm tr}{\cal D}^{(1)}_{\rm s}=1-\frac{1}{2} \sigma^2 \frac{N_c^2-1}{N_c} \tau= 1-2 C_F \langle r^2\rangle_0 \tau=1-\Gamma_0 t,
\eeq
where $\Gamma_0=C_F \kappa \langle r^2\rangle_0$, with  $\langle r^2\rangle_0=\sigma^2/2$, plays the role of a damping rate.\footnote{This is in agreement with the definition (\ref{eq:dampingratedef}), with here $\L_{2a}=-2C_F \Gamma(r)$ and $\Gamma(r)=\kappa r^2/2$.} This leading order result, however, does not exponentiate into $\rme^{-\Gamma_0 t}$. At second order we have indeed 
\begin{equation}
{\rm tr}\mathcal{D}_{\rm s}^{(2)}(t) =
1-2 C_F \langle r^2 \rangle_0 \tau + (2 C_F^2+\frac{C_F}{N_c})   \langle r^4 \rangle_0 \tau^2 .
\end{equation}
The term in $\tau^2$ obviously differs from $(1/2) \Gamma_0^2 t^2$ coming from the expansion of $\rme^{-\Gamma_0 t}$. There are two effects responsible for this. One is a ``correction'' coming from the coupling between the singlet and octet sectors, represented by the term $C_F/N_c$ which adds to the term $2C_F^2$ in the coefficient of $\tau^2$. Such a coupling term is absent in the term linear in $\tau$ if $\rho_{\rm o}=0$ at the start of the evolution, which is the considered case. The second effect comes from the fact that $\langle r^4\rangle_0\ne \langle r^2\rangle_0^2$, where $\langle \cdot\rangle_0$ denotes an average over the initial  density $\rho_0(r)$. For the Gaussian density (\ref{eq:gaussianwp2}), $\langle r^4\rangle_0= 3 \langle r^2\rangle_0^2$. Thus, even ignoring the back-contribution of the octet sector, there is a deviation from the exponential behavior that stems simply from the fact that the average of the exponential is not the exponential of the average. The comparison between the leading order terms and the exact behavior (\ref{eq:trDsL2alonegaussian}) reveals the intricate coupling between the color dynamics and the evolution of the singlet and octet densities. Note in particular how the color prefactor in Eq.~(\ref{eq:trDs0}) contributes to change the time scale of the evolution, from $(N_c\kappa \langle r^2\rangle_0)^{-1}$ to $(C_F\kappa \langle r^2\rangle_0)^{-1}$.  Note also that, leaving aside the color dynamics that couple strongly  the singlet and octet sectors, the calculation above requires only ``classical'' information, meaning that it involves only the diagonal components of the coordinate space density matrices ${\cal D}_{\rm s}$ and ${\cal D}_{\rm o}$, i.e. the densities of singlet and octet $Q \bar Q$ pairs. We shall consider later in this appendix quantities that are sensitive to the non-diagonal matrix elements ${\cal D}(s,s')$. 

\begin{figure*}[t]
\centering
\includegraphics[width=0.5\linewidth]{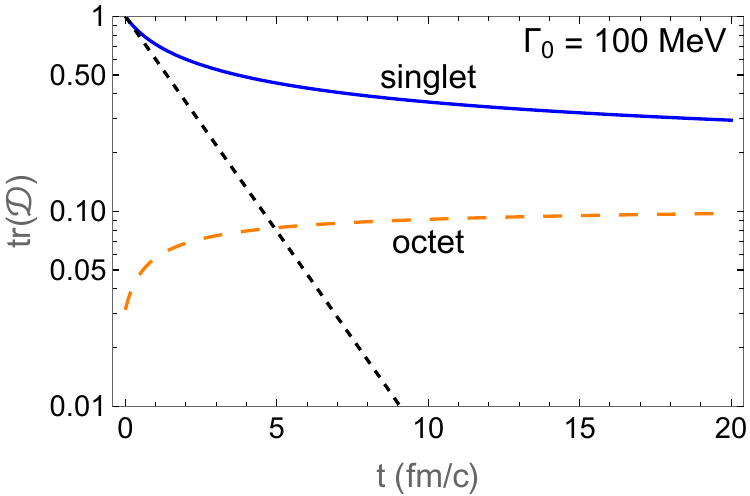}
\caption{Same as Fig.~\ref{fig:singletoctetvstime} for the evolution performed solely with $\mathcal{L}_2$. $\Gamma_0$ in the harmonic approximation of $W$ has been set to $100\,{\rm MeV}$, corresponding to the decay-width of a 1S in-medium state  -- $\langle r^2 \rangle_0 \approx 0.125\,{\rm fm}^2$ -- at $T=300$ MeV. The slope of the dashed black line is minus this decay rate.}
\label{fig:singletoctetvstimeL2}
\end{figure*}

A plot of the singlet and octet weights as a function of time is given in  Fig.~\ref{fig:singletoctetvstimeL2}. 
The octet weight is obtained form the trace of the full density matrix: ${\rm tr} {\cal D}=1={\rm tr} {\cal D}_{\rm s}+(N_c^2-1){\rm tr} {\cal D}_{\rm o}$. The qualitative behavior is quite similar to that observed in the left panel of Fig.~\ref{fig:singletoctetvstime}. However it appears that color equilibration is taking here a much longer time. This can, to a large extent, be attributed to the fact that, in contrast to the situation in Fig.~\ref{fig:singletoctetvstime},  the role of the continuum states is here hindered  by the fact that the entire density stays localized in the region defined by the initial density $\rho_0(r)$.  In comparison with the system considered in Sec.~\ref{sec:2.A}, a $Q\bar Q$ pair in a box of size 20 fm, the typical values of $r$ involved in the present calculation are ``small'', meaning that the damping rate is small, hence the longer time scale observed for color equilibration.

Another characterization of the variation of the densities with time is provided by the variation of the expectation value of $r^2$ (see e.g. Fig.~\ref{fig:p2oftime}). For the singlet density, we obtain 
\beq\label{eq:rmsL2}
\frac{\langle r^2\rangle_{\rm s}}{\langle r^2\rangle_0}=
\frac{1+\frac{N_c^2-1}{\left(1+ N_c \sigma^2 \tau \right)^{3/2}}}{1+\frac{N_c^2-1}{\left(1+ N_c \sigma^2 \tau \right)^{1/2}}}.
\eeq
The variation of $\langle r^2\rangle_{\rm s}$ with time is a genuine non-abelian effect, i.e. an effect which results from the coupling between the singlet and octet sectors. Indeed, in an abelian theory the density would remain constant under the action of $\L_2$ alone. To see that, consider the abelian version of Eqs.~(\ref{eq:evolF3}) 
\beq
\dot {\cal D}_{_{\rm QED}}=\left[ -C_F \left(\Gamma(s)+\Gamma(s') \right) + 2 C_F \left(\Gamma(\frac{s+s'}{2})-\Gamma(\frac{s-s'}{2})\right)\right] {\cal D}_{_{\rm QED}}(s,s',t).
 \eeq
 The solution, in the harmonic approximation, reads simply
 \beq\label{eq:DQEDssp}
{\cal D}_{_{\rm QED}}(s,s',t)=\rm e^{-\frac{C_F\kappa t}{2} (s-s')^2 }{\cal D}_{_{\rm QED}}(s,s',0)\,,
\eeq
which shows that the main effect of collisions in the abelian case is momentum broadening (see below). It  follows from this equation that $\rho_{_{\rm QED}}(r,t)={\cal D}_{_{\rm QED}}(r,r,t)=\rho_0(r)$ is a constant, and so is therefore $\langle r^2\rangle$. The equation $\rho_{_{\rm QED}}(r,t)=\rho_0(r)$ reflects the local conservation of the total number of $Q\bar Q$ pairs. The same conservation law in the case of QCD applies to the combination $\rho_{\rm s}+(N_c^2-1) \rho_{\rm o}$, as we have already emphasized.  What happens in the QCD case is that the singlet to octet transitions start to populate the octet states at large $r$ (where $\Gamma(r)$ is the largest). This naturally entails a variation of the singlet density, reflected in the decrease of $\langle r^2\rangle_t$ according to the formula above. At small time, we have approximately ${\langle r^2\rangle_{\rm s}}/{\langle r^2\rangle_0}\simeq 1-2\Gamma_0 t$. At late time however, the populations equilibrate, as we have seen earlier, and in particular $\rho_{\rm s }(r,t)\simeq \rho_0(r)/N_c^2$. When this situation is reached, $\langle r^2\rangle_{\rm s}\approx \langle r^2\rangle_0$ again. Note that the variation of $\langle r^2\rangle_{\rm s}$ concerns only the singlet density. Because the total density of $Q\bar Q$ pairs is locally conserved, it remains independent of time, and so does the corresponding rms radius. As for the octet, its rms radius starts to grow at early time, reaches some maximum and then decreases toward $\langle r^2\rangle_0$ for $t\to\infty$.

\begin{figure*}[h]
\centering
\includegraphics[width=0.49\linewidth]{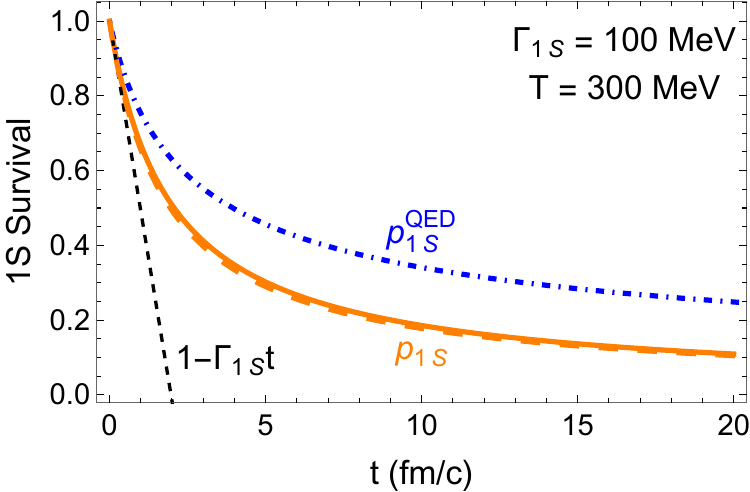}
\includegraphics[width=0.49\linewidth]{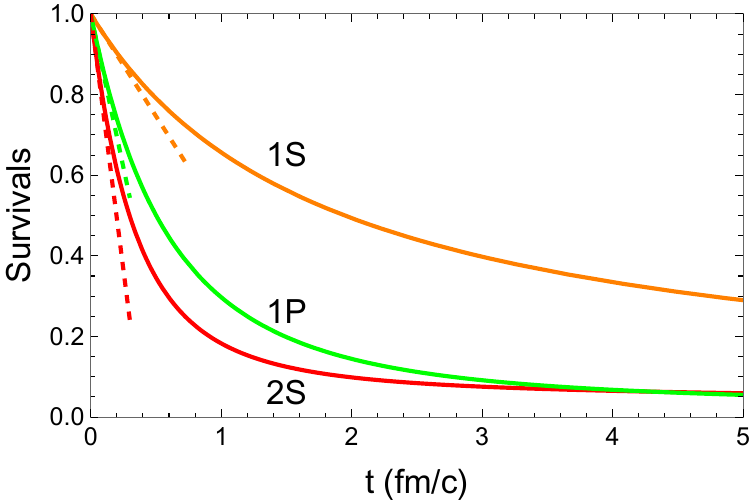}
\caption{Left: Same as Fig.~\ref{fig:Proba1Ssingletthermal300} (left panel) for the evolution performed solely with $\mathcal{L}_2$. $\sigma_{\rm 1S}$ has been taken as 0.55~fm while $\kappa\approx 500$~MeV~fm$^{-2}$ has been adjusted to get $\Gamma_{\!_{\rm 1S}}= 100\,{\rm MeV}=\Gamma_{\!_{\rm 1S}}^{\rm in-medium}$ for the ``exact" decay width at $T=300\,{\rm MeV}$. The dashed black line indicates the initial linear decay. The thick-solid orange line represents the QCD survival probability obtained numerically, and the dashed orange line is  the approximation Eq.~(\ref{eq:QCDp1SL2alone}), while the blue dot-dashed line illustrates the equivalent abelian law (\ref{eq:psurvivalQEDbis}). Right: Survival probabilities of the 1S, 1P and 2S states, calculated from Eqs.~(\ref{eq:QCDp1SL2alone}) and corresponding laws for excited states; same conditions. The dashed lines represent the linear approximation $1-\Gamma_{\!{n \rm {S/P}}}\,t$.}
\label{fig:probvstimeL2} 
\end{figure*}

We turn now to the  solution for the general (non-diagonal) case. This will allows us to exhibit the general structure of ${\cal D}(s,s')$ and calculate some quantities which require non-diagonal information, such as the average momentum square, or survival probabilities. In the harmonic approximation, Eqs.~(\ref{eq:evolF3}) read
\begin{equation}
\left(\begin{array}{c} \rmd{\mathcal{D}}_{\rm s}/\rmd\tau \\
\rmd{\mathcal{D}}_{\rm o}/\rmd\tau 
\end{array} \right)\approx  
 \left(
\begin{array}{cc} 
- C_F\left(s^2+{s'}^2  \right) &   2 C_F  s s'   \\
\frac{1}{N_c}\, s s'  & -\frac{1}{2N_c}\left(s^2+{s'}^2\right)   -\frac{\left(N_c^2-4\right) (s-s')^2}{4 N_c} 
\end{array} \right)
\left(\begin{array}{c} \mathcal{D}_{\rm s} \\
{\mathcal{D}}_{\rm o}
\end{array}\right),
\end{equation}
These equations admit explicit solutions. We focus on solutions corresponding to an initial condition such that 
$\mathcal{D}_{\rm o}(s,s',t=0)=0$. One then gets:
\begin{equation}\label{eq2:evolutionDUss}
\mathcal{D}_{\rm s}(s,s',\tau) = U^{\rm ss}(s,s',\tau) {\mathcal{D}_{\rm s}(s,s',0)} 
\end{equation}
with
\begin{equation}
U^{\rm ss}(s,s',\tau)=e^{-\left(\frac{N_c}{8} (s+{s'})^2
+\frac{N_c^2-2}{4 N_c}(s-s')^2\right)\tau  }
   \left(\cosh
   \left(\alpha \tau\right)-\left(
   \frac{N_c^2-2}{2}(s+s')^2+ (s-s')^2\right) \frac{\sinh
   \left(\alpha  \tau\right)}{4 N_c\alpha}\right),
\label{eq:L2alonefullsol}
\end{equation}
where
\begin{equation}
\alpha(s,s') =
\frac{1}{8} \sqrt{N_c^2 (s+s')^4-16 s s' (s-s')^2}.
\end{equation}
In the long time limit, $U^{\rm ss}(s,s',\tau)$ shrinks along the $s-s'$ direction and becomes 
\begin{equation}\label{eq:usslongtime}
U^{\rm ss}(s,s',\tau) \approx \frac{1}{N_c^2} \exp\left(-\left(\frac{N_c}{2} r^2 +\frac{N_c^2-2}{4 N_c}(s-s')^2- \alpha\right) \times \tau  \right)\underset{\mathrm{finite}\; r}{\approx} \frac{1}{N_c^2}
\rme^{-\frac{N_c^2-1}{4 N_c}(s-s')^2\,\tau }. 
\end{equation}
The factor $(s-s')^2 \tau$ in the exponent of this expression suggests  that at late time, the density matrix   $\mathcal{D}_{\rm s}(s,s',t)$ becomes diagonal,  $\propto \rho_0(s) \delta(s-s')$, an expected feature of classicalization.\footnote{For $r=0$, one has $\alpha=\frac{y^2}{4}$ and $U^{\rm s s'}\approx \frac{1}{N_c^2} \exp(-\frac{N_c^2+N_c-2}{4 N_c}(s-s')^2)$, confirming that the shrinking along the $s-s'$ direction acts $\forall r$.}
\begin{figure}[h]
    \centering
\includegraphics[width=.48\linewidth]{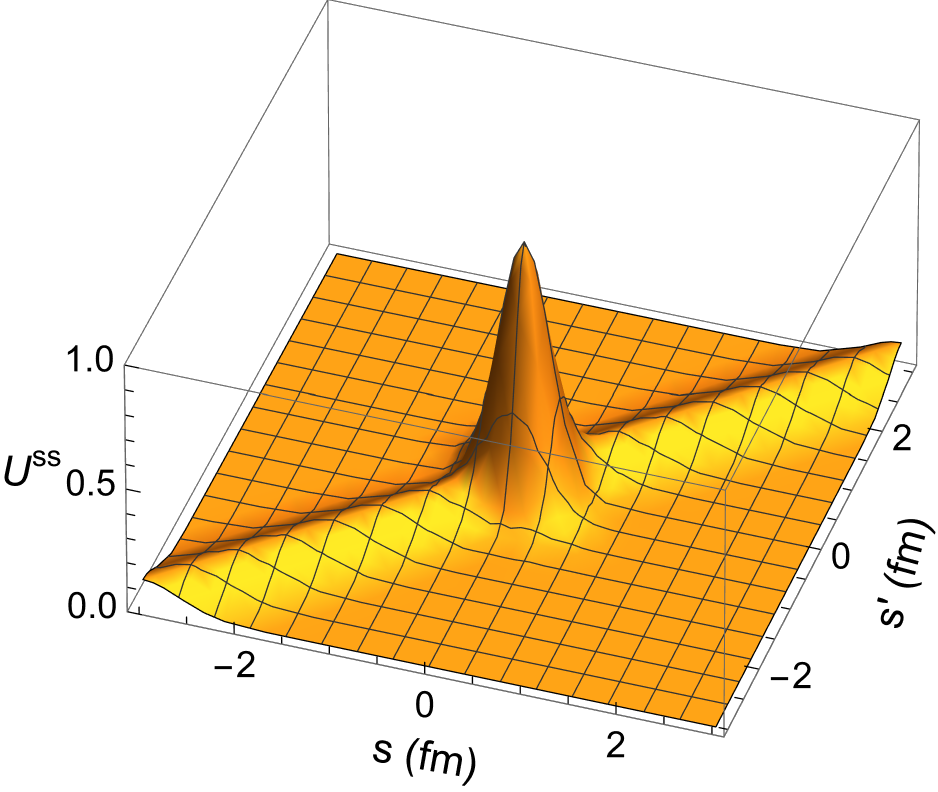}
\includegraphics[width=.48\linewidth]{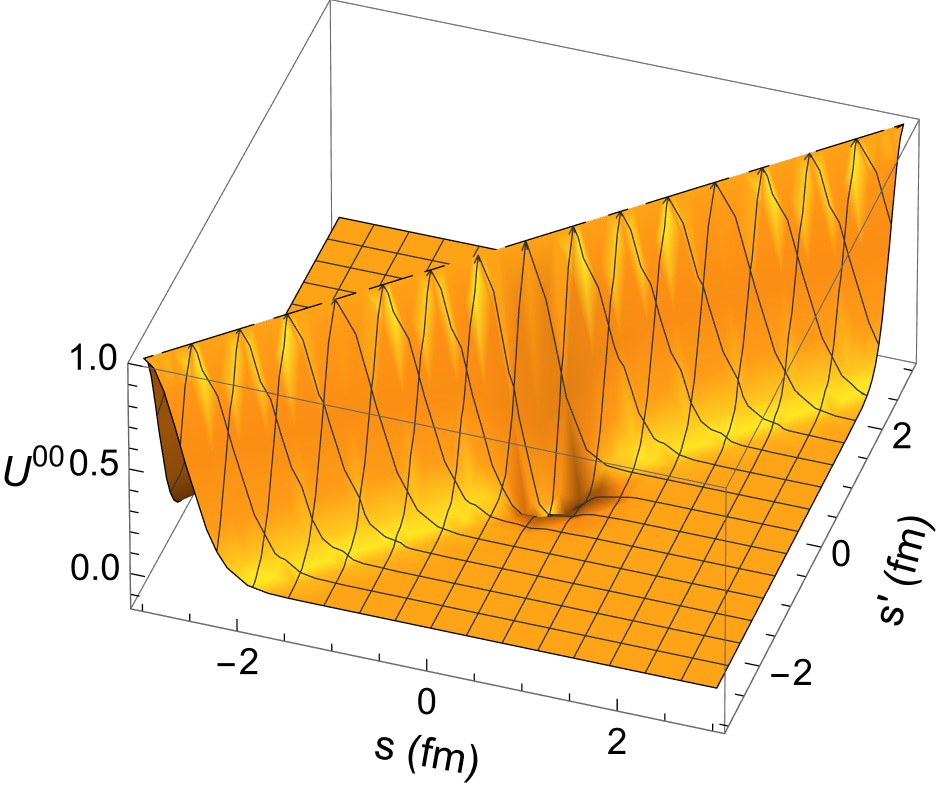}
\caption{Illustration of $U^{\rm ss}(s,s',\tau)$ for $\tau=5\,{\rm fm}^{-2}$. For $\tau=0$, $U^{\rm ss}(s,s',\tau)=1$, while at late time $U^{\rm ss}(s,s',\tau)$ is given by Eq.~(\ref{eq:usslongtime}). The damping of the non-diagonal elements is obvious. Near the origin, i.e. for small values of $s$ and $s'$, the damping is slower, as anticipated from the color transparency phenomenon encoded in the harmonic approximation of $W$. This manifests itself explicitly here by the fact that $U^{\rm ss}(s,s',\tau)$ depends on $s,s'$ and $\tau$ only through the combinations $s\sqrt{\tau}$ and $s'\sqrt{\tau}$. At late time the peak conserves its height, but it extends on a small region of size $1/\sqrt{\tau}$ and its weight becomes negligible. Right: Same for the $U^{00}$ evolution superoperator in the $\mathcal{D}_0-\mathcal{D}_8$ basis. }
\label{fig:Uss}
\end{figure}
This can be indeed verified in Fig.~\ref{fig:Uss}, which reveals, in addition to the diagonal structure the presence of a peak in the vicinity of the origin. To better encompass this overall structure, it is convenient to use the following approximation of $U^{\rm ss}$:
\begin{equation}
U^{\rm ss}(s,s',\tau) \approx 
\frac{N_c^2-1}{N_c^2} \rme^{-{N_c}\left(r^2+\frac{y^2}{4}\right)\,\tau} +\frac{1}{N_c^2}
\rme^{-\frac{N_c^2-1}{4 N_c}y^2\,\tau },
\label{eq:Ussapprox}
\end{equation}
where we have set $y=s-s'$ and $r=(s+s')/2$.
When $s'=s$, this expression coincides with that of the singlet density given in Eq.~(\ref{eq:densitiesos}),  and it is equal to unity for $\tau=0$ as it should.   We have checked numerically that this approximate expression provides  a reasonably  accurate representation of $U^{\rm ss}$.\footnote{Although the $N_c$ dependence of the leading order expansion in $y$ of the expression (\ref{eq:Ussapprox}) is not correctly reproduced, the ensuing error is quantitatively negligible (see Eq.~(\ref{eq:p0wrong}) and Fig.~\ref{fig:probvstimeL2}).}  The formula (\ref{eq:Ussapprox}) clearly exhibits the peak near $s=s'=0$. The existence of this peak, which is also present in the density $\rho_{\rm s}(r,t)={\cal D}_{\rm s}(r,r,t)$, reflects the fact that the damping rate $\Gamma(r)$ vanishes for $r=0$, so that there are not singlet to octet transitions at this point. The height of the peak stays therefore constant as time increases ($U^{\rm ss}(0,0,\tau)=1$), but it extends over a smaller and smaller spatial region so that its contribution to physical quantities becomes eventually negligible at late time. Note finally that a peak exists only in the singlet density, or in ${\cal D}_{\rm s}$. As already emphasized, the total density of $Q\bar Q$ pairs is independent of $r$ and presents no peak at small $r$. To illustrate this point, it is convenient to consider an alternative representation of the density matrix \cite{Blaizot:2017ypk}, namely ${\cal D}_0=\frac{1}{N_c^2}({\cal D}_{\rm s}+(N_c^2-1) {\cal D}_{\rm o} )$, for which ${\cal D}_0(s,s',t)=U^{00}(s,s',t) {\cal D}_0(s,s',0)+U^{08}(s,s',t) {\cal D}_8(s,s',0)$.  The evolution superoperator $U^{00}(s,s',t)$ is plotted in the right panel of Fig.~\ref{fig:Uss}. As expected, it reveals a uniform ridge in the $r$ direction, with a non trivial structure only in the $y$ direction. 

With the full non-diagonal density matrix, there are several interesting quantities that can be calculated. One is the average momentum squared of the full distribution of heavy quarks. This can be obtained from the relation 
\beq
\langle p^2 \rangle =- \left. \partial_y^2 \int {\rm d}r \left[\mathcal{D}_{\rm s}(r+\frac{y}{2},r-\frac{y}{2},t)+ (N_c^2-1) \mathcal{D}_{\rm o}(r+\frac{y}{2},r-\frac{y}{2},t)\right]\right|_{y=0}
\eeq 
where both singlet and octet components are taken into account. It turns out that an accurate estimate of $\langle p^2 \rangle$ can be obtained from the leading order expansion of ${\cal D}$ at small time,  which are given in Eqs.~(\ref{eq:Dfirstordertau}). 
The singlet contribution writes
\beq
\frac{d\langle p^2 \rangle_{\rm s}}{dt}=\left.-\frac{\kappa}{2}
 \int {\rm d}r \, \partial_\tau\partial_y^2 \left[1-
C_F (s^2+{s'}^2) \tau \right]\mathcal{D}_{\rm s}(0)\right|_{y=0}=\frac{C_F \kappa}{2}, 
\eeq
where we have used $s^2+{s'}^2=2 r^2+y^2/2$ and $\left.\int {\rm d}r \mathcal{D}_{\rm s}(0)\right|_{y=0} ={\rm tr}{\cal D}_{\rm s}(0)=1$. As for the octet contribution, one has (with $ss'=r^2-\frac{y^2}{2}$)
\beq
\frac{d\langle p^2 \rangle_{\rm o}}{dt}=\left.-\frac{N_c^2-1}{N_c}\frac{\kappa}{2}
 \int {\rm d}r \, \partial_\tau\partial_y^2 \left[
 ss' \tau \right]\mathcal{D}_{\rm s}(0)\right|_{y=0}=\frac{C_F \kappa}{2}. 
\eeq
Adding both contributions leads to $\frac{d\langle p^2 \rangle}{dt}=C_F \kappa$,
which is also the result found in the abelian case. The linear relation $\langle p^2 \rangle=C_F\kappa t=\hat q t$ remains approximately valid till late time. It represents one of the important effects of collisions, namely the diffusion in momentum space, also referred to as momentum broadening. Since the momentum $p$ is conjugate to the variable $y=s-s'$ under Fourier transform, this growth of $\langle p^2 \rangle$ is correlated with the squeezing of the non-diagonal matrix elements of ${\cal D}$ in coordinate space, a phenomenon that we have associated with collisional decoherence. 

A further manifestation of collisional decoherence is seen in the time dependence of the survival probability of initial states, as can be exemplified by taking, as initial state, the 1S state
\beq\label{eq:psi1S0}
\psi_{\rm 1S}(s)=
\frac{1}{\sqrt{\sqrt{\pi}\sigma_{\rm 1S}}}\exp\left(-\frac{s^2}{2\sigma_{\rm 1S}^2}\right),
\eeq
and let ${\cal D}_{_\mathrm{1S}}(s,s')=\psi_{_{\rm 1S}}(s)\psi_{_{\rm 1S}}^*(s')$ be the density matrix associated to  this initial state. We have, for QED,   
\beq
p_{_{\rm 1S}}^{_{\rm QED}}(t)= \int \rmd s \rmd s' \, \psi_{_{\rm 1S}}^*(s)\bra{s}{\cal D}^{_{\rm QED}}_{_\mathrm{1S}}(t)\ket{s'}\psi_{_{\rm 1S}}(s') = \int \rmd s \rmd s' \,\rho_{_{\rm 1S}}(s) \rme^{-\tau (s-s')^2} \rho_{_{\rm 1S}}(s'),
\label{eq:psurvivalQED}
\eeq
where we have used Eq.~(\ref{eq:DQEDssp}) for $\bra{s}{\cal D}_{_{\rm 1S}}^{_{\rm QED}}(t)\ket{s'}$. Note that the squeezing  along the $y=s-s'$ direction caused by the exponential factor entails a power law decrease of $p_{_{\rm 1S}}$ at late time,  $p_0 \propto 
\int dr \rho_{_{\rm 1S}}^2(r) \times \int dy \rme^{-\tau y^2} \propto \tau^{-\frac{1}{2}}$. The explicit evaluation yields
\beq
p_{_{\rm 1S}}^{_{\rm QED}}(t)= \frac{1}{\sqrt{1+C_F \kappa \sigma_{_{\rm 1S}}^2 t}}.
\label{eq:psurvivalQEDbis}
\eeq 
Note that we can write $C_F \kappa \sigma_{_{\rm 1S}}^2 t=\hat q t/(2\langle p^2\rangle_{_{\rm 1S}})$, with   $\hat q=C_F \kappa$ and $\langle p^2\rangle_{_{\rm 1S}}=1/(2 \sigma_{_{\rm 1S}}^2)$. 
Thus, in the abelian case,  the decay of the survival probability $p_{_{\rm 1S}}^{_{\rm QED}}(t)$ is tightly connected with momentum broadening. The characteristic time scale for the decay is $2\langle p^2\rangle_0/\hat q$, and the ``suppression'' of the initial state becomes significant when the average momentum squared gained by collisions, $\hat q t$, becomes comparable to the intrinsic  momentum squared in the initial wave function. The survival probability $p_{_{\rm 1S}}^{_{\rm QED}}(t)$ is displayed as a function of time in the left panel of Fig.~\ref{fig:probvstimeL2}. For the parameters used, we have $\hat q\approx 0.13\,{\rm GeV}^2$/fm and $\langle p^2\rangle_{_{\rm 1S}}\approx 0.07\,{\rm GeV}^2$. It follows that the linear behavior predicted by Eq.~(\ref{eq:psurvivalQEDbis}) holds only for a time of the order of $t=\langle p^2\rangle_{_{\rm 1S}}/\hat q\approx 0.5$ fm/$c$.

While in the abelian case, the decay of the survival probability arises solely from momentum space diffusion, or momentum broadening, in QCD, an additional factor comes from the color dynamics. In order to get a simple estimate of $p_{_{\rm 1S}}(t)$ in he QCD case, we use the approximation (\ref{eq:Ussapprox}) and get:
\begin{equation}
p_{\rm 1S}(t)
\approx\frac{N_c^2-1}{N_c^2} \times \frac{1 }{ \left(1+ \bar{\tau} \right)}+ 
\frac{1}{N_c^2} \times \frac{1}{\sqrt{1+\hat{\tau}}},
\label{eq:QCDp1SL2alone}
\end{equation} 
where we have set $\hat \tau=\Gamma_{\!_{\rm 1S}} t$, with $\Gamma_{\!_{\rm 1S}} = C_F \kappa \langle r^2\rangle_{\!_{\rm 1S}}$, and   $\bar{\tau}=\frac{1}{2}N_c\kappa \langle r^2\rangle_{\!_{\rm 1S}} t$. This estimate is compared to the exact result obtained with the full expression (\ref{eq:L2alonefullsol}) in Fig.~\ref{fig:probvstimeL2}. As one can see, the agreement is quite good.\footnote{As mentioned earlier,  a drawback of the approximation (\ref{eq:Ussapprox}) is that it does not treat accurately the color dynamics at early time and in particular it does not yield the correct $N_c$ dependence of the small time expansion of the survival probability $p_{{\rm 1S}}$. We have indeed from Eq.~(\ref{eq:QCDp1SL2alone})
\beq\label{eq:p0wrong}
p_{{\rm 1S}}(t)\simeq 1-\Gamma_{\rm 1S} t \left(1+\frac{1}{2 N_c^2}  \right),
\eeq 
whereas the correct behavior is $p_{{\rm 1S}}(t)\simeq 1-\Gamma_{\rm 1S} t$, as can be easily deduced from the exact expression at linear order in $\tau$. The additional term in Eq.~(\ref{eq:p0wrong}) is a spurious artifact of the approximation (\ref{eq:Ussapprox}). The comparison with the exact calculation shown in Fig.~\ref{fig:probvstimeL2} indicates that this spurious contribution is of little quantitative consequence.} Fig.~\ref{fig:probvstimeL2} also shows that the time dependence of $p_{\rm 1S}$ follows that of the abelian case, with however the additional damping that comes from the singlet-octet transitions. 

The early time behavior $p_{_{\rm 1S}}(t)\simeq 1-\Gamma_{\!_{\rm 1S}} t$ coincides with the leading expression of the singlet weight, Eq.~(\ref{eq:trDs0}). This is because, at early time, the density matrix is dominated by the initial singlet state. Similar results hold for other choices of initial states. Let us consider 
the states 1P and 2S, with the following harmonic wave functions 
\begin{equation}
\psi_{_{\rm 1P}}(s)=\frac{\sqrt{2}}{\pi^{1/4}} \sigma_{_{\rm 1S}}^{3/2}\,s\,\rme^{-\frac{s^2}{2 \sigma_{_{\rm 1S}}^2}},
\qquad
\psi_{_{\rm 2S}}(s)=
\frac{1}{\sqrt{2\sqrt{\pi}\sigma_{_{\rm 1S}}}}\left(1-\frac{2s^2}{\sigma_{_{\rm 1S}}^2}\right)e^{-\frac{s^2}{2\sigma_{_{\rm 1S}}^2}}.  
\label{eq:1P2Sharmonic}
\end{equation} 
These wave functions are orthogonal and also orthogonal to the 1S wave function (\ref{eq:psi1S0}). It is easy to verify, using Eq.~(\ref{eq:trDs0}) that the early time linear behavior of $p_{_{\rm 2S}}$ and $p_{_{\rm 1P}}$ is the same as that of $p_{_{\rm 1S}}$, once the proper sizes ($\langle r^2\rangle_{_{\rm 2S}}=5 \langle r^2\rangle_{_{\rm 1S}}$ and $\langle r^2\rangle_{_{\rm 1P}}=3\langle r^2\rangle_{_{\rm 1S}}$) are used in the respective expressions of the decay constant, i.e. $\Gamma_{\,_{\rm 1P}}=3\Gamma_{\,_{\rm 1S}}$ and $\Gamma_{\,_{\rm 2S}}=5\Gamma_{\,_{\rm 1S}}$. As can be seen in Fig.~\ref{fig:probvstimeL2}, this expected linear behavior is well reproduced, while at late time  $p_{_{\rm 2S}}$ and $p_{_{\rm 1P}}$ behave similarly as $p_{_{\rm 1S}} \propto t^{-1/2}$.

\begin{figure*}[h]
\centering
\includegraphics[width=0.48\linewidth]{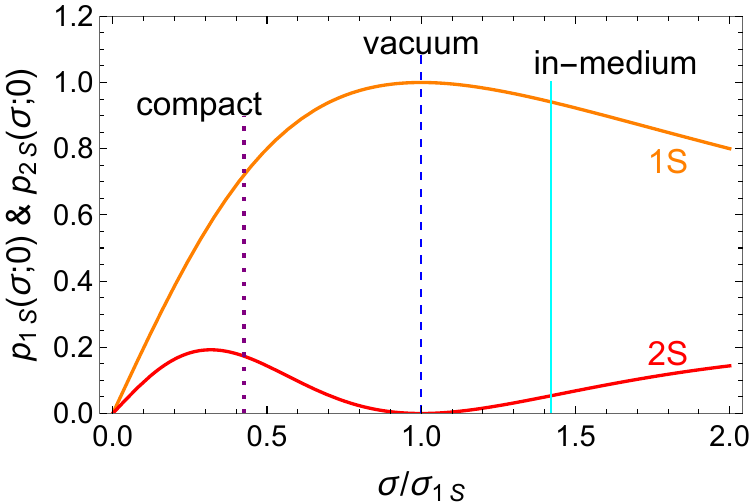}
\caption{Initial 1S and 2S weights in a Gaussian wave packet -- Eq.~(\ref{eq:gaussianwp2}) -- as a function of the rescaled width $\sigma/\sigma_{_{\rm 1S}}$. The 3 vertical lines illustrate the 3 initial states studied in Sec.~\ref{sec:2.C}: vacuum 1S ($\sigma=\sigma_{_{\rm 1S}}\equiv 0.387\,{\rm fm}$), in-medium  1S state ($\sigma \equiv$~0.55~fm), and compact Gaussian state ($\sigma =$~0.165~fm).}
\label{fig:2Sprobappendix}
\end{figure*}

We have seen  that at large time, the density matrix becomes rapidly diagonal in coordinate space (see Eq.~(\ref{eq:DQEDssp})), a property that we have associated to collisional decoherence. To emphasize the dependence on the choice of basis, and the fact that quantum coherence may survive in other basis, we return to the case of an initial wave packet considered in Sec.~\ref{sec:2.C} and write the corresponding initial singlet density matrix ${\cal D}_{\rm s}^\sigma(0)=|G_\sigma \rangle \langle G_\sigma|$ where $\sigma$ characterizes the width of the Gaussian. Depending on the value of $\sigma$, the probabilities $p_{_{\rm 1S}}$ and $p_{_{\rm 2S}}$ have different values at $t=0$. These are given explicitly by 
\beq
p_{_{\rm 1S}}(\sigma;t=0)=
\frac{2 \sigma \sigma_{_{\rm 1S}}}{\sigma ^2+ \sigma_{_{\rm 1S}}^2},\qquad p_{_{\rm 2S}}(\sigma;t=0)=\frac{\sigma \sigma_{_{\rm 1S}}(\sigma_{_{\rm 1S}}^2- \sigma^2 )^2}{(\sigma_{_{\rm 1S}}^2 +\sigma^2 )^3},
\label{eq:evolp1Sp2SgaussianL2alone}    
\eeq
and their dependence on $\sigma$ is illustrated in Fig.~\ref{fig:2Sprobappendix}. Note that $p_{_{\rm 2S}}$ vanishes for $\sigma=\sigma_{_{\rm 1S}}$. In this case, $\ket{G_\sigma}=\ket{\rm 1S}$ is orthogonal to $\ket{\rm 2S}$. We shall be interested in the range of $\sigma$ values delineated in Fig.~\ref{fig:2Sprobappendix} by $\sigma_{\rm compact}$ and $\sigma_{\rm in-medium}$. In this range, the initial density matrix is well accounted for by its matrix elements in the subspace spanned by the $\ket{\rm 1S}$ and $\ket{\rm 2S]}$ states. For instance, in the range considered, $p_{_{\rm 1S}}+p_{_{\rm 2S}}\gtrsim 0.9$ (and equals 1 for $\sigma=\sigma_{_{\rm 1S}}$), while the non diagonal element $\langle {\rm 2S} |D_{\rm s}^\sigma (t=0) | {\rm 1S}  \rangle$ is given by 
\beq\label{eq:2SD1S} 
\langle {\rm 2S} |D_{\rm s}^\sigma(0) | {\rm 1S}  \rangle = \frac{\sqrt{2} \sigma \sigma_{_{\rm 1S}}(\sigma_{_{\rm 1S}}^2-\sigma^2)}{(\sigma_{_{\rm 1S}}^2+\sigma^2)^2}.
\eeq
It is then interesting to ask to which extent the evolution of the density matrix can be captured by that of these two states. 

The evolution of the initial density matrix is given by Eq.~(\ref{eq2:evolutionDUss}), 
from which we can extract the probability $p_{_{\rm 2S}}(t)=\int_{ss'} \psi_{_{\rm 2S}}^*(s) \mathcal{D}_{\rm s}^\sigma(s,s',t) \psi_{_{\rm 2S}}(s')$. To within the normalization constants we get 
\beq
p_{_{\rm 2S}}(t) \propto \iint \rmd s \rmd s' e^{-(\frac{1}{2\sigma^2}+\frac{1}{2\sigma_{_{\rm 1S}}^2})(s^2+{s'}^2)} \left[\vphantom{\sum}\right.\underbrace{1}_{A} - \underbrace{\frac{2(s^2+{s'}^2)}{\sigma^2}}_{B} + \underbrace{\frac{4 s^2 {s'}^2}{\sigma^4}}_{C} \left.\vphantom{\sum}\right]\, U^{\rm ss}(s,s',\tau).
\label{eq:ovelap2S}
\eeq
As a result of the evolution induced by $U^{\rm ss}(s,s',\tau)$, both $\langle (s-s')^2\rangle$ and $\langle (s+s')^2\rangle$ shrink with time, producing a differential evolution of the three contributions labeled  $A$, $B$ and $C$ in Eq.~(\ref{eq:ovelap2S}). 
Depending on the value of $\sigma$ this can lead to a  reduction or to an increase of $p_{_{\rm 2S}}$.  Such an effect can be anticipated from Fig.~\ref{fig:2Sprobappendix}, by noticing that the shrinking of $\langle (s-s')^2\rangle$ and $\langle (s+s')^2\rangle$ with time can be seen as an effective  decrease of $\sigma$, producing a decrease of $p_{_{\rm 2S}}$ if initially $\sigma>\sigma_{_{\rm 1S}}$ and an increase in the opposite case. 

Further insight can be gained by  focusing on the short time evolution. At order $\mathcal{O}(t)$,  using Eq.~(\ref{eq:Dfirstordertau}), one obtains  
\beq
p_{_{\rm 1S}}(\sigma;t)=
\frac{2 \sigma \sigma_{_{\rm 1S}}}{\sigma ^2+ \sigma_{_{\rm 1S}}^2}  \left(1- C_F \frac{\sigma ^2
    \sigma_{_{\rm 1S}}^2}{\sigma ^2+ \sigma_{_{\rm 1S}}^2}\,\kappa t  \right). 
\label{eq:evolp1SgaussianL2alone}    
\eeq
Thus, whatever the value of $\sigma$, $p_{_{\rm 1S}}$ decreases linearly at small time. Note however that the decay constant is the harmonic average of the decay constants $\Gamma_{\!_{\rm 1S}}$ and $\Gamma_\sigma$. It reduces to $\Gamma_{\!_{\rm 1S}}$ only when $\sigma\to \sigma_{_{\rm 1S}}$.  For the 2S, we get
\beq
p_{_{\rm 2S}}(\sigma;t)=\frac{\sigma \sigma_{_{\rm 1S}}}{(\sigma_{_{\rm 1S}}^2 +\sigma^2 )^3}
\left[(\sigma_{_{\rm 1S}}^2- \sigma^2)^2  - 
C_F\, \frac{\sigma^2 \sigma_{_{\rm 1S}}^2(\sigma_{_{\rm 1S}}^2- \sigma^2)(\sigma_{_{\rm 1S}}^2-5 \sigma^2)}{\sigma_{_{\rm 1S}}^2 +\sigma^2} \,\kappa t + \cdots \right]. 
\label{eq:evolp2SgaussianL2alone}
\eeq
In the relevant range of $\sigma$-values, the time derivative of  $p_{_{\rm 2S}}$ is  negative for $\sigma\ge\sigma_{_{\rm 1S}}$ and positive for $\sigma\le \sigma_{_{\rm 1S}}$, confirming the previous heuristic argument.  This result can also be understood from a different perspective, relying on the expression of the density matrix on the $\{\ket{\rm 1S},\ket{\rm 2S}\}$ subspace. By projecting the density matrix at any time $t$ on the basis $\{|{\rm 1S} \rangle \langle {\rm 1S} |,|{\rm 2S} \rangle \langle {\rm 2S} |,|{\rm 1S} \rangle \langle {\rm 2S} |,|{\rm 2S} \rangle \langle {\rm 1S} |\}$, one can then follow the time evolution of corresponding matrix elements. At leading order in $t$ the expression of $p_{_{\rm 1S}}$ is reproduced, while for $p_{_{\rm 2S}}$ one finds
\beq\label{eq:evolp2SgaussianL2aloneb}
p_{_{\rm 2S}}(\sigma;t) &=& p_{_{\rm 2S}}(\sigma;0) 
- C_F \frac{\sigma  \sigma_{_{\rm 1S}}^3
   (\sigma_{_{\rm 1S}}^2-\sigma^2)  \left(3 \sigma_{_{\rm 1S}}^2-7 \sigma^2\right)}{2
   \left(\sigma ^2+\sigma_{_{\rm 1S}}^2\right)^3} \kappa t\nn 
   & \approx & p_{_{\rm 2S}}(\sigma;0) + C_F 
   \frac{\sigma_{_{\rm 1S}}^2-\sigma^2}{4} \kappa t
\eeq
This relation is not the same as  (\ref{eq:evolp2SgaussianL2alone}). However, both relations yield the same time dependence when $\sigma\simeq \sigma_{_{\rm 1S}}$, as indicated in the second line of Eq.~(\ref{eq:evolp2SgaussianL2aloneb}) above. The difference can be attributed to the fact that the initial wave packet $\ket{G_\sigma}$ contains an admixture of a $\ket{\rm 3S}$ state, and this can feed the 2S (but not the 1S). This admixture does not affect $p_{_{\rm 1S}}$, and is negligible when $\sigma \simeq \sigma_{_{\rm 1S}}$. Once this is included, one indeed reproduces (\ref{eq:evolp2SgaussianL2alone}). 

\begin{figure*}[t]
\centering
\includegraphics[width=0.49\linewidth]{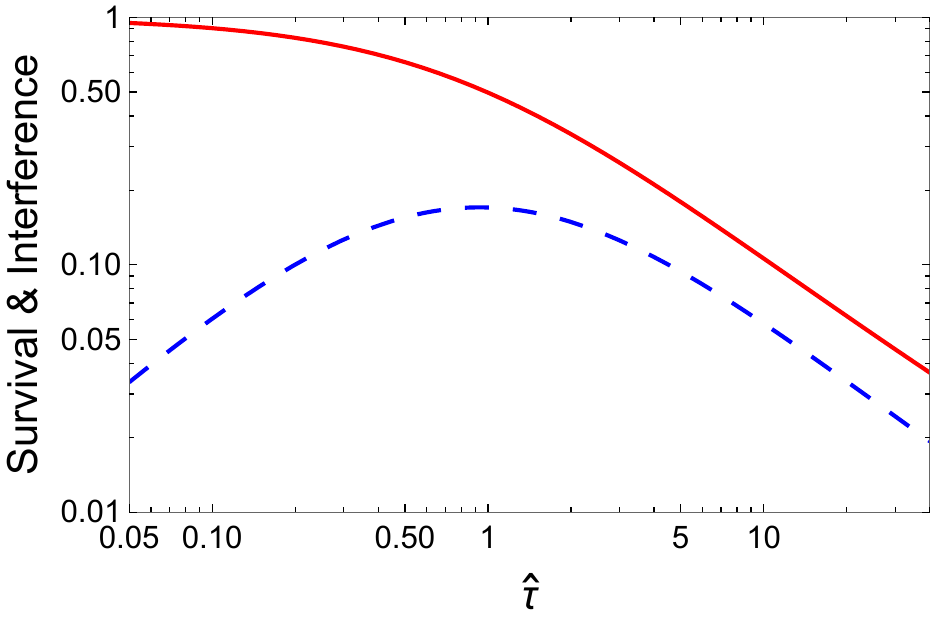}
\includegraphics[width=0.49\linewidth]{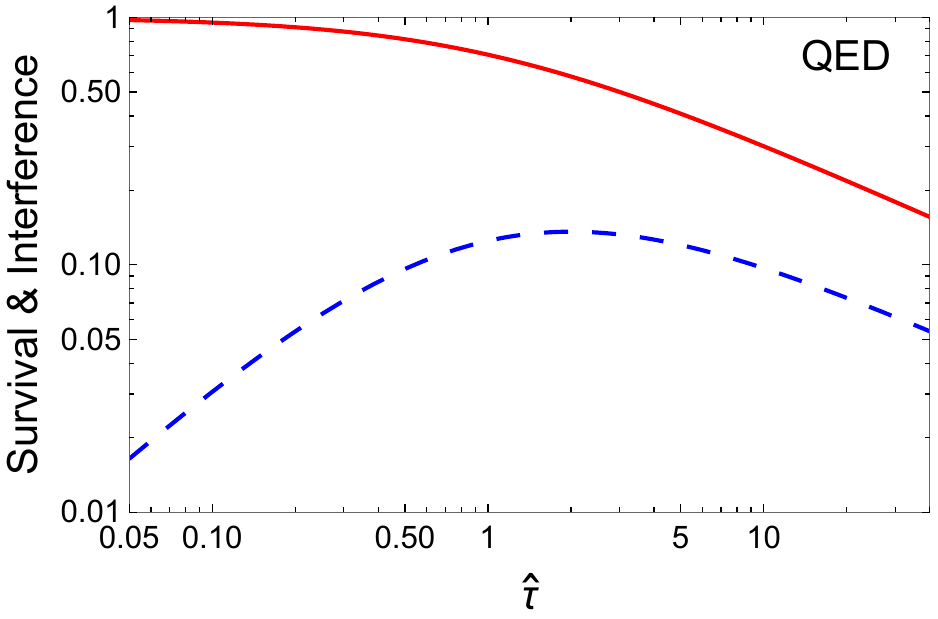}
\caption{Left: 1S survival Eq.~(\ref{eq:QCDp1SL2alone}) as a function of the dimensionless time $\hat{\tau}=\Gamma_{\!_{\rm 1S}} t$ (solid red), and the matrix element $\langle {\rm 2S}|\mathcal{D}_{\rm s}(t)|{\rm 1S}\rangle$, for the  initial conditions $\mathcal{D}_{\rm s}(t=0)=|{\rm 1S}\rangle \langle {\rm 1S}|$ (long-dashed blue). Right: Same for the QED case (which can be easily obtained by setting $N_c=1$ in Eq.~(\ref{eq:1S2Scoherence})).}
\label{fig:survandinterf}
\end{figure*}

Consider now the non diagonal matrix element  $\langle {\rm 2S}|\mathcal{D}_{\rm s}(t)|{\rm 1S}\rangle$, and choose as initial state $\mathcal{D}_{\rm s}(0)=|{\rm 1S}\rangle \langle {\rm 1S}|$. The density matrix at any time $t$ is then obtained  from the approximate expression (\ref{eq:Ussapprox}). The resulting expression is given  by
\beq\label{eq:1S2Scoherence}
\langle {\rm 2S}|\mathcal{D}^{_{\rm (1S)}}_{\rm s}(t)|{\rm 1S}\rangle=\frac{N_c^2-1}{N_c^2} \; \frac{\bar{\tau}/\sqrt{2}}{ (1+ \bar{\tau})^2}
+ \frac{1}{N_c^2} \; \frac{\hat{\tau}/\sqrt{2}}{2  (1+\hat{\tau})^{3/2}}, 
\eeq 
where the ``(1S)" superscript recalls the initial state. 
The expression (\ref{eq:1S2Scoherence}) shows that the matrix element $\langle {\rm 2S}|\mathcal{D}^{_{\rm (1S)}}_{\rm s}(t)|{\rm 1S}\rangle$ evolves in time with the same time scales as the 1S-survival probability. In particular, it decreases at late time as $\tau^{-1/2}$ ,what is in fact true irrespective of the initial condition. This behavior is illustrated on Fig.~\ref{fig:survandinterf}.  The comparison with the QED calculation shown in the right panel indicates that the same basic mechanisms are at work in the non-abelian case. The slow power law decay shows that in the basis of the vacuum eigenstates states, effects due to quantum coherence can survive on a long time scale. This is to be contrasted with the exponential squeezing of the non-diagonal elements of the density matrix in coordinate space. 

\section{Equilibrium limit of the momentum distribution}
\label{app:equilibrium}
In this appendix, we study the equilibrium limit of the momentum distribution. To keep the presentation simple, we consider the abelian case. We emphasize however that identical results are obtained in QCD (as long as the energy gap between bound states can be ignored). We  assume that once equilibrium is reached (and the temperature is large enough), the bound states play a minor role, the quark and the antiquark occupying continuum states. In this regime, one expects a semi-classical approximation to be valid, and accordingly we treat the quantity  $y=s'-s \sim \lambda_{\rm th}=\frac{1}{\sqrt{M T}}$ as being small. Going back to the various Lindblad superoperators in App.~\ref{app:transition_operators},  one then expands those of the form $\tilde{W}^{(n)}(\frac{y}{2})$ in powers of $y$, while others such as $\tilde{W}^{(n)}(s)$, $\tilde{W}^{(n)}(s')$ or $\tilde{W}^{(n)}(\frac{s+s'}{2})$, where $s\sim s'\sim r\gg \lambda_{\rm th} $ are typically suppressed by powers of $\lambda_{\rm th}/r$ and can be neglected. Besides, we only retain the derivatives of $\mathcal{D}$ with respect to $y$ since $\mathcal{D}$ is quasi uniform along $r\gg \lambda_{\rm th}$  once equilibrium is reached. The following expressions for the minimal set of Lindblad superoperators are then obtained, up to quadratic terms in $y$:
\begin{equation}
    \mathcal{L}_2 \approx - \frac{\tilde{W}''(0)}{4}\,y^2\,\mathcal{D}
\label{eq:equilL2MS}
\end{equation}
\begin{equation}
    \mathcal{L}_3 \approx  -\frac{\tilde{W}''(0) }{2MT} y \partial_y \mathcal{D} 
\label{L3MS}
\end{equation}
\begin{equation}
    \mathcal{L}_4 \approx  -\frac{\tilde{W}^{(4)}(0)}{64 M^2T^2}   y^2 \partial_y^2 \mathcal{D}.
\label{eq:equilL4MS}
\end{equation}
If one solely considers the effect of $\mathcal{L}_2 $ and $\mathcal{L}_3$, the unique non trivial solution of $(\mathcal{L}_2 + \mathcal{L}_3)\mathcal{D}_{\rm asymp}=0 $ is $\mathcal{D}_{\rm asymp}\propto \exp(-\frac{M T y^2}{4})$, from which on deduces that the  Wigner transform of ${\cal D}_{\rm asymp}$ is proportional to $\exp(-\frac{p^2}{MT})$, with the expected average $\langle p^2 \rangle = \frac{M T}{2}$. By also including the contribution of $\L_4$, one finds that the equation $(\mathcal{L}_2 + \mathcal{L}_3 + \mathcal{L}_4)\mathcal{D}_{\rm asymp} =0 $ is  equivalent to a modified Bessel equation. This leads to a more complicated solution for the  Wigner transform, implying deviations from the Maxwell-Boltzmann distribution at large momenta. Note however that for the mere evaluation of $\langle p^2 \rangle$, knowing the small-$y$ behavior is enough since $\langle p^2 \rangle=-\frac{\mathcal{D}''(0)}{\mathcal{D}(0)}$. Writing $\mathcal{D}_{\rm asymp}(0)=1-\frac{a_2}{2} y^2 + o(y^2)$ and injecting this expression in the $(\mathcal{L}_2 + \mathcal{L}_3 + \mathcal{L}_4)\mathcal{D}_{\rm asymp} =0 $ equation, one obtains
\begin{equation}
a_2 = \frac{M T}{2+\gamma}\quad\text{with}\quad \gamma =  \frac{\tilde{W}^{(4)}(0)}{16 MT\tilde{W}''(0)}.
\end{equation}
We have checked that these deviations from the Maxwell-Boltzmann distribution, represented by the factor $\gamma$ in the above equation,\footnote{We may also interpret this correction as an effective modification of the temperature, $T\mapsto T_{\rm eff}= 2T/(2+\gamma)$.} are not due to the restriction to the minimal set of superoperators that we have used, but they hold more broadly for the QME used in this paper and in~\cite{Blaizot:2017ypk}.\footnote{Such deviations are indeed also present in the equilibrium limit of the Lindblad equation with the $\mathcal{L}_2$ and $\mathcal{L}_3$ superoperators defined in App.~B of~\cite{Blaizot:2017ypk}, where the action of $\mathcal{L}_3$ on $\mathcal{D}$ is evaluated as $\approx -\frac{1 }{2MT} \left(W''(0) y \partial_y \mathcal{D} + \frac{W^{(4)}(0)}{8} y^2 \mathcal{D}\right) $, with an extra term $\propto W^{(4)}(0)$ as compared to Eq.~(\ref{L3MS}).}

At this point one may note that the 2nd and 4th derivatives of the imaginary potential at the origin are crucial ingredients for the determination of the asymptotic value of $\langle p^2 \rangle$. For the HTL potential defined in  Eq.~(\ref{eq:WHTL}), the 2nd derivative at the origin is divergent and the integral adopted in the definition (\ref{eq:defphi}) of $\phi$ needs to be regulated. Adopting a UV cut-off $\Lambda\gg 1$, one gets 
\begin{equation}
{\phi}(\tilde{r};\Lambda)=2\int_0^{\Lambda} dz \frac{z}{(z^2+1)^2}\, \left(1 - \frac{\sin(\tilde{r}\,z)}{\tilde{r} z} \right)=\frac{c_2(\Lambda)}{3!} {\tilde r}^2 -\frac{c_4(\Lambda)}{5!} {\tilde r}^4  + \cdots,
\label{eq:defphireg}
\end{equation}
where $\tilde r=r m_D$, $z=q/m_D$ and $\Lambda=q_{\rm max}/m_D$. Then, recalling that $W=\alpha \phi$ (see  Eq.~(\ref{eq:WHTL})) we obtain the derivatives of $W$ w.r.t. $r$:
\begin{equation}
\mathcal{H}(0)=W''(0) = \frac{\alpha T m_D^2}{3}\, c_2(\Lambda)\quad \text{and}\quad 
W^{(4)}(0) = -\frac{\alpha T m_D^4}{10}\, c_4(\Lambda)
\end{equation}
where
$$
c_2(\Lambda)=2\int_0^{\Lambda} dz \frac{z^3}{(z^2+1)^2}\approx \ln(\Lambda^2) \quad \text{and}\quad c_4(\Lambda)=2\int_0^{\Lambda} dz \frac{z^5}{(z^2+1)^2}\approx \Lambda^2, $$
leading to 
\begin{equation}
\mathcal{H}(0)=\frac{2\alpha T m_D^2}{3}\ln\frac{q_{\rm max}}{m_D},\quad \text{and}\quad 
W^{(4)}(0) = -\frac{\alpha T m_D^2 q_{\rm max}^2}{10}\,.
\label{eq:W2W4}
\end{equation}
Due to the quartic dependence of $W^{(4)}(0)$ on $q_{\rm max}$, the choice of the value of $q_{\rm max}$ can affect significantly the value of $\langle p^2 \rangle$ at equilibrium. Let us notice that this is true even in the weak coupling case as the $\frac{\tilde{W}^{(4)}(0)}{\tilde{W}''(0)}$ ratio that appears to be responsible for $\gamma \neq 0$ does not depend on the coupling between the heat bath and the subsystem.

Besides the regularization issues just discussed, there is another potential source of deviation between the QME steady state and the canonical Maxwell-Boltzmann law: a potential mismatch between the fluctuation and the dissipation terms (respectively denoted $\mathcal{L}_2$ and $\mathcal{L}_3$ in our work). Quite generally, the structures of $\mathcal{L}_2\propto W$ and $\mathcal{L}_{3,{\rm lin}}\propto \nabla W \cdot \nabla$ conspire to reach an exact Maxwell-Boltzmann law in the semi-classical approximation. However, adding a $\mathcal{L}_{3,{\rm cst}}$ contribution is equivalent to shifting the imaginary potential in $\mathcal{L}_2$ from $W\to W + \frac{\Delta W}{4MT}$. It is to be viewed effectively as a fluctuation term. If such shift is not  ``transferred" to the terms $\propto \nabla W$ in the $\mathcal{L}_{3,{\rm lin}}$ structure, this may induce  deviations with respect to the Maxwell-Boltzmann law. This source of deviation is absent in the present work because of our restriction to the ``minimal set'' of superoperators. However, it may be at the origin of the  discrepancies $\propto \frac{T}{M}$ that are for instance observed in App.~C of \cite{Miura:2019ssi} (more specifically Eq.~(C2) in \cite{Miura:2019ssi}  illustrates the imbalance between the $\mathcal{L}_{\rm cst}$ and $\mathcal{L}_{\rm lin}$ structures, respectively based on $F_1$ and $\vec{F}_2$).

\bibliographystyle{hieeetr}
\bibliography{main}

\end{document}